\documentstyle[amsfonts,amssymb,epsfig,12pt]{article}

\makeatletter

\renewcommand \thesection {\@arabic\c@section}
\renewcommand\thesubsection   {\thesection.\@arabic\c@subsection}
\renewcommand\thesubsubsection{\thesubsection .\@arabic\c@subsubsection}
\renewcommand\theparagraph    {\thesubsubsection.\@arabic\c@paragraph}
\renewcommand\section{\@startsection {section}{1}{\z@}%
                                   {-3.5ex \@plus -1ex \@minus -.2ex}%
                                   {1.9ex \@plus.2ex}%
                                   {\normalfont\large\bfseries\centering}}
\renewcommand\subsection{\@startsection{subsection}{2}{\z@}%
                                     {-2ex\@plus -1ex \@minus -.2ex}%
                                     {1.2ex \@plus .2ex}%
                                  {\normalfont\normalsize\bfseries\centering}}
\renewcommand\subsubsection{\@startsection{subsubsection}{3}{\z@}%
                                     {-2ex\@plus -1ex \@minus -.2ex}%
                                     {.5ex \@plus .2ex}%
                                     {\normalfont\normalsize\em}}
\renewcommand\paragraph{\@startsection{paragraph}{4}{\z@}%
                                    {3.25ex \@plus1ex \@minus.2ex}%
                                    {-1em}%
                                    {\normalfont\normalsize\em}}
\renewcommand\subparagraph{\@startsection{subparagraph}{5}{\parindent}%
                                       {3.25ex \@plus1ex \@minus .2ex}%
                                       {-1em}%
                                      {\normalfont\normalsize\em}}
\makeatother

\newcounter{subequation}
        \newenvironment{subequation}%
        {\addtocounter{equation}{-1}%
        \stepcounter{subequation}%
        \begin{equation}}%
        {\end{equation}%
}

\newcommand{\resetseq}{\setcounter{subequation}{0}}

\newcommand{\beq}{\begin{equation}}
\newcommand{\eeq}{\end{equation}}
\newcommand{\bseq}{\begin{subequation}}
\newcommand{\eseq}{\end{subequation}}
\newcommand{\bea}{\begin{eqnarray}}
\newcommand{\eea}{\end{eqnarray}}

\newcommand{\refeq}[1]{(\ref{#1})}

\newcommand{\cf}{{\textrm{cf.\ }}}

\newcommand{\ie}{{\textrm{i.e.\ }}}

\newcommand{\wrt}{{\textrm{w.r.t.\ }}}

\newcommand{\rhs}{{\textrm{right-hand\ side\ }}}

\newcommand{\supp}{{\mathrm{supp}\, }}

\newcommand{\Id}{{\mathrm{Id}\,}}

\newcommand{\bL}{{\mathrm{bL}}}
\newcommand{\bLd}{{\mathrm{bL}^{\!*}}}
\newcommand{\Ld}{{\mathrm{L}^{\!*}}}

\newcommand\abs[1]{{| #1 |}}
\newcommand\Abs[1]{{\left| #1 \right|}}
\newcommand\norm[1]{{\| #1 \|}}
\newcommand\Norm[1]{{\left\| #1 \right\|}}
\newcommand\dLIPnorm[1]{\norm{ #1}_{\Ld}}
\newcommand\dLIPnormT[1]{\norm{ #1}_{\widetilde\Ld}}

\newcommand\LipSnorm[1]{{{\mathrm{Lip}}\left( #1 \right)}}

\newcommand\LipSnormSMALLitera[1]{{{\mathrm{Lip}}^{\circ 2}( #1 )}}
\newcommand\LipSnormITERA[1]{{{\mathrm{Lip}}^{\circ 2}\!\left( #1 \right)}}

\newcommand\HalphaSnorm[1]{{{\mathrm{H\ddot{o}l}}_\alpha\!\left( #1 \right)}}

\newcommand\dbL[1]{{{\mathrm{dist}}_{\bLd}\!\left( #1\right)}}
\newcommand\dLip[1]{{{\mathrm{dist}}_{\Ld}\!\left( #1\right)}}
\newcommand\dKRc[1]{{{\mathrm{dist}}_{\mathrm{KRc}}\!\left( #1\right)}}
\newcommand\dKR[1]{{{\mathrm{dist}}_{\mathrm{KR}}\!\left( #1\right)}}

\newcommand{\dotH}{{{\dot H}^1}}
\newcommand{\dotHH}{{{\dot H}^2}}
\newcommand{\dotHk}{{{\dot H}^k}}

\newcommand{\probL}{{{{L}}^1_{+,1}}}
\newcommand{\probLpx}{{L^{1,1}_{+,1}}}

\newcommand{\wM}{{\widetilde{M}_1}} 
\newcommand{\wMplus}{{{\widetilde{M}_{1,+}}}}

\newcommand{\Cw}{C_{\mathrm{w}}}

\newcommand{\wPpx}{{\widetilde{P}_1}}

\newcommand{\GaB}{\Gamma_{\mathrm{B}}}
\newcommand{\GaBpx}{\Gamma_{\mathrm{B},1}}
\newcommand{\wGaBpx}{{\widetilde\Gamma}_{\mathrm{B},1}}

\newcommand{\inlawto}{{\stackrel{{\cal L}}{\longrightarrow}}}
\newcommand{\statelimWIGGLE}{{\stackrel{\wGaBpx}{\longrightarrow}}}

\newcommand{\crprod}{{\times}} 

\newcommand{\veps}{\varepsilon}

\newcommand{\cC}{{\cal C}}
\newcommand{\cD}{{\cal D}}
\newcommand{\cE}{{\cal E}}
\newcommand{\cF}{{\cal F}}
\newcommand{\cG}{{\cal G}}
\newcommand{\cH}{{\cal H}}
\newcommand{\cJ}{{\cal J}}

\newcommand{\cM}{{\cal M}}

\newcommand{\cP}{{\cal P}}
\newcommand{\cS}{{\cal S}}

\newcommand{\EE}{{\Bbb E}}
\newcommand{\II}{{\Bbb I}}
\newcommand{\NN}{{\Bbb N}}

\newcommand{\RR}{{\Bbb R}}
\newcommand{\SS}{{\Bbb S}}

\newcommand{\sympJ}{{\mathbf{J}}}

\newcommand{\pr}{\prime}
\newcommand{\prpr}{{\prime\prime}}

\newcommand{\QED}{$\quad$\textrm{Q.E.D.}}

\newcommand{\fe}{\varrho}

\newcommand{\iint}{{\int\!\!\!\int}}

\newcommand{\SSint}{{-\!\!\!\!\!\!\int_{\SS^2} }}

\newcommand{\supN}{\scriptscriptstyle{(N)}}

\newcommand{\dd}{{\mathrm{d}}}
\newcommand{\ddt}{\frac{\dd}{\dd t}}
\newcommand{\ddtsq}{\frac{\dd^2 }{ \dd t^2}}

\newcommand{\zetaFREE}{\zeta^{\mathrm{free}}}

\newcommand{\pdt}{{\partial_t^{\phantom{0}}}}
\newcommand{\pdx}{{\partial_x^{\phantom{0}}}}
\newcommand{\pdy}{{\partial_y^{\phantom{0}}}}
\newcommand{\pdz}{{\partial_z^{\phantom{0}}}}

\newcommand{\pdp}{{\partial_p^{\phantom{0}}}}

\newcommand{\pdtsq}{{\partial_t^2}}
\newcommand{\pdxsq}{{\partial_x^2}}

\newcommand{\wcrit}{{\underline{w}}}

\newcommand{\Ccrit}{{\underline{C}}}

\newcommand{\olF}{{\overline{F}}}

\newtheorem{defn}{Definition}[section] 

\newtheorem{Coro}[defn]{\sc Corollary}

\newtheorem{Lemm}[defn]{\sc Lemma}
\newtheorem{Prop}[defn]{\sc Proposition}
\newtheorem{Rema}[defn]{\sc Remark}
\newtheorem{Theo}[defn]{\sc Theorem}

\newcommand\bz{{\mathbf{z}}}
\newcommand\bzN{\bz^{\supN}}
\newcommand\bzj{\bz^{(j)}}
\newcommand\bzu{\bz^{(1)}}
\newcommand\bzd{\bz^{(2)}}
\newcommand\bzuu{\tilde\bz{}^{(1)}}
\newcommand\bzud{\tilde{\tilde\bz}{}^{(1)}}

\newcommand\VXi{{\vec{\Xi}}}
\newcommand\zett{{\mathfrak{z}}}
\newcommand\Zett{{\mathfrak{Z}}}
\newcommand\zet{{{z}}}
\newcommand\Vzet{{\vec{z}}}
\newcommand\Zet{Z}
\newcommand\VZet{{\vec{Z}}}
\newcommand\Vt{{\vec{t}}}

\newcommand\opQ{{\stackrel{\leftrightarrow}{Q}}}

\begin{document}

\title{The Vlasov limit and its fluctuations for a system of
       particles which interact by means of a wave field}

\author{
    Y. Elskens$^a$,
    M.K.-H. Kiessling$^{a,b}$,
    V. Ricci$^c$\\
  $^a$ \textit{Equipe turbulence plasma, case 321, PIIM,}\\
       \textit{UMR 6633 CNRS-universit\'e de Provence,
               campus Saint-J\'er\^ome,}\\
       \textit{F-13397 Marseille cedex 13}\\
       {\tt elskens@up.univ-mrs.fr}\\
  $^b$ On sabbatical leave from:
       \textit{Department of Mathematics,}\\
       \textit{Rutgers, The State University of New Jersey,}\\
       \textit{110 Frelinghuysen Rd., Piscataway, NJ 08854}\\
       {\tt miki@math.rutgers.edu}\\
  $^c$ \textit{Dipartimento di Metodi e Modelli Matematici},\\
       \textit{Universit\`a  di Palermo}\\
       \textit{Viale delle Scienze Edificio 8, 90128 Palermo}\\
       {\tt vricci@mat.uniroma1.it}\\  
 }
\date{Version of: \today .}

\maketitle

\begin{abstract}
\noindent
    In two recent publications,
                [Commun. PDE \textbf{22}, 307--335 (1997),
                Commun. Math. Phys. \textbf{203},  1--19 (1999)],
A. Komech, M. Kunze and H. Spohn studied the joint dynamics
of a classical point particle and a wave type generalization of
the Newtonian gravity potential, coupled in a regularized way.
    In the present paper the many-body dynamics of this model
is studied.
    The Vlasov continuum limit is  obtained in form equivalent
to a weak law of large numbers.
    We also establish a central limit theorem for the fluctuations
around this limit. 
\end{abstract}

\vfill
\hrule
\smallskip
\noindent \copyright{2005} The authors. Reproduction of this
article, in its entirety, for non-commercial purposes is permitted.

\newpage

\section{Introduction}
\label{intro}

    In recent years, significant progress has been
made on the Cauchy problem of relativistic kinetic 
theory.\footnote{Beside these physical Vlasov models, 
       		 also the ``relativistic Vlasov--Poisson equations'' 
                \cite{GlasseySchaefferA}
	        and more recently the Vlasov--Nordstr\"om equations
		\cite{CalogeroReinI,CalogeroReinII} 
		have been studied.}
  The special-relativistic Vlasov--Maxwell equations
\cite{LuchinaVlasov,
        VlasovBOOK},
with applications in theories of astrophysical plasma waves
  \cite{SchindlerJanicke,
        Janicke},
are treated with rigor in
\cite{HorstHABIL,
        GlasseyStraussA,
        GlasseyStraussB,
        GlasseyStraussC,
        GlasseySchaefferB,
        diPLioB,
        ReinA,
        BoGoPaI,
	BoGoPaII,
        KlaSta};
the general-relativ\-istic Vlasov--Einstein equations
\cite{EhlersA,
        EhlersB},
which play a r\^{o}le in models of cosmological evolutions
  \cite{Bernstein},
have been treated rigorously in
\cite{ReinRendall,
        Rendall,
        RRS,
        ReinHABIL};
see also 
\cite{AnguigeB,
        AnguigeC}.
 The relativistic Vlasov--Maxwell and Vlasov--Einstein
equations both reduce in their strictly non-relativistic limits
to corresponding Vlasov--Poisson equations
         \cite{VlasovA},
for which the classical Cauchy problem has been settled
\cite{PfaffDISS,
        Schaeffer,
        Pfaff}.
        Much of the special-relativistic material is reviewed in
\cite{Glassey}, 
the non-relativistic results in
\cite{ReinB}.

  Progress on the microscopic foundations of all these
Vlasov models has been lagging behind in comparison.
 Regularized Vlasov--Poisson equations have been derived
through a continuum limit for a family of classical $N$-body problems 
with regularized Coulomb and Newton interactions, see
\cite{NeunzertA, NeunzertB} and \cite{BraunHepp}.
        In
\cite{BraunHepp} also a law of large numbers (LLN) 
and a central limit theorem (CLT) for the
fluctuations around this Vlasov limit are proven; see also 
	\cite{SpohnBOOKa, CerIllPul}
for further discussions.
 The global regularity results of \cite{PfaffDISS} should definitely
allow one to remove the regularization after the Vlasov limit of
the regularized $N$-body dynamics has been taken and to obtain the
proper (\ie non-regularized) Vlasov--Poisson equations
          \cite{VlasovA},
but we are not aware of work where this has been done explicitly.
 In any event, while mathematically clean, physically such a derivation
of the proper Vlasov--Poisson equations is still far from
satisfactory, for it uses the wrong order of limits, physically speaking.
 The regularization should actually be removed while taking the
Vlasov limit for the regularized dynamical system, which likewise seems
feasible with current techniques, but as far as we know has not yet
been done either; however, see
   \cite{KunzeRendallA, KunzeRendallB}
for relevant work on the expected radiation-reaction corrections to 
Vlasov--Poisson and other Vlasov models.
 Another open question is whether one can obtain the proper Vlasov--Poisson
equations directly from the classical Newtonian $N$-body problem for
Newton or Coulomb interactions without any regularization, essentially 
because the classical $N$-body problem is still not well-controlled.
 For a derivation of the classical Vlasov--Poisson equations from 
a regularized quantum mechanical $N$-particles model, see 
   \cite{NarSew};
we also mention a derivation of a Schr\"odinger--Poisson model from an $N$-particles
quantum model without regularization, see
   \cite{BardosEtal}.
 While there is thus plenty of mathematical work left to be done 
on the microscopic foundations of the non-relativistic Vlasov--Poisson 
equations, their status is moderately well established.
 The microscopic foundations of the relativistic Vlasov--Maxwell 
and Vlasov--Einstein equations 
by contrast seem not to have been established with rigor in any form.

 To bring about a modest change in the state of affairs of the 
microscopic foundations
of relativistic Vlasov theory, in this paper we prove a LLN and 
a CLT for a regularization of the following (almost) special-relativistic 
generalization of the Vlasov--Poisson equations for a self-gravitating 
system,\footnote{We use natural dimensionless units to avoid burdening 
	the equations with irrelevant dimensional constants.           
	Conversion of equations \refeq{VfEQ}--\refeq{rhoASfINT} to the
	more conventional Gaussian units for a ``gravitational'' system
	       is effected by replacing
	       $t\mapsto Nct$,
	       $x\mapsto Nx$,
	       $v\mapsto v/c$,
	       $p\mapsto p/(mc)$,
	       $\phi\mapsto\phi/c^2$,
	       $\rho\mapsto 4\pi G m \rho/(Nc^2)$,
	       $f\mapsto 4\pi G c m^4 f/N$;
	       here, $c$ is the speed of light,
	       $G$ is Newton's constant of universal gravitation,
	       $N$ is the total number of particles in the system,
	       and $m$ is the empirical mass of a particle.
	       Note that $\rho$ and $f$ retain their normalization 
	       as probability densities on $\RR^3$ and $\RR^6$.}
comprising the continuity equation 
\beq
        \biggl(
\pdt + v \cdot\pdx - \partial_{x}\phi(x,t)\cdot\partial_{p}
        \biggr) 
	       f(x,p,t)
                       = 0  
\label{VfEQ}
\eeq
on $x,p$ phase space, where
\beq
  v = \frac{p}{\sqrt{1+|p|^2}}
\label{vOFp} 
\eeq 
is the velocity of a generic particle with momentum $p$ 
and empirical mass of unity,
and the inhomogeneous wave equation
\beq
     \square {\phi}(x, t) =  \rho(x,t)
\label{VphiEQ} 
\eeq 
on $x$ space, where $\square = -\pdtsq +  \pdxsq$ is the 
d'Alembertian,\footnote{We write $\pdxsq\equiv$ div grad rather
		than $\Delta$, for $\Delta$ is used with a different 
		meaning later on.}
and where
\begin{equation}
  \rho(x,t)
  =
  \int_{\RR^3}\! f(x,p,t)\dd{p} 
\label{rhoASfINT}
\end{equation}
is the normalized density of particles 
attributed to the space point $x\in\RR^{3}$ at time $t\in\RR$.
 Clearly, $\phi(x,t)$ is a wave-like generalization of the 
Newtonian gravity potential 
generated by $\rho(x,t)$, and $f(x,p,t)$ in turn is the normalized
density of particles attributed to the phase-space point
$(x,p)\in\RR^{3}\times\RR^3$ at time $t\in\RR$.
 We remark that although a normalized density $f(\,.\,,\,.\,,t)$ 
formally satisfies the definition of a probability density function, 
$f(\,.\,,\,.\,,t)$ is more properly thought of as (an approximation to) the 
actual empirical phase space density of particles for an individual system.

 It is to be noted that our set of equations 
\refeq{VfEQ}, \refeq{vOFp}, \refeq{VphiEQ}, \refeq{rhoASfINT}
is not meant to be taken physically seriously in itself; 
in particular, the model is not manifestly Lorentz-covariant 
(more on that in a moment).
 Its derivation from a microscopic model mainly serves as a simpler 
primer for the derivation of the special-relativistic Vlasov--Maxwell 
equations, which we undertake in a sequel to this paper.
 Indeed, the model 
\refeq{VfEQ}, \refeq{vOFp}, \refeq{VphiEQ},  \refeq{rhoASfINT} 
is a simple truncation of the usual set of special-relativistic 
Vlasov--Maxwell equations for a single species of (say, positive unit 
charge) particles, obtained as follows:\footnote{To obtain this 
       ``truncated Vlasov--Maxwell system'' in the conventional 
	Gaussian units, replace
	       $t\mapsto Nct$,
	       $x\mapsto Nx$,
	       $v\mapsto v/c$,
	       $p\mapsto p/(mc)$,
	       $\phi\mapsto -e\phi/(mc^2)$,
	       $\rho\mapsto 4\pi e^2 \rho/(Nmc^2)$,
	       $f\mapsto 4\pi e^2 c m^2 f/N$
	in our dimensionless equations 
        \refeq{VfEQ}, \refeq{vOFp}, \refeq{VphiEQ}, \refeq{rhoASfINT};
	here, $c$, $m$, and $N$ have the same meaning as for the 
	``gravitational'' system, while $e$ is the empirical unit 
	of charge of a particle.}
in the Vlasov--Maxwell equations, the electromagnetic fields $E$ and $B$ 
are expressed in terms of the electromagnetic potentials $\phi$ and $A$
as $B=\nabla\times A$ and $E= -\partial_t A -\nabla \phi$, gauged by the 
Lorentz--Lorenz condition $\partial_t\phi +\nabla\cdot A =0$; 
one then purges the inhomogeneous vector wave equation for $A$ 
and all terms involving $A$ (or rather its derivatives) in the Lorentz force.
 Curiously, and contrary to what one might have naively thought, 
this mutilation of the Vlasov--Maxwell equations does not result 
in a model which approximates quasi-electrodynamical behavior without 
magnetic fields, but in one which rather mimics some quasi-gravitational 
system, for in the strictly non-relativistic limit the model formally 
reduces to the Vlasov--Poisson equations for a Newtonian gravitational system.

 We remark that the replacement 
$\partial_{x}\phi(x,t)\to \partial_{x}\phi(x,t)/ \sqrt{1+|p|^2}$ 
in \refeq{VfEQ} results in an essentially Lorentz-covariant model 
with scalar interaction field $\phi$.
 We say `essentially' because this modification of equations 
\refeq{VfEQ}--\refeq{rhoASfINT} is still not manifestly Lorentz-covariant
when $\phi$ is interpreted as a Lorentz scalar field, for the right-hand 
side of  \refeq{VphiEQ} when taken face value is the time component of a 
Minkowski vector.
 However, the model becomes manifestly Lorentz-covariant when this set of 
equations is supplemented by the constraint 
$V\equiv\int_{\RR^3}\!j(x,t)\dd{x} =0$, where
\begin{equation}
  j(x,t)
  =
  \int_{\RR^3}\!v f(x,p,t)\dd{p} 
\label{jASfINT}
\end{equation}
is the mass current vector density, and the right-hand side of \refeq{VphiEQ} 
is interpreted as the Minkowski scalar $M\rho - Vj$ in the center-of-mass 
frame, in which $V =0$ and $M\equiv \int_{\RR^3}\!\rho(x,t)\dd{x} =1$.
 As does our truncated Vlasov--Maxwell model, the Vlasov model with a 
factor $1/ \sqrt{1+|p|^2}$ multiplying $\partial_{x}\phi(x,t)$ in 
\refeq{VfEQ} formally reduces to the Vlasov--Poisson equations for a 
Newtonian gravitational system in the strictly non-relativistic limit.
 While this model has a number of interesting features, we will not 
pursue it further here because it is less relevant to the
Vlasov--Maxwell equations.

 Ideally, we would like to prove that the kinetic equations 
\refeq{VfEQ}--\refeq{rhoASfINT}
constitute a LLN for the dynamics of an atomistic
system of $N$ classical point particles that interact by means of a 
wave gravity field.
 The natural candidate for this atomistic system is suggested by
``atomizing'' the characteristic system for  \refeq{VfEQ}, which reads
\bea
&&
        \frac{\dd{q}}{\dd{t}}
=
        \frac{p}{\sqrt{1 +|p|{}^2}}
\,,
\label{CHqDOT}\\
&&
        \frac{\dd{p}}{\dd{t}}
=
        - \nabla\phi\left(q,t\right)
\,, \label{CHpDOT} 
\eea 
with $\phi(x,t)$ the wave field for \refeq{VfEQ}--\refeq{rhoASfINT}.
 Thus, interpreting $f$ as an empirical atomic measure of $N$ 
classical point particles, having positions $q_i^{\supN}(t)$ and 
momenta $p_i^{\supN}(t)$ at time $t$, these particle motions satisfy
the characteristic equations of motion, viz.
\bea
&&
        \dot{q}_i^{\supN}(t)
=
        \frac{p_i^{\supN}(t) }{ \sqrt{1 +\bigl|p_i^{\supN}(t)\bigr|{}^2}}
\,,
\label{qDOT}
\\
&&
        \dot{p}_i^{\supN}(t)
=
        - \nabla\phi^{\supN}\!\left({q}_i^{\supN}(t),t\right)
\,,
\label{pDOT}
\eea
for a $\phi^{\supN}$ which satisfies the inhomogeneous wave equation
\beq
        \square\phi^{\supN}(x, t)
=
        {\textstyle{\frac{1}{N}\sum_{i=1}^{N}}}\delta_{q_i^{\supN}(t)}(x)
\,.
\label{WAVEeqNdelta}
\eeq
 Unfortunately, this system of equations has only a symbolic character, 
at best.  
        Since $\phi^{\supN}$ is a distributional solution of
\refeq{WAVEeqNdelta} which is not in $H^1(B)$ for any open ball
$B$ containing the location of a point particle, each particle is
surrounded by an infinite field energy which equips the particles
with an infinite inertia (via Einstein's $E=mc^2$); furthermore,
the right-hand side of \refeq{pDOT} is not well-defined.
  Infinite self-interaction terms are encountered also if one applies the
above line of reasoning in the context of the microscopic foundations of 
the strictly non-relativistic Vlasov--Poisson equations, but in that case 
the self-interactions are not dynamical, and simply  discarding them formally 
yields a locally well-defined and consistent dynamical system.
 In a local relativistic theory such a formal omission of the self-interaction
terms is not justified because of the dynamical radiation-reaction.
 Hence, before any classical microscopic derivation based on point particles 
can be attempted for the relativistic Vlasov--Maxwell, Vlasov--Einstein, and 
for that matter also for the simpler Vlasov equations considered here,
one first has to overcome the even harder conceptual problem of 
setting up a well-defined microscopic relativistic model with
point particles.
 While this is being sorted out,\footnote{For recent progress on
	relativistic microscopic classical electromagnetic theory, 
	see  \cite{Kie}.}
it is still of interest to study the obvious amelioration of the 
infinite self-interactions dilemma by regularizing the ill-defined
point particle models \cite{SpohnBOOKb}.

 In this vein, we follow
   \cite{KoSpKu,
          KoKuSp},
who discussed a regularized version of the symbolic equations 
\refeq{qDOT}, \refeq{pDOT}, \refeq{WAVEeqNdelta} with $N=1$.
 They ``smear'' the instantaneous location
$q(t)\in {\RR}^3$ of a particle with a probability density function
$\fe(\,.\,):\RR^3\to\RR_+$.
 Consistency requires that in Newton's equation, 
the gradient of $\phi$ for point sources is replaced by the 
$\fe$-average of the gradient of a $\phi_\fe$ for $\fe$-averaged point sources.
 The global existence and uniqueness of the dynamics for the
regularized microscopic model with $N=1$ 
 \cite{KoSpKu, KoKuSp}
is readily extended to arbitrary $N$, with uniform control in $t$.
 It should be noted that the regularization just described is non-relativistic.

 Interestingly, one of the caveats of the similarly regularized 
electromagnetic models discussed in
\cite{KomechSpohn,
        KunzeSpohnA,
        KunzeSpohnB,
        KunzeSpohnC,
	BauerDuerr}
that was pointed out in 
\cite{KiePLA}
does {not} occur in the regularized  scalar model of 
   \cite{KoSpKu, KoKuSp}. 
 Namely, in contrast to what is the case in 
the electromagnetic models, the a-priori density function 
$\fe$ does \emph{not} act as a ``source or sink'' for the conventional 
scalar-field angular momentum.
 Thus, conservation of angular momentum holds in its conventional form 
and does not need to be rescued through the cosmetical 
surgery of associating to each particle a spin variable 
(\cf Appendix A.3 of \cite{AppKieAOP} for the electromagnetic models).

  Our main objective in this paper then is to show that the corresponding
regularization of the Vlasov model \refeq{VfEQ}--\refeq{rhoASfINT}
governs a LLN for the regularization of the 
microscopic $N$ particles model with wave gravity interactions
\refeq{qDOT}, \refeq{pDOT}, \refeq{WAVEeqNdelta}.
 To achieve this goal we adapt the strategies of
\cite{NeunzertA,
        NeunzertB},
and
           \cite{BraunHepp}
from the Vlasov--Poisson to our system of equations;
see 
     \cite{SpohnBOOKa}
for an account of Neunzert's proof, and 
     \cite{FirpoElskens}
for an application to a wave modes truncation of the Vlasov 
equations of plasma physics.
 The limit $N\to\infty$ not only yields a LLN for
the regularized Vlasov equations, but also their well-posedness globally 
in time.
 By adapting the strategy of 
        \cite{BraunHepp} 
we also establish a CLT for the fluctuations around the Vlasov limit.
 It goes without emphasis that these ``adaptations'' involve plenty of 
technical and conceptional innovations.

 The removal of the regularization has to be addressed at a later time.
 We expect violations of Lorentz symmetry 
caused by the finite support of $\fe$ to vanish when the regularization 
is removed, either after the Vlasov limit has been taken or along with it.
 Should this expectation turn out to be unfounded, it would become
pointless to try to derive the relativistic Vlasov--Maxwell equations 
along the lines developed here.

\section{The regularized field \&\ $N$-body problem}

  Let $ C_{\rm c}^\infty(\RR^3)$ denote
the infinitely many times continuously differentiable functions 
with compact support.
        In the following, it is assumed that 
$\fe(\,.\,)\in C_{\rm c}^\infty(\RR^3)$.
 For convenience we will also demand that $\fe$ is 
radially symmetric and decreasing.
   For technical reasons \cite{KoSpKu}
a Wiener condition (positive Fourier transform) needs to be imposed on $\fe$.

 We introduce the abbreviation
$
\int = \int_{\RR^3}
$
and the convolution notations
\bea
(\fe * g)(x)
&&\!\!\!\!\!\!\!\!
=
{\int}\! \fe(y -x)g(y) \dd{y}
\,,
\label{EPSconvA}
\\
(\fe * \nabla{g})(x)
&&\!\!\!\!\!\!\!\!
=
{\int}\! \fe(y -x) \pdy{g}(y) \dd{y}
\,,
\label{EPSconvB}
\\
  (\fe\Id *\crprod  \nabla{g})(x) &&\!\!\!\!\!\!\!\!
 = 
{\int}\! \fe(y -x)(y-x) \crprod\pdy{g}(y) \dd{y}
\,, 
\label{EPSconvC}
\eea
where $g: \RR^3 \to \RR$ is any scalar function the derivative 
of which is in $L^2(\RR^3)$.

\subsection{The dynamical system}

 We begin by listing the first-order evolution equations which define the
regularized microscopic dynamical model. 
 Incidentally, the model can be viewed as a Hamiltonian system, on 
which we briefly comment at the end of the next subsection.

 Regularizing the inhomogeneous wave equation for the
microscopic wave gravity potential with point particle sources gives
an inhomogeneous wave equation for the regularized wave gravity
potential.
 Recast as a first-order system for the canonically conjugate 
scalar field variables\footnote{We recall that the homogeneous
		Sobolev spaces $\dotHk(\RR^d)$ are
	        defined	as the closure of $C^k_c(\RR^d)$ \wrt
   $\norm{u}_{\dotHk}^2 =\sum_{|\alpha|=k} \norm{D^\alpha u}_{L^2}^2$, 
		where $C^k_c(\RR^d)$ in turn denotes the $k$ times
		classically differentiable functions with compact support,
		and $\alpha$ is a multi-index
		\cite{GilbargTrudinger}.
		The reason for why we do not work with $H^1(\RR^3)$ is
		\refeq{NphiNULLepsASYMP}:
		functions in $\dotH(\RR^3)$ satisfying 
		\refeq{NphiNULLepsASYMP} are not in $L^2(\RR^3)$.
		However, alternatively we could work with the affine Sobolev
		space 
	$\{\psi:\psi +\textstyle{\frac{1}{4\pi}}\fe*|\,.\,|^{-1}
	\in L^2(\RR^3)\ \&\ \nabla\psi\in L^2(\RR^3)\}$
		with seminorm 
	$|\!|\!|\psi|\!|\!|^2 = \|\nabla\psi\|^2_{L^2} +
	\|\psi +\textstyle{\frac{1}{4\pi}}\fe*|\,.\,|^{-1}\|^2_{L^2}$.}
$\psi^{\supN}(\,.\,,t)\in \dotH(\RR^3)$ and 
$\varpi^{\supN}(\,.\,,t)\in L^2(\RR^3)$ satisfying
\bea
\psi^{\supN}(x,0) 
\!\!\!&=& \!\!\!
-{1}/{4\pi |x|}
\label{NphiNULLepsASYMP}
\\
 \varpi^{\supN}(x,0) 
\!\!\!&=& \!\!\!
0
\label{NvarpiNULLepsASYMP}
\eea
\newpage

\noindent
outside a closed ball $B_R\subset \RR^3$ which contains the 
initial locations of the $N$ particles and the supports of their 
regularizations, the inhomogeneous wave equation becomes
\bea
&&
        \pdt{\psi^{\supN} (x, t)}
=
        \varpi^{\supN}(x, t)
\,,
\label{NphiEQeps}
\\
&&
   \pdt{\varpi^{\supN} (x, t)}
=
   \pdxsq{\psi^{\supN}}(x, t) - \bigl(\fe *\rho^{\supN}_t\bigr) (x)
\,,
\label{NvpiEQeps}
\eea
with
\beq
        \rho^{\supN}_t (\,.\,)
=
{\textstyle{\frac{1}{N}\sum_{i=1}^{N}}}\delta_{q_i^{\supN}(t)}(\,.\,)
\, ;
\label{NrhoEQeps}
\eeq
derivatives are understood in the sense of distributions.
 Note that for given trajectories $t\mapsto {q_i^{\supN}(t)}, i=1,...,N$,
we have just ${\psi^{\supN} (x, t)} = (\fe*{\phi^{\supN}(\,.\,,t)) (x)}$ 
with ${\phi^{\supN}}$ solving \refeq{WAVEeqNdelta}.
 For $i=1,...,N$, the evolution equations for the $i$-th particle's 
canonically conjugate positions $q_i^{\supN}(t)\in\RR^3$ and
momenta $p_i^{\supN}(t)\in\RR^3$ at time $t$, 
are Einstein's law relating relativistic momentum to velocity, 
\beq
    \dot{q}_i^{\supN}(t)
=
    \frac{p_i^{\supN}(t) }{ \sqrt{1+ |p_i^{\supN}(t)|{}^2}}
\,,
\label{NqDOTeps}
\eeq
and Newton's law of motion,
\beq
        \dot{p}_i^{\supN}(t)
=
-\left(\fe *\nabla\psi^{\supN}(\,.\,,t)\right)
\left(q_i^{\supN}(t)\right)
\,.
\label{NpDOTeps}
\eeq

  A complete specification at time $t\in\RR$ of all the first-order
evolutionary variables 
$\left(
        q_1^{\supN}(t),p_1^{\supN}(t);...;q_N^{\supN}(t),p_N^{\supN}(t);
        \psi^{\supN}(\,.\,,t), \varpi^{\supN}(\,.\,,t)
\right)$
constitutes a \textit{physical state} in this model.
 To shorten the notation, we frequently write $z_k^{\supN}(t)$ for 
the particle variables $(q_k^{\supN}(t),p_k^{\supN}(t))$ 
and $\bzN_t$ for $(z_1^{\supN}(t),...,z_N^{\supN}(t))$; furthermore
$\zeta^{\supN}_t$ for the wave field variables
$(\psi^{\supN}(\,.\,,t), \varpi^{\supN}(\,.\,,t))$, yet sometimes
$\zeta[{\bzN_0}]$ rather than $\zeta^{\supN}_0$ for the initial fields
when we want to emphasize their dependence on the initial data 
$\bzN_0$ rather than merely on $N$; finally, we frequently write 
$\zett^{\supN}_t$ for the physical state at time $t$, viz.
\beq 
\zett^{\supN}_t := \left(\bzN_t, \zeta^{\supN}_t \right)
\,.
\label{stateEPS}
\eeq
  The space of all possible physical states is known as the
\textit{system phase space}. 
  To conveniently adapt some results of 
   \cite{KoSpKu}, 
$\Gamma^{\supN}$ is given Hilbert space topology by taking
the Hilbert space direct sum of the particle and 
the field Hilbert spaces,
\beq
  \Gamma^{\supN}
  =
  \underbrace{\RR^3\oplus \dots \oplus \RR^3}_{2N\ \mathrm{terms}}
  \oplus\,  \dotH(\RR^3)
  \oplus L^2 (\RR^3)
\,,
\label{PHASEspaceN}
\eeq
equipped with the conventional Hilbert space inner product
$\langle\,.\,,\,.\,\rangle$ implied by (\ref{PHASEspaceN}).
 The subset of $\Gamma^{\supN}$ on which  \refeq{NphiNULLepsASYMP},
\refeq{NvarpiNULLepsASYMP} is satisfied is denoted $\GaB^{\supN}$.

  The Hilbert space topology of  $\Gamma^{\supN}$ is of course equivalent 
to the Banach space topology for \refeq{PHASEspaceN} interpreted as a Banach 
space direct sum, but the Hilbert space topology is indeed more natural for 
the $N$-body plus field dynamics.
 In contrast, a Banach space topology is the natural one for 
the Vlasov model which we discuss in section \ref{secVlasovIntro}.

 We remark that, while $\dotH(\RR^3)$ and $L^2 (\RR^3)$ allow quite
rough fields $\psi^{\supN}(\,.\,,t)$ and $\varpi^{\supN}(\,.\,,t)$,
any roughness would be inherited from the initial data. 
 To have strong solutions of the wave equation in our case, we demand
$\psi(\,.\,,0)\in (\dotH\cap\dotHH)(\RR^3)$, rather than the 
usual $\psi(\,.\,,0)\in {H}^2(\RR^3)$; cf. \cite{Ikawa}.
 Higher regularity, e.g. as for classical solutions, can also be 
obtained by the usual bootstrapping, if desired.

\subsection{The conservation laws}

  The conventional conservation laws for mass, momentum, angular momentum,
and energy are satisfied for sufficiently regular solutions of
the dynamical system.
 To state the conservation laws, we introduce several functionals
on the system phase space of  generic states 
$(z_1,...,z_N,\zeta)=(\bzN,\zeta) =: \zett^{\supN}\in\Gamma^{\supN}$.

        The mass functional, for $\rho^{\supN}$ given in (\ref{NrhoEQeps}) 
with generic $q_i$, is given by
\begin{equation}
  \cM\left(\zett^{\supN}\right)
  =
  \int  \fe*\rho^{\supN} \dd{x}
\,, 
\label{EPSfuncM}
\end{equation}
the momentum functional by
\begin{equation}
  \cP\left(\zett^{\supN}\right)
  =
  \frac{1}{N}{\textstyle\sum\limits_{i=1}^N}  p_i
  - \int \varpi \pdx\psi \dd{x}
\,, 
\label{EPSfuncP}
\end{equation}
the angular momentum functional by
\begin{equation}
  \cJ\left(\zett^{\supN}\right)
  =
  \frac{1}{N}{\textstyle\sum\limits_{i=1}^N}   
q_i \crprod p_i                                  
  - \int (x \crprod\pdx\psi) \varpi \dd{x}
\,,
\label{EPSfuncJ}
\end{equation}
and the energy functional by
\begin{equation}
  \cH\left(\zett^{\supN}\right)
  =
  \frac{1}{N}{\textstyle\sum\limits_{i=1}^N}
  \Bigl(\!\sqrt{1 + \abs{p_i}^2 }
  \!+\!
  \left(\fe *\psi\right)\!\left(q_i\right)
  \Bigr)
  +
  \frac{1}{ 2} \int \left(\abs{\pdx\psi}^2 +
  \abs{\varpi}^2\right)\!\, \dd{x}
\,.
\label{EPSfuncH}
\end{equation}
 We note that $\cM,\cP,\cH$ are well-defined on all of $\Gamma^{\supN}$, while
$\cJ$ is well-defined only on a subset of $\Gamma^{\supN}$; in particular, 
$\cJ$ is well-defined on $\GaB^{\supN}$.

\begin{Rema}
 The energy functional \refeq{EPSfuncH} 
furnishes the Hamiltonian for the regularized dynamical system.
 It is readily verified that the  Hamiltonian system,
$\ddt{q}_i^{\supN}= \partial  \cH/\partial{p}_i^{\supN}$ and
$\ddt{p}_i^{\supN}= -\partial  \cH/\partial{q}_i^{\supN}$, 
together with
$\pdt{\psi^{\supN} } = \delta \cH/\delta \varpi^{\supN}$
and
$\pdt{\varpi^{\supN}} = -\delta \cH/\delta \psi^{\supN}$,
coincides with the evolution equations for the wave gravity potential and 
the particles.
\end{Rema}

 A map $t\mapsto \zett^{\supN}_t\in C^1(\RR,\Gamma^{\supN})$ satisfying
our microscopic scalar wave gravity equations will be called a 
$\Gamma^{\supN}$-strong solution.

\begin{Prop} \label{propClawsNeps} 
        {For any sufficiently regular (in particular, a
	$\Gamma^{\supN}$-strong) solution $t\mapsto \zett^{\supN}_t$ of
        the  microscopic scalar wave gravity system, we have}
\bea
\cM\left(\zett^{\supN}_t\right)  \!\!&=&\!\! M
\,,
\label{EPSconstM}
\\
 \cP\left(\zett^{\supN}_t\right) \!\!&=&\!\! P
\,,
\label{EPSconstP}
\\
\cH\left(\zett^{\supN}_t\right) \!\!&=&\!\! E
\,,
\label{EPSconstE}
\eea
        {with $M$, $P$, $E$ independent of time; in particular, $M=1$.
	  If $\zett^{\supN}_0\in \GaB^{\supN}$,
	then also}
\beq
\cJ\left(\zett^{\supN}_t\right) =J
\,,
\label{EPSconstJ}
\eeq
        {with $J$ independent of time.}
\end{Prop}

\noindent
        \textit{Proof of Proposition \ref{propClawsNeps}.}
        Proposition \ref{propClawsNeps} is proved in the appendix
as a special case of the conservation laws in Theorems 
\ref{theoVAGUEsolVeps} and \ref{theoKRsolVeps}
of subsubsection \ref{secVAGUEform}.\QED

\begin{Rema}
 One may contemplate attaching also an Euler spin variable 
$s_i^{\supN}(t)\in\RR^3$
at time $t$ to the $i$-th particle, the (non-relativistic) 
evolution equations for $s_i^{\supN}(t)$ being 
just Euler's equations for a degenerate gyroscope, viz.
\beq
\qquad    \dot{s}_i^{\supN}(t)
=
-\left(\fe\Id *\crprod\nabla\psi^{\supN}(\,.\,,t)\right)
\left(q_i^{\supN}(t)\right)
\,.
\label{NsDOTeps}
\eeq
 However, standard identities of vector analysis and the radial
symmetry of $\fe$ yield for the (negative of the) field torque 
on the $i$-th particle
\beq
\left(\fe\Id *\crprod\nabla\psi^{\supN}(\,.\,,t)\right)(x)
=\int\pdy\crprod[\fe(y-x)(y-x) \psi^{\supN}(y,t)]\dd{y}\equiv 0,
\eeq
the vanishing as a result of one of Green's theorems and 
the compact support of $\fe$.
 Hence, each $s_i^{\supN}(t)$ is itself a constant of the motion.
 Moreover, since  there is no feedback loop from  $s_i^{\supN}(t)$ 
to the particle-field dynamics, the introduction of spin into this
model is uncalled for.
\end{Rema}
\newpage

\subsection{Global existence and uniqueness}
\label{secNexist}

  In this subsection we extend the single particle global
existence and regularity results of \cite{KoSpKu}, \cite{KoKuSp} 
to the many-body problem.
 To get started, one needs decent a-priori bounds on the norms of 
the various dynamical quantities.

\subsubsection{A-priori bounds without invoking conservation laws}
\label{secAprioriNboundsINt}

 We begin with  a-priori bounds that can be obtained
without invoking the conservation laws.
 It is trivially clear by the upper bound 1 on 
their speeds that  the positions of the particles are bounded
above linearly in $t$.
 In the following we recall the familiar linear in $t$ a-priori estimate 
for the field norms, and a bound on the momenta quadratic in $t$.

\begin{Lemm}\label{AprioriBOUNDSnormZETA}
{Let 
$\left(q_1^{\supN}(\,.\,),...,q_N^{\supN}(\,.\,)\right)
\in C^{0,1}(\RR,\RR^{3N})$
be a given Lipschitz-continuous curve,
its components having Lipschitz constant $<1$, 
and let $\zeta_{.}\in C^1(\RR,(\dotH\oplus L^2)(\RR^3))$ be a
strong solution of 
  \refeq{NphiEQeps}, \refeq{NvpiEQeps}, satisfying conditions
  \refeq{NphiNULLepsASYMP} \refeq{NvarpiNULLepsASYMP}. 
    Then at any $t\in\RR$,
\beq
 \max\lbrace\norm{\psi^{\supN}(\,.\,,t)}_{\dotH}\,,
          \,\norm{\varpi^{\supN}(\,.\,,t)}_{L^2}
     \rbrace
\leq 
 (2\cE_{\mathrm{W}}(\zeta^{\supN}_0))^{1/2} +  \norm{\varrho}_{L^2} |t|,
\label{fieldNORMSboundINt}
\eeq
where 
  $\cE_{\mathrm{W}}(\zeta_{t}) = \frac{1}{2} \int 
   \bigl( \abs{\pdx\psi(\,.\,,t)}^2 + |\varpi(\,.\,,t)|^2\bigr)
				\dd{x}$ 
is the wave field energy at time $t$.}
\end{Lemm}

\begin{Rema} 
The a-priori bound \refeq{fieldNORMSboundINt} extends to the strong 
solution of the wave equation for \textit{any subluminal} source 
$\fe*\rho \in C^0(\RR,C^\infty_0(\RR^3))$.
\end{Rema} 

\noindent
\textit{Proof of Lemma \ref{AprioriBOUNDSnormZETA}:}
  By hypothesis, 
$t\mapsto \zeta^{\supN}_{t}\in C^1(\RR,(\dotH \oplus L^2)(\RR^3))$ 
is a strong solution of the wave equation with source 
  $t\mapsto\varrho*\rho^{\supN}_t\in C^0(\RR,C^\infty_c(\RR^3))$
  moving at speeds less than light;
  hence, $t\mapsto \cE_{\mathrm{W}}(\zeta^{\supN}_t)$ is differentiable. 
  We have
\bea
\ddt \cE_{\mathrm{W}}(\zeta^{\supN}_t) 
=
 - \int \varpi^{\supN}(t,x) (\varrho * \rho^{\supN}_t)(x) \dd{x}
 \!\!\!&\leq&\!\!\!
\norm{\varpi^{\supN}(\,.\,,t)}_{L^2}
\norm{\varrho * \rho^{\supN}_t}_{L^2}
 \cr
 \!\!\!&\leq&\!\!\!
\norm{\varrho}_{L^2}
\left(2\cE_{\mathrm{W}}(\zeta^{\supN}_t)\right)^{1/2},
\label{fieldENERGYdiffineqINt}
\eea
the first inequality by Cauchy--Schwarz, 
while for the second one we used the estimate
$\norm{\varrho * \rho^{\supN}_t}_{L^2}^2
\leq
\sup_{x\in\RR^3}\{(\varrho*\varrho)(x)\} = \norm{\varrho}_{L^2}^2$, 
as well as the obvious estimate
$\norm{\varpi^{\supN}(\,.\,,t)}_{L^2} \leq 
\left(2\cE_{\mathrm{W}}(\zeta^{\supN}_t)\right)^{1/2}$
implied by the definition of $\cE_{\mathrm{W}}$.
 Dividing \refeq{fieldENERGYdiffineqINt} by
$\left(2\cE_{\mathrm{W}}(\zeta^{\supN}_t)\right)^{1/2}$ 
and integrating over $t$ gives 
\beq
 (2\cE_{\mathrm{W}}(\zeta^{\supN}_t))^{1/2} 
\leq 
 (2\cE_{\mathrm{W}}(\zeta^{\supN}_0))^{1/2} +  \norm{\varrho}_{L^2} t.
\label{fieldENERGYROOTboundINt}
\eeq
  The definition of $\cE_{\mathrm{W}}$ given in Lemma
\ref{AprioriBOUNDSnormZETA} now shows that
\refeq{fieldENERGYROOTboundINt} implies \refeq{fieldNORMSboundINt}.\QED

\begin{Lemm}\label{AprioriBOUNDSnormP}
{Let 
$\left(q_1^{\supN}(\,.\,),...,q_N^{\supN}(\,.\,)\right)
\in C^{0,1}(\RR,\RR^{3N})$
be a given Lipschitz-continuous curve,
its components having Lipschitz constant $<1$, and let
	$\zeta_{.}\in C^1(\RR,(\dotH\oplus L^2)(\RR^3))$ 
  be a strong solution of the wave equation with source 
	$t\mapsto\varrho*\rho^{\supN}_t\in C^0(\RR,C^\infty_c(\RR^3))$.
    Suppose $t\mapsto\left(p_1^{\supN}(t),...,p_N^{\supN}(t)\right)$
is a classical solution of \refeq{NpDOTeps}.
 Then the momenta  at $t\in\RR$, $p_k^{\supN}(t)$, $k=1,...,N$,
are bounded by
\beq
 \max_{1\leq k\leq N}
\lbrace\norm{p_k^{\supN}(t)}\rbrace
\leq 
 \max_{1\leq k\leq N}
\lbrace\norm{p_k^{\supN}(0)}\rbrace
+  \norm{\varrho}_{L^2}  (2\cE_{\mathrm{W}}(\zeta^{\supN}_0))^{1/2}|t| 
+  \textstyle{\frac{1}{2}} \norm{\varrho}_{L^2}^2 t^2
\,.
\label{pNORMSboundINt}
\eeq
}
\end{Lemm}

\noindent
\textit{Proof of Lemma \ref{AprioriBOUNDSnormP}:}
 Use $p(t) = p(0) + \int_0^t \dot{p}(\tilde{t})\dd\tilde{t}$,
take absolute values, use the triangle inequality, then
invoke \refeq{NpDOTeps} and estimate
\beq
\abs{\left(\fe *\nabla\psi^{\supN}(\,.\,,t)\right)
\left(q_i^{\supN}(t)\right)} 
\leq
 \norm{\varrho}_{L^2} 
\norm{\psi^{\supN}(\,.\,,t)}_{\dotH}
\,,
\label{forceBOUND}
\eeq
then recall Lemma \ref{AprioriBOUNDSnormZETA}.\QED

\begin{Rema}
 We note that \refeq{fieldENERGYROOTboundINt} is far from optimal, 
which is evident from the fact that no details of the time dependence
of $\rho^{\supN}_t$ enter \refeq{fieldENERGYROOTboundINt}; 
in any event, squaring gives an upper  bound on the 
wave field energy quadratic in $t$.
 Similarly, \refeq{pNORMSboundINt} is far from optimal; in any
event, its \rhs provides a quadratic-in-$t$ upper bound on the kinetic
energy, ``l.h.s.\refeq{EPSboundEkin}$-1$.'' 
 These bounds together with the bound \refeq{forceBOUND} and the 
asymptotics \refeq{NphiNULLepsASYMP} now also imply that 
$\bigl|\frac{1}{N} \sum_{i=1}^N
 \bigl(\fe*\psi^{\supN}(\,.\,,t)\bigr)\left(q_i^{\supN}(t)\right)\bigr|$,
and therefore finally also the 
total energy, are both bounded above by $a + b|t| + ct^2$.
\end{Rema}

 This does not yet exhaust our list of bounds that obtain without invoking
conservation laws. 
 The next such bound is nevertheless given its own subsection, for the special
role it plays subsequently.

\subsubsection{A lower bound on the Hamiltonian functional}
\label{secHlowerBOUND}

  To state our lower bound on the Hamiltonian functional, 
we first define
\beq
E_\bot
:=
1
-
\frac{1}{ 8\pi} \iint \frac{\fe(x) \fe(y)}{ |x-y|}\dd{x}\dd{y}
\,.
\label{groundE}
\eeq
  Note that the energy value $E_\bot$ 
depends only on the regularization but not on $N$.
\newpage

\smallskip
\begin{Prop}\label{propHboundN} 
        {The Hamiltonian functional is bounded below by
\beq
\cH\left(\zett^{\supN}\right)
\geq
E_\bot
\,,
\label{EPSboundH}
\eeq
independently of $N$.
 The bound is attained when $\zett^{\supN}$ is any 
translation in space of $\zett_{\bot}^{\supN}$, 
the state in which for all $k=1,...,N$ we have
$q_k = 0$, $p_k = 0$, and furthermore $\varpi \equiv 0$ 
and $\psi\equiv \psi_{\fe}$, with
\beq
\psi_{\fe}(x)
=
- {\frac{1}{4\pi}} \left(\abs{\,.\,}^{-1}*\fe\right)(x).
\eeq
 (However, 
note that only the standard ground state satisfies \refeq{NphiNULLepsASYMP}.)
}
\end{Prop}

 The state $\zett_{\bot}^{\supN}$ will be called the standard ground state 
of the regularized dynamical system, and \refeq{groundE} will be called the 
ground state energy.
\smallskip

\noindent
   \textit{Proof of Proposition \ref{propHboundN}}.
    For later purposes, we will prove the bound \refeq{EPSboundH}
as an upper limit of a one-parameter family of bounds to 
$\cH\left(\zett^{\supN}\right)$.
   Thus, let $\kappa\in (0,1]$.
   Then
\begin{eqnarray}
  \cH\left(\zett^{\supN}\right)
  - {\textstyle\frac{1-\kappa}{2}}\norm{\psi}_{\dotH}^2
=
\textstyle{\frac{1}{N}\sum\limits_{i=1}^N}
\left(\sqrt{1 + \abs{p_i}{}^2} + \left(\fe*\psi\right)\left(q_i\right)\right)
\!\!+
  {\textstyle\frac{1}{2}}\norm{\varpi}_{L^2}^2
  +
{\textstyle\frac{\kappa}{2}}\norm{\psi}_{\dotH}^2
 .
\label{kappaHidentity}
\end{eqnarray}
  Discarding the manifestly positive momentum contributions we obtain
\beq 
\cH\left(\zett^{\supN}\right) -
(1-\kappa){\textstyle\frac{1}{2}}\norm{\psi}_{\dotH}^2 \geq 1+
\kappa\textstyle{\frac{1}{2}}
\norm{\psi}_{\dotH}^2 +
\textstyle{\frac{1}{N}\sum\limits_{i=1}^N} 
	\left(\fe *\psi\right)\left(q_i\right) 
. 
\label{EPSboundHa} 
\eeq
        Minimizing the \rhs of \refeq{EPSboundHa} 
with respect to $\psi$ now gives
\beq 
\cH\left(\zett^{\supN}\right) -
(1-\kappa){\textstyle{\frac{1}{2}}}\norm{\psi}_{\dotH}^2 \geq 1
-
{\textstyle{\frac{1}{N^2}
\sum\limits_{i=1}^N
\sum\limits_{k=1}^N}}
\frac{1}{8\pi\kappa}
\iint
\frac{\fe\!\left(q_i\!-x\!\right)\fe\!\left(q_{k}\!-y\!\right)}
                {|x-y|}
\dd{x}\dd{y}
\,.
\label{EPSboundHb}
\eeq
  The right-hand side of \refeq{EPSboundHb} can be minimized \wrt
the $\left\{q_i\right\}_{i=1}^N$ by equi-measurable, radially symmetric 
rearrangement of $\sum_{n=1}^N \fe(\,.\,-q_{n})$ centered at the origin.
  Since $\fe$ is itself radially symmetric and decreasing, this 
is achieved by simply translating all $q_n$ to the same 
position, in particular to the origin.
  This gives, for all $\kappa\in (0,1]$,
\beq \cH\left(\zett^{\supN}\right) -
(1-\kappa){\textstyle\frac{1}{2}}\norm{\psi}_{\dotH}^2 \geq 1
-
\frac{1}{ 8\pi\kappa} \iint \frac{\fe(x) \fe(y)}{ |x-y|}\dd{x}\dd{y}
.
\label{EPSboundHc}
\eeq

  The bound \refeq{EPSboundH} now obtains by taking $\kappa = 1$ 
in \refeq{EPSboundHc} and recalling \refeq{groundE}.
  Straightforward computation of $\cH\left(\zett_{\bot}^{\supN}\right)$
proves that \refeq{EPSboundH} is attained at $\zett_{\bot}^{\supN}$ and, 
by the translation invariance in position space of 
$\cH\left(\zett^{\supN}\right)$, also at any translate of 
$\zett_{\bot}^{\supN}$.\QED

\subsubsection{Bounds invoking conservation laws}
\label{secBOUNDSfromClaws}

 Using energy conservation of sufficiently regular solutions, 
we next bootstrap from the proof of Proposition \ref{propHboundN} 
to uniform bounds in $t$ and $N$ on the four major 
additive contributions to $\cH\left(\zett^{\supN}_t\right)$.

\begin{Lemm}\label{lemmNbounds} 
  {Let $t\mapsto \zett^{\supN}_t$ be a sufficiently
          regular (e.g. $\Gamma^{\supN}$-strong) solution 
	  of the dynamical system
          \refeq{NphiEQeps}--\refeq{NpDOTeps} conserving energy.
          Then, uniformly in $t$ and $N$, we have
\bea
\norm{\psi^{\supN}(\,.\,,t)}_{\dotH}^2
\hskip-.5truecm
&&
\leq
4 + 4E  - 8E_\bot
\,,
\label{EPSboundGRADphiZWEInorm}
\\
\norm{\varpi^{\supN}(\,.\,,t)}_{L^2}^2
\hskip-.5truecm
&&
\leq
2E  - 2E_\bot
\,,
\label{EPSboundDOTphiZWEInorm}
\\
\frac{1}{N}{\textstyle\sum\limits_{i=1}^N}\sqrt{1 + \abs{p_i^{\supN}(t)}{}^2 }
\hskip-.5truecm&&
\leq
1 + E - E_\bot
\,,
\label{EPSboundEkin}
\eea
\beq
6E_\bot - 3E  - 3
\leq
\frac{1}{N}{\textstyle\sum\limits_{i=1}^N}
 \bigl(\fe*\psi^{\supN}(\,.\,,t)\bigr)\left(q_i^{\supN}(t)\right)
\leq
E  - 1
\label{EPSboundsEpot}
\eeq
}
\end{Lemm}

\noindent
        \textit{Proof of Lemma \ref{lemmNbounds}}.
  Since $\cH\left(\zett^{\supN}_t\right) = E$ is fixed by the Cauchy data
$\zett^{\supN}_0$, a simple rewriting of \refeq{EPSboundHc} 
with $\kappa = 1/2$, using the definition \refeq{groundE}, 
gives us \refeq{EPSboundGRADphiZWEInorm}.
  As to \refeq{EPSboundDOTphiZWEInorm} and \refeq{EPSboundEkin}, 
$\cH\left(\zett^{\supN}_t\right) = E$ and the definition \refeq{EPSfuncH} 
of $\cH\left(\zett^{\supN}\right)$ give us the identity
\bigskip
\bea
\frac{1}{N}{\textstyle\sum\limits_{i=1}^N}\sqrt{1 + \abs{p_i^{\supN}(t)}^2 }
+
&&\hskip-.5truecm
{\textstyle\frac{1}{2}}\norm{\varpi^{\supN}(\,.\,,t)}_{L^2}^2
=
\cr
&& \hskip-1.3truecm
E - {\textstyle\frac{1}{2}}\norm{\psi^{\supN}(\,.\,,t)}_{\dotH}^2 
- 
\frac{1}{N}{\textstyle\sum\limits_{i=1}^N} 
\left(\fe*\psi^{\supN}(\,.\,,t)\right)\left(q_i^{\supN}(t)\right)
\,.
\label{HequalErewritten}
\eea
 Recalling the minimization steps that lead from \refeq{EPSboundHa} 
to \refeq{EPSboundHc} (here with $\kappa =1$), and the definition 
\refeq{groundE}, we see that the right-hand side of 
\refeq{HequalErewritten} is bounded above, giving
\beq
\frac{1}{N}{\textstyle\sum\limits_{i=1}^N}\sqrt{1 + \abs{p_i^{\supN}(t)}^2 }
+
{\textstyle\frac{1}{2}}\norm{\varpi^{\supN}(\,.\,,t)}_{L^2}^2
\leq
1 + E - E_\bot
\,.
\label{EPSdominanceE}
\eeq
        Now \refeq{EPSboundDOTphiZWEInorm} follows at once
from  \refeq{EPSdominanceE} by estimating $\abs{p_i^{\supN}(t)}\geq 0$;
to get \refeq{EPSboundEkin}, 
we instead use $\norm{\varpi^{\supN}(\,.\,,t)}_{L^2}^2\geq 0$
in  \refeq{EPSdominanceE}.
 Finally, to obtain \refeq{EPSboundsEpot}, rewrite the definition 
\refeq{EPSfuncH} of $\cH\left(\zett^{\supN}\right)$ into an identity for
$\frac{1}{N}{\textstyle\sum_{i=1}^N} 
\bigl(\fe*\psi^{\supN}(\,.\,,t)\bigr)\left(q_i^{\supN}(t)\right)$,
then use $\cH\left(\zett^{\supN}_t\right) = E$; 
now the bounds 
\refeq{EPSboundGRADphiZWEInorm},
\refeq{EPSboundDOTphiZWEInorm},
\refeq{EPSboundEkin}
give the first,  the positivity of
$\norm{\psi^{\supN}(\,.\,,t)}_{\dotH}^2$, 
$\norm{\varpi^{\supN}(\,.\,,t)}_{L^2}^2$,
and $\abs{p_i^{\supN}(t)}$ the second inequality in 
\refeq{EPSboundsEpot}.\QED

\begin{Rema} 
  The bounds \refeq{EPSboundDOTphiZWEInorm} and 
\refeq{EPSboundEkin} happen to be asymptotically sharp when 
$E\downarrow{E_\bot}$, in which case they correctly imply that 
$\norm{\varpi^{\supN}(\,.\,,t)}_{L^2}\downarrow{0}$
and 
$\abs{p_i^{\supN}(t)}\downarrow{0}$ for all $i=1,...,N$.
 It is to be doubted though that \refeq{EPSboundDOTphiZWEInorm} and 
\refeq{EPSboundEkin} are sharp for $E > {E_\bot}$; in any event, certainly
\refeq{EPSboundGRADphiZWEInorm} and \refeq{EPSboundsEpot} are not sharp 
(for instance, \refeq{EPSboundGRADphiZWEInorm} misses the correct ground 
state value by a factor 2).
 Of course, it is a straightforward matter to improve on 
\refeq{EPSboundGRADphiZWEInorm} 
and \refeq{EPSboundsEpot} by optimizing w.r.t. $\kappa$ 
	(N.B.: $\kappa= 1/2$ is 
the optimizer for $E=1$), and while this does lead to asymptotically sharp
upper bounds as $E\downarrow{E_\bot}$ (in which case $\kappa\uparrow{1}$), 
for $E > {E_\bot}$ these bounds are still not sharp, but now 
 more cumbersome than 
\refeq{EPSboundGRADphiZWEInorm} and \refeq{EPSboundsEpot}.
 Fortunately, for our purposes any a-priori bounds uniform in $t$ and $N$
will do; hence, we gain by sticking to the simple ones given in Lemma 
\ref{lemmNbounds}.
\end{Rema}

  As a corollary to \refeq{EPSboundEkin} the particle momenta are 
bounded above in magnitude.
 This has an easy but important corollary for the particle speeds, 
which we state explicitly.

\begin{Coro}\label{coroSPEEDlimit} 
     {The particle speeds are bounded away from the speed of light,
viz.
\beq
\max_{i\in\{1,...,N\}} \abs{\dot{q}_i^{\supN}(t)}
 \leq 
\sqrt{1- \left(1+ N(E  - E_\bot)\right)^{-2}}
\,,
\label{EPSveloBOUND}
\eeq
 uniformly in $t$. 
 In particular, when $E = E_\bot$, then $\abs{\dot{q}_i^{\supN}(t)}=0$ 
for all $i$ and $N$.
}
\end{Coro}

\noindent
        \textit{Proof of Corollary \ref{coroSPEEDlimit}.}
  We rewrite \refeq{EPSboundEkin} as
\beq
\frac{1}{N}{\textstyle\sum\limits_{i=1}^N}\left(\sqrt{1 + \abs{p_i^{\supN}(t)}^2 } -1\right)
\leq
E - E_\bot
\,.
\label{EPSboundEkinNEW}
\eeq
  Since $\sqrt{1+|p|^2} -1 \geq 0$, the bound \refeq{EPSboundEkinNEW} 
now implies that for all $i$, 
\beq
\sqrt{1+|p_i^{\supN}(t)|^2}  -1
\leq
 N(E  - E_\bot)
\,,
\label{EPSboundEkinSUP}
\eeq
and solving for $\abs{p_i^{\supN}(t)}$ gives, uniformly in $t$,
\beq
\max_{i\in\{1,...,N\}} \abs{p_i^{\supN}(t)}
\leq
\sqrt{\left(1+ N(E  - E_\bot)\right)^2-1}
\,,
\label{EPSmomentumBOUND}
\eeq
which now yields \refeq{EPSveloBOUND} by inverting the monotone
map $|p|\mapsto |v|$ given in \refeq{vOFp}.\QED

 Note that for any $E> E_\bot$, \refeq{EPSveloBOUND} 
does not imply boundedness away 
from the speed of light of the $\abs{\dot{q}_i^{\supN}(t)}$ uniformly in $N$;
only $\max_{i} \abs{\dot{q}_i^{\supN}(t)}\leq{1}$ holds uniformly in $N$.
\smallskip

\subsubsection{Global existence and uniqueness of solutions}
\label{secExUnTHM}

 Lemma \ref{lemmNbounds} and Corollary \ref{coroSPEEDlimit} imply
that any energy-conserving solution is represented by a 
point moving in a weakly compact subset of $\Gamma^{\supN}$,
and such solutions do exist.

\begin{Theo}\label{theoKKSforN}
     {For every $\zett^{\supN}_0\in \GaB^{\supN}$ there
exists a unique, global strong solution $t\mapsto \zett^{\supN}_t\in
C^1(\RR,\Gamma^{\supN})$ of the Hamiltonian field \&\ $N$-body problem
\refeq{NphiEQeps}--\refeq{NpDOTeps}, satisfying
$\lim_{t \to  0}\zett^{\supN}_t = \zett^{\supN}_0$,
and conserving mass, energy, momentum and angular momentum as 
stated in Proposition \ref{propClawsNeps}.
 For more regular initial data one can bootstrap to correspondingly
higher regularity of $t\mapsto \zett^{\supN}_t$.}
\end{Theo}

\noindent
        \textit{Proof of Theorem \ref{theoKKSforN}}.
 The proof is a largely straightforward adaption to our many-body problem 
of the proof for a single particle system in
    \cite{KoSpKu}.
 We remark that our Wiener condition for $\fe$ is only needed to adapt 
their proof. 
 The strategy is to first construct local weak solutions conserving
energy, then to use the uniform bounds on the norms of the various 
dynamical quantities that follow from energy conservation (see Lemma
\ref{lemmNbounds} and its Corollary) to continue to all times.
 Strong solutions obtain by restricting 
$\psi(\,.\,,0)\in (\dotH\cap\dotHH)(\RR^3)$.
 Proofs of the conservation laws, which are stated without proof in 
    \cite{KoSpKu}, 
are provided in our appendix, for the convenience of the reader.\QED

\begin{Rema}
\label{remaMOREbounds}
 For the proof of Theorem \ref{theoKKSforN} the a-priori bounds
in Lemma \ref{lemmNbounds} based on energy conservation suffice.
 The other conservation laws provide additional bounds that
may be useful in different contexts. 
 For instance, momentum conservation and the Cauchy--Schwarz 
inequality give us the uniform bound in $t$,
\beq
\frac{1}{N}\abs{{\textstyle\sum\limits_{i=1}^N} p_i^{\supN}(t)} 
\leq
\abs{P} 
+ 
\norm{\psi^{\supN}(\,.\,,t)}_{\dotH}
\norm{\varpi^{\supN}(\,.\,,t)}_{L^2}
\,,
\label{PmomBOUND}
\eeq
while angular momentum conservation, the Cauchy--Schwarz inequality, 
and the finite wave speed give us the linear bound in $t$,
\beq
\frac{1}{N}\abs{{\textstyle\sum\limits_{i=1}^N} 
p_i^{\supN}(t)\crprod q_i^{\supN}(t)} 
\leq
\abs{J} 
+ 
\left(R+|t|\right)
\norm{\psi^{\supN}(\,.\,,t)}_{\dotH}
\norm{\varpi^{\supN}(\,.\,,t)}_{L^2}
\,,
\label{JposmomBOUND}
\eeq
where $R$ is the radius of the ball $B_R$ containing the initial 
positions of all particles and the supports of their regularizations, 
and outside of which \refeq{NphiNULLepsASYMP} 
and \refeq{NvarpiNULLepsASYMP} hold.
\end{Rema}

\begin{Rema}
 As is the case for the regularized Vlasov--Poisson equations
 \cite{SpohnBOOKa},
the solutions described by Theorem \ref{theoKKSforN} map one-to-one
into generalized solutions of the regularized wave gravity Vlasov model
in which derivatives of $f$ are meant in the sense of distributions.
\end{Rema}

\section{The regularized Vlasov model}
\label{secVlasovIntro}

 In this section we discuss the regularized wave gravity Vlasov model.
 First, we present the Vlasov equations formally as a continuum
model.
 Next we recall the concept of generalized (distributional)
solutions, for which we introduce two suitable topologies, one 
based on the vague and one on a strong Banach space topology for 
distributions. 
 The solutions to the field \&\ $N$-body model of the 
previous section furnish particular generalized solutions of our 
Vlasov model in either of the just mentioned topologies.
 We then prove global existence and  uniqueness in the strong 
Banach space topology of generalized solutions to our 
regularized wave gravity Vlasov model.

\subsection{The dynamical continuum system}

 As first order system, the inhomogeneous wave equation for the 
regularized wave gravity potential $\psi(\,.\,,t)\in \dotH(\RR^3)$ 
and its conjugate variable $\varpi(\,.\,,t)\in L^2(\RR^3)$ now reads
\bea
&&
        \pdt{\psi (x, t)}
=
        \varpi(x, t)
\label{VphiEQeps}
\\
&&
        \pdt{\varpi (x, t)}
=
        \pdxsq{\psi}(x, t) -
        \bigl(\fe * \rho(\,.\,,t) \bigr)(x) \, .
\label{VvpiEQeps}
\eea
 The initial data 
$\psi(\,.\,,0) \equiv \psi_0(\,.\,)\in(\dotH\cap\dotHH)(\RR^3)$ and
$\varpi(\,.\,,0) \equiv \varpi_0(\,.\,)\in L^2(\RR^3)$ satisfy
\bea
&&
\psi_0(x) 
= 
-{1}/{4\pi |x|}
\,,
\label{phiNULLepsASYMP}
\\
&& 
\varpi_0(x) 
=
0
\label{varpiNULLepsASYMP}
\eea
outside some closed ball $B_R\subset \RR^3$.
 The density $\rho (x,t)$ on the r.h.s. in \refeq{VvpiEQeps} is given by
\begin{equation}
  \rho (x,t)
  =
  \int f(x,p,t) \dd p  \, ,
\label{defRHOeps}
\end{equation}
where $f(\,.\,,\,.\,,t)$ is the normalized particle density function
at time $t$, satisfying the following (continuity)
equation on time-position-momentum space $\RR\times\RR^3\times\RR^3$,
\beq
\pdt 
f(x,p,t) 
= 
- \left(\pdp  \sqrt{1+|p|^2} \cdot\pdx -
\pdx \left(\fe\,{\stackrel{\phantom{.}}{*}}\,\psi(\,.\,,t)\right)(x)\cdot\pdp 
	\right)
f(x,p,t)
\,, 
\label{VfEQeps} 
\eeq 
with $x\in{\RR}^3$ being the space and ${p}\in{\RR}^3$ the momentum variable.
 Initial data $f(\,.\,,\,.\,,0) \equiv f_0(\,.\,,\,.\,)$ for \refeq{VfEQeps} 
are restricted by the requirement that $\fe*\rho(\,.\,,0)$ is supported
in $B_R$. 
 As to the appropriate function space, 
we re-emphasize that in the form stated above, one should think of 
Vlasov's $f(\,.\,,\,.\,,t)$ as a continuum approximation to the empirical 
$x,p$ phase space density of particles for an actual individual $N$-body 
system in the large $N$ regime, when fine details of the particles' 
behaviors become irrelevant on the ``macroscopic'' scales so that the 
empirical atomic measure can be well approximated by a function 
$f(\,.\,,\,.\,,t)\in L^1_{+,1}(\RR^6)$ --- 
the subset of $L^1(\RR^6)$ consisting of the Radon--Nikodym derivatives $f$ 
of  Borel probability measures $\mu^{f} (\dd x\dd p)$ which are absolutely 
continuous w.r.t. Lebesgue measure.
 In fact, such functions $f(\,.\,,\,.\,,t)$, 
the fields $\psi(\,.\,,t)$ and their formal 
time derivatives $\varpi(\,.\,,t)$, would even be expected to 
have time and space derivatives in the classical sense for 
all time whenever their initial data are chosen sufficiently regular.

\begin{Rema} 
  It is known, but perhaps not well-known, that for sufficiently regular
solutions $f$ (say, classical with rapid decay at infinity), the
continuity equation \refeq{VfEQeps} can readily be associated with 
a Hamiltonian ${\cH}_{\mathrm{C}}$, given  $\psi(\,.\,,t)$ for all $t$.
 To obtain the Hamiltonian ${\cH}_{\mathrm{C}}$ for $f$ given 
$\psi(\,.\,,t)$, multiply r.h.s.\refeq{VfEQeps}
with a test function $g(\,.\,,\,.\,,t)\in C^1(\RR^6)$ of at most
polynomial growth in $x,p$ whose $t$-dependence
is yet to be determined, and integrate over $\RR^6$; $f$ and $g$ 
can now be viewed as conjugate variables, with
$\pdt f = \delta{\cH}_{\mathrm{C}}(f,g)/\delta{g}$, 
$\pdt g = -\delta{\cH}_{\mathrm{C}}(f,g)/\delta f$.
 Interestingly, the equation for $g$ is just \refeq{VfEQeps} with
$g$ in place of $f$, and in this sense \refeq{VfEQeps}
already is the Hamiltonian system, given the fields. 
 The inhomogeneous wave equation \refeq{VphiEQeps}, 
\refeq{VvpiEQeps} for the fields $\psi(\,.\,,t)$ and 
$\varpi(\,.\,,t)$ is a Hamiltonian dynamical system, 
given $f$. 
 The full set of equations 
\refeq{VphiEQeps},
\refeq{VvpiEQeps},
\refeq{VfEQeps}
becomes a Hamiltonian system with the help of non-canonical Lie brackets, cf. 
   \cite{Morrison, WeinMor, MarMorWein}.
\end{Rema}

 Our goal is to validate the continuum approximation to the microscopic 
atomistic dynamics by means of a continuum limit in $x,p$ space 
(the ``Vlasov limit''), supplemented by a law of large numbers and a 
central limit theorem.
  To pave the way for the continuum validation, we next
recall the concept of generalized solutions.

\subsection{Distributional form of the regularized Vlasov model}
\label{secDISTRIBUTIONALform}

  In order to think of $f(\,.\,,\,.\,,t)$  as the actual atomic 
measure of an individual $N$-body system, one has to interpret 
the derivatives in the sense of distributions.
 Thus,  for given
$\psi(\,.\,,t)\in \dotH(\RR^3)$ and
$\varpi(\,.\,,t)\in L^2 (\RR^3)$, 
we implement the idea of distributional derivatives of $f$
in the usual way by multiplying (all of)
\refeq{VfEQeps} with any real test function $g(\,.\,,\,.)\in
C^1_0(\RR^6)$ and integrate over $\RR^6$ by parts to transfer the
partial derivatives \wrt $x,p$ onto the smooth $g$; also, the
partial derivative \wrt $t$ is pulled out of the integral.
 So far, $f(\,.\,,\,.\,,t)$ had to be a sufficiently regular function, 
but nothing now prevents us from allowing $f\in\probL(\RR^6)$,
the Radon--Nikodym derivative of an absolutely continuous measure $\mu^{f}$.
 The so integrated and manipulated form of \refeq{VfEQeps} remains
well-defined even if we replace $\mu^{f}(\dd{x}\dd{p})$ with any 
Borel probability measure $\mu_t(\dd{x}\dd{p})$.
 Indeed, let $Dg$ denote any of the partial derivatives of $g$. 
 Then $Dg\in C^0_0(\RR^6)$, where  $C_0^0(\RR^6)$ is equipped with
the uniform norm (a.k.a. sup-norm) 
$\norm{Dg}_{\mathrm{u}} = \sup_{z\in\RR^6}|Dg(z)|$.
 On the other hand, the Borel probability measures
$P(\RR^6)$ are a subset of $M(\RR^6)$, 
the Banach space of finite signed Radon measures $\sigma$
(which on $\RR^6$ coincide with the finite regular signed 
Borel measures $\sigma$) equipped with the total variation (TV) norm 
$\|\sigma\|_{\mathrm{TV}} = (|\sigma_+| + |\sigma_-|)(\RR^6)$, and
$M(\RR^6)$ is isometrically isomorphic to $C_0^0(\RR^6)^*$, the dual
space for (real) $C_0^0(\RR^6)$.

 In the above we used $z$ to denote a generic point $(x,p)\in \RR^6$.
 In the same vein, we sometimes write $\zeta$ for the generic wave variables
$(\psi, \varpi)$.

 A \textit{physical generalized state} of the regularized Vlasov model 
constitutes a complete specification of all its first-order evolutionary 
variables.  
 We accordingly define the set $\Gamma$ of all possible physical generalized 
states at time $t\in\RR$ to be the subset of points
\beq
  \Zett_t :=
  (\mu_t,\zeta_t) \in
M(\RR^6) \oplus\,  \dotH(\RR^3) \oplus L^2 (\RR^3) 
\label{gammaSPACE}
\eeq
for which $\mu_t\in P(\RR^6)$.
 The subset  $\GaB$ of $\Gamma$ denotes those physical generalized states
which satisfy \refeq{phiNULLepsASYMP} and \refeq{varpiNULLepsASYMP}, and
for which $\supp(\mu_0(\dd{x}\times\RR^3))\subset B_{R}$.

 In \refeq{gammaSPACE}, the first direct sum is clearly in the sense
of Banach spaces while the second may be either in Banach or Hilbert
space sense (with the understanding then that Hilbert binds stronger than 
Banach on its left); since it is a little awkward to have two direct 
sum symbols with different meanings in a single expression, 
\emph{for the Vlasov model} we use the Banach space meaning throughout.

 In this paper we are only interested in systems with finite 
energy, momentum, and angular momentum (the mass of a system
is finite  by default, namely unity). 
 Thus, for a suitable subset of \textit{generic} physical 
generalized states $\Zett$ in $\GaB$ we formally define 

\noindent
the mass functional
\beq
\cM\left(\Zett\right)
=
\int\! \mu(\dd{z})
\,,
\label{MfuncVeps}
\eeq

\noindent
the momentum functional 
\beq
\cP\left(\Zett\right)
=
 \int\! p \, \mu(\dd{z})
- \int\! \varpi \pdx\psi \, \dd{x} \,,
\label{PfuncVeps}
\eeq

\noindent
the angular momentum functional 
\beq
 \cJ\left(\Zett\right)
 =
 \int\! x \crprod{p}\, \mu(\dd{z})
 - \int\! \varpi \, x \crprod\pdx\psi \, \dd{x} \,,
\label{JfuncVeps} 
\eeq

\noindent
and the energy functional 
\beq
\cE\left(\Zett\right)
=
\!
\int\!\!
\left( \sqrt{1+ |p|^2} + (\fe *\psi)  \right) \mu(\dd{z})
+
\frac{1}{ 2}
\int\!\! \left(\abs{\pdx\psi}^2 + |\varpi|^2\right)\dd{x}
\,.
\label{HfuncVeps}
\eeq
 Here, in keeping with our already stipulated abbreviations, 
$\dd z$ denotes the Lebesgue measure
$\dd x\dd p $ on $\RR^6$ and $\int ... \mu(\dd{z})$ stands for 
$\iint ... \mu(\dd{x}\dd{p})$.
 We restrict the set of physical generalized states
to measures with finite expected values of  $|x|$ and $|p|$.
 Now $|x|$ and $|p|$, understood as functions on $\RR^6$, are 
not in $C^0_0(\RR^6)$,
but they are lower semi-continuous and therefore Radon measurable; 
hence, our condition of finite expected values of $|x|$ and $|p|$ 
(equivalently, of $|z|$) defines a proper subset ${P_1}(\RR^6)$ 
of the Borel probability measures.
 The corresponding subset of the physical states $\GaB$ is denoted by $\GaBpx$;
the energy, momentum, and angular momentum are well-defined on $\GaBpx$.

 It remains to stipulate a suitable topology on $\GaBpx$ in which
the maps $t\mapsto{\Zett_t}$ for $t\in\RR$ are continuous curves in $\GaBpx$ 
that qualify as \textit{generalized solutions}  of \refeq{VfEQeps} (given 
the fields).
 Unfortunately, the Banach space topology which $\GaBpx$ naturally inherits 
as a subset of $C_0^0(\RR^6)^* \oplus\,  \dotH(\RR^3) \oplus L^2 (\RR^3)$ is 
too strong to study families of empirical atomic measures, for any two 
(atomic)  empirical measures with disjoint supports  are always at a 
distance 2 from each other in the metric induced by the TV topology.

 A more suitable topology that immediately comes to mind 
is the \textit{vague} (a.k.a. weak$^*$) topology on
$M(\RR^6)\equiv C_0^0(\RR^6)^*$ induced by $C^0_0(\RR^6)$. 
 The set $\GaBpx$ with the vague topology on  $M(\RR^6)$ in place 
of the TV topology is denoted by $\GaBpx^v$.
 However, since we are interested only in the subset 
${P_1}(\RR^6)\subset M(\RR^6)$, we can do somewhat better and
equip ${P_1}(\RR^6)$ with the the standard Kantorovich--Rubinstein 
topology\footnote{The relationship between the various 
		  topologies is summarized in Appendix A.1.}
induced by the  dual Lipschitz distance in ${P_1}(\RR^6)$ 
(a map on ${P_1}(\RR^6)\times{P_1}(\RR^6)$), 
  \begin{equation}
    \dLip{\mu_1,\mu_2}
    :=
    \sup_{g \in C^{0,1}(\RR^6)}
      \left\{
\Abs{ \int g\, \dd(\mu_1-\mu_2)} : \LipSnorm{g}\leq 1
      \right\}\,.
  \label{defLdist}
  \end{equation}
 We write $\mu_n\leadsto\mu$ if $\dLip{\mu_n,\mu}\to 0$.

 Since it is convenient for the presentation to have a Banach space, 
we note that the metric $\dLip{\,.\,,\,.\,}$ defines a norm
on $P_1-P_1$ by $\dLIPnorm{\sigma} := \dLip{\sigma_+,\sigma_-}$ 
for $\sigma\in (P_1-P_1)(\RR^6)$. 
 As described in Appendix A.1, $\dLIPnorm{\,.\,}$ 
can be extended\footnote{This extension will not be needed for
				any of our technical estimates.}
to a norm $\dLIPnormT{\,.\,}$ on the linear span of ${P_1}(\RR^6)$, 
such that $\dLIPnormT{\sigma}= \dLIPnorm{\sigma}$ whenever $\sigma(\RR^6)=0$.
 The completion of the linear span of ${P_1}(\RR^6)$ w.r.t. 
$\dLIPnormT{\,.\,}$, denoted $\wM(\RR^6)$, 
is a Banach space.
 We also write $\wPpx(\RR^6)$ for ${P_1}(\RR^6)\hookrightarrow\wM(\RR^6)$.
 By $\wGaBpx$ we denote the closed subset of generic phase space points 
\beq
  \Zett =   (\mu;\zeta) \in
  \wM(\RR^6) \oplus\, ( \dotH \oplus L^2 )(\RR^3) 
\eeq
for which $\mu\in\wPpx(\RR^6)$, and for which $\zeta$
satisfies \refeq{phiNULLepsASYMP}, \refeq{varpiNULLepsASYMP}; once again,
the  Banach space direct sum is meant on the right-hand side, 
so that in particular the norm of $\Zett_t$ reads
\beq
\norm{\Zett}
: = 
\dLIPnormT{\mu} +  \norm{\zeta}_{HL}
\eeq
with 
$\norm{\zeta_t}_{HL} 
= \norm{\psi}_\dotH + \norm{\varpi}_{L^2}$.
  

\subsubsection{Generalized solutions w.r.t. to the vague topology for $\mu$}
\label{secVAGUEform}
 
 Considering first 
the field variables $\zeta_{.}\in C^1(\RR,(\dotH\oplus L^2) (\RR^3))$ 
as given, we will call \textit{$M$-vague solution} of \refeq{VfEQeps}
a $C^1$ map $t\mapsto\mu_t$ satisfying \refeq{VfEQeps} 
with $\mu_t$ in place of $f$, for all $t\in\RR$
integrated against any test function 
$g \in C_0^1(\RR^6)$, and with the $\pdt$ pulled in front of
the corresponding integral.
 Accordingly,  a map
$t\mapsto \Zett_t=(\mu_t;\zeta_t)$ $\in C^1(\RR,\GaBpx^v)$
will be called \textit{$M$-vague $HL$-strong solution} of
\refeq{VphiEQeps}, \refeq{VvpiEQeps}, \refeq{VfEQeps}.

 With the help of the concept of the $M$-vague $HL$-strong solution of
\refeq{VphiEQeps}, \refeq{VvpiEQeps}, \refeq{VfEQeps},
we can now immediately reformulate Theorem \ref{theoKKSforN} 
into an existence result for what we call 
$M$-vague $N$-body solutions of the regularized Vlasov model 
\refeq{VphiEQeps},
\refeq{VvpiEQeps},
\refeq{VfEQeps}.

\begin{Theo} 
\label{theoVAGUEsolVeps} 
        {Let
$t\mapsto(\bzN_t,\zeta^{\supN}_t)=\zett^{\supN}_t\in C^1(\RR,\GaB^{\supN})$,
with 
${\displaystyle{\lim_{t\to 0}}}\,\zett^{\supN}_t = (\bzN_0,\zeta[{\bzN_0}])$,
be the unique strong solution of the Hamiltonian field \&\ $N$-body problem
  \refeq{NphiEQeps}--\refeq{NpDOTeps}, and denote the empirical measure 
associated to $\bzN_t$ by}
\begin{equation}
  \veps[{\bzN_t}] (\dd x\dd p )
  =
  {\textstyle{\frac{1}{N}}}{\textstyle\sum}_{k=1}^N
  \delta_{q_k^{\supN}(t)}(\dd x)
  \times
  \delta_{p_k^{\supN}(t)}(\dd p)
\, 
\label{empirMEASt}
\end{equation}
        {Then 
$(t\mapsto (\veps[{\bzN_t}];\zeta^{\supN}_t) =  \Zett^{\supN}_t)$ 
$\in C^1(\RR,\GaBpx^v)$  is an $M$-vague $N$-body solution
of the regularized wave gravity Vlasov equations 
\refeq{VphiEQeps}, \refeq{VvpiEQeps}, \refeq{VfEQeps},
  satisfying the Cauchy data
${\displaystyle{\lim_{t\to 0}}}\, \Zett^{\supN}_t = \Zett^{\supN}_0$, 
and conserving mass, momentum, angular momentum, and energy:}
\beq
  \cM\left(\Zett^{\supN}_t\right) = M 
\,,
\label{MconstVeps}
\eeq
\beq
 \cP\left(\Zett^{\supN}_t\right) =P
\,,
\label{PconstVeps}
\eeq
\beq
\cJ\left(\Zett^{\supN}_t\right) = J
\,,
\label{JconstVeps}
\eeq
\beq
\cE\left(\Zett^{\supN}_t\right) = E
\,,\label{EconstVeps}
\eeq
        {with $M,P,J,E$ independent of time; in particular, $M=1$.}
\end{Theo}

 Note that by Thm. \ref{theoVAGUEsolVeps} the set 
$\GaB^{\supN}$ becomes identified with a subset of $\GaBpx^v$.
\newpage

\subsubsection{Generalized solutions w.r.t. the Kantorovich--Rubinstein 
	topology for $\mu$}
\label{secKRform}

 Since $\wGaBpx$ is equipped with a Banach space topology, a map 
$t\mapsto \Zett_t=(\mu_t;\zeta_t)$ $\in C^1(\RR,\wGaBpx)$
satisfying \refeq{VphiEQeps}, \refeq{VvpiEQeps}, \refeq{VfEQeps}
is properly called a \textit{$\wGaBpx$-strong generalized solution} 
of our regularized Vlasov model.
 Such solution satisfy the conventional conservation laws.
 Particular $\wGaBpx$-strong generalized solutions,
called \textit{$\wGaBpx$-strong $N$-body solutions},
are generated by the solutions of the field \&\ $N$-body model of section 2.
 We summarize this in 

\begin{Theo} 
\label{theoKRsolVeps} 
        {Let $t\mapsto \Zett_t=(\mu_t;\zeta_t)$ $\in C^1(\RR,\wGaBpx)$
be a {$\wGaBpx$-strong generalized solution} of
\refeq{VphiEQeps}, \refeq{VvpiEQeps}, \refeq{VfEQeps}
with Cauchy data $\lim_{t\to  0} \Zett_t = \Zett_0$.
   Then mass, momentum, angular momentum, and energy are conserved;
i.e. \refeq{MconstVeps}, \refeq{PconstVeps}, \refeq{JconstVeps}, 
\refeq{EconstVeps} hold.
 In particular, let
  $t\mapsto \zett^{\supN}_t\in C^1(\RR,\GaB^{\supN})$
  and
  $\veps[{\bzN_t}] (\dd x\dd p )$ be given as in Theorem
 \ref{theoVAGUEsolVeps}.
  Then 
$t\mapsto \Zett^{\supN}_t=(\veps[{\bzN_t}];\zeta^{\supN}_t)$ 
$\in C^1(\RR,\wGaBpx)$
  is a $\wGaBpx$-strong $N$-body solution
of the regularized wave gravity Vlasov equations 
\refeq{VphiEQeps},\refeq{VvpiEQeps},\refeq{VfEQeps},
  with Cauchy data
  ${\displaystyle{\sup_{t\to 0}}}\, \Zett^{\supN}_t = \Zett^{\supN}_0$.}
\end{Theo}

 Note that by Theorem \ref{theoKRsolVeps} the set 
$\GaB^{\supN}$ becomes identified with a subset of $\wGaBpx$.

 We next show that arbitrary initial data $\Zett_0\in \GaBpx$ 
launch a unique $\wGaBpx$-strong generalized solution 
$t\mapsto \Zett_t\in C^1(\RR,\wGaBpx)$ of our Vlasov model.
 Since the vague topology on ${P_1}$ is controlled by the 
standard Kantorovich--Rubinstein topology, solutions of the 
type $C^1(\RR,\wGaBpx)$ are automatically solutions of the 
type $C^1(\RR,\GaBpx^v)$.

\subsection{The Cauchy problem for $\wGaBpx$-strong solutions}
\label{secVlasovCauchy}

 To study the general Cauchy problem for
(\ref{VphiEQeps}), (\ref{VvpiEQeps}), (\ref{VfEQeps}) in
the $\wGaBpx$-strong topology, we rewrite 
(\ref{VphiEQeps}), (\ref{VvpiEQeps}), (\ref{VfEQeps}) 
together with their Cauchy data as a fixed point problem,
\beq
\Zett_{.} = F_{.\,,0}(\Zett_{.}|\Zett_0),
\label{VlasovWaveEQasFP}
\eeq
where $F_{.\,,0}$ is a continuous map 
from $C^0(\RR,\wGaBpx)$ into $C^0(\RR,\wGaBpx)$, 
conditioned on $\Zett_0\in \GaBpx$.
 We will show that, w.r.t. a suitably weighted sup-norm,
a truncated version of $F$ is a Lipschitz map, with Lipschitz constant $<1$, 
from a closed subset of weighted $C^0(\RR,\wGaBpx)$ into itself.
 Existence of a unique fixed point of the truncated $F$
then follows from the standard contraction mapping theorem. 
 By bootstrapping regularity, fixed points of the full $F$ will then 
be shown to exist and to be in $C^1(\RR,\wGaBpx)$, thus furnishing 
unique $\wGaBpx$-strong solutions
of (\ref{VphiEQeps}), (\ref{VvpiEQeps}), (\ref{VfEQeps}) that
conserve mass \refeq{MfuncVeps}, momentum \refeq{PfuncVeps},
angular momentum \refeq{JfuncVeps}, and energy \refeq{HfuncVeps}.

\subsubsection{Definition of the fixed point map}
\label{subsubsecFPmap}

 Given any $\Zett_0=(\mu_0;\zeta_0)\in \wGaBpx$, for each $t$ the map 
$F_{t,0}(.\,;\,.\,|\mu_0;\zeta_0)$ is given by
\beq
F_{t,0}(\mu_{.};\zeta_{.}|\mu_0;\zeta_0)
  \equiv
  \left(\Pi_{t,0}^\dagger[\zeta_{.}](\mu_0)\,;\,
  \Phi_{t,0}^{\phantom{0}}[\mu_{.}](\zeta_0) \right) \, ,
\label{Fcomponents}
\eeq
where 
$\Pi_{t,0}^\dagger[\zeta_{.}](\mu_0) \equiv
\mu_0 \circ \Pi_{0,t}^{\phantom{0}}[\zeta_{.}]$, 
and where $\Pi_{.,.}$ and $\Phi_{.,.}$
are two-parameter groups, \textit{flows} on the phase subspaces
of the particles and the fields, respectively.
  \textit{Given} a trajectory $\zeta_{.}$ in field space, 
$\Pi_{.,.}[\zeta_{.}]$ is the particle phase space flow, and 
\textit{given} a trajectory $\mu_{.}$ in probability measure space, 
 $\Phi_{.,.}[\mu_{.}]$ is  the field phase space flow.

 As to the flow on particle phase space,
let $t\mapsto \zeta_t\in C^0_b(\RR,(\dotH\oplus L^2)(\RR^3))$ be a generic,
bounded continuous curve in $(\dotH\oplus L^2)(\RR^3)$.
 Given $t\mapsto \zeta_t$, the characteristic equations for \refeq{VfEQeps} 
are the Hamiltonian equations for test particle motion
$\dd{z}/\dd{t} = \sympJ\cdot\pdz\cH(z,\zeta_{t})$,  with $z=(q,p)$, 
where $\sympJ$ 
is the symplectic matrix, and $\cH(z,\zeta_{t})$ is the Hamiltonian 
\refeq{EPSfuncH} for $N=1$ and with $(z,\zeta_{t})$ substituted for 
$\zett^{(1)}$. 
 Explicitly,
\beq
\sympJ\cdot\pdz\cH(z,\zeta_{t})
= \left( \pdp{\sqrt{1 + |p |^2}} \,,\,
   - \pdx\!\left(\fe\, {\stackrel{\phantom{.}}{*}}\, 
\psi(\,.\,,t)\right)\! (q) \right)
\,;
\label{charVecFIELD}
\eeq
note that 
only the $\psi$ part of $\zeta$ enters in \refeq{charVecFIELD}.
 The particle phase space flow $\Pi_{.,.}[\zeta_{.}]$ is now defined 
implicitly as follows:  given $t\mapsto\psi(\,.\,,t)$, for each solution 
$z_. \in C^1(\RR,{\RR^6})$ of the characteristic equations 
the integrated characteristic equations give the identity
\beq
    z_t =  z_{t^\prime} + \int_{t^\prime}^t  
\sympJ\cdot\pdz\cH(z_\tau,\zeta_{\tau}) 
\dd \tau
    =:  \Pi_{t,t^\prime}[\zeta_.] \left(z_{t^\prime}\right)
, \label{defPARTICLEflow}
\eeq
the r.h.s. of which being the transition function from some $z$ at time 
$t^\prime$ to another $z$ at time $t$ for \textit{all} $t$, $t^\prime$; 
considering the totality of all $t$, $t^\prime$ gives the particle flow. 

  Similarly, to define the flow on field phase space, suppose
 $t\mapsto\mu_t \in C^0(\RR,\wPpx({\RR^6}))$ is given, and let
$\rho_t(\dd x) = \int \mu_t(\dd x\dd p)$, and 
$(\fe*\rho_t)(x) = \int\fe(y-x)\rho_t(\dd y)$.
 Then $\frac{1}{2 T}\int_{-T}^T \norm{\fe * \rho_t}^2_{L^2(\RR^3)} \dd t
  \leq 
\norm{\fe*\fe}_{\mathrm{u}} \left(\equiv \norm{\fe}_{L^2}^2\right)$.    
  Given such $\fe * \rho_{.} \in L^2([-T,T], L^2(\RR^3))$ for any $T > 0$, 
the solution $\zeta_{.} = (\psi,\varpi)(\,.\,,.\,)$ 
to the wave equation with field source  $\fe * \rho_{.}$ 
defines the flow $\Phi_{.,.}[\mu_.]$ on field space through
the transition function
\begin{equation}
  \zeta_t =: \Phi_{t,t^\prime}[\mu_.] (\zeta_{t^\prime})
\,.
\label{waveEqFlow}
\end{equation}
 An explicit representation of \refeq{waveEqFlow} in terms of
Fourier \& Laplace transforms is available. 
 However, by the higher regularity of $\fe * \rho_{.}$, the
$\dotH$ and $L^2$ estimates of $\Phi_{.,.}[\mu_.]$ 
are conveniently obtained from Kirchhoff's explicit pointwise 
expressions for classical solutions.
 In components,  $\!\Phi_{.,.}[\mu_.]\! 
 \equiv\! 
 (\Phi_{.,.}^\psi[\mu_.],\Phi_{.,.}^\varpi[\mu_.])$
reads (\cite{BrezisBOOK}, \cite{Ikawa}, \cite{ShatahStruweBOOK})
\begin{eqnarray}
  \psi(x,t)
\!\!\!  & = &\!\!\!
  \SSint
\Bigl(\left[1 + (t-t^\prime) \Omega\cdot\nabla\right] \psi(x^\prime,{t^\prime})
               +(t-t^\prime) \varpi(x^\prime,{t^\prime})
     \Bigr.
  \cr
  &&
\qquad \hskip+2.7truecm
-\Bigl.
  \int_{t^\prime}^t (t-t^\prpr) (\fe* \rho_{t^{\prpr}})(x^{\prpr}) 
  \dd t^{\prpr}
     \Bigr)
  \dd \Omega 
\cr
  & =: &\!\!\!  
\Phi_{t,t^\prime}^\psi[\mu_.] (\zeta_{t^\prime})(x)
  \label{ea05}
  \\
  \varpi(x,t)
\!\!\! & = &\!\!\!
 \SSint \Bigl(\bigl[1 + 2(t-t^\prime) \Omega\cdot\nabla\bigr]
               \Omega\cdot\nabla \psi(x^\prime,{t^\prime})
+\bigl[1+(t-t^\prime)\Omega\cdot\nabla\bigr]\varpi(x^\prime,{t^\prime})\Bigr. 
  \cr
  &&
\qquad \hskip+3.7truecm
-\Bigl.
              \int_{t^\prime}^t [1 + (t-t^{\prpr}) \Omega\cdot\nabla]
             (\fe* \rho_{t^{\prpr}})(x^{\prpr})  \dd t^{\prpr}
	\Bigr) {\dd \Omega}
\cr
  & =: &\!\!\! 
 \Phi_{t,t^\prime}^\varpi[\mu_.] (\zeta_{t^\prime})(x)
\, ,
  \label{ea06}
\end{eqnarray}
where 
$x^{\prime\,\mathrm{or}\,\prpr} 
=x + (t-t^{\prime\,\mathrm{or}\,\prpr})\Omega$, 
where $\Omega\in\SS^2$, and where  $\SSint$ is short for 
$\frac{1}{4\pi}\int_{\SS^2}$.

 Having defined $ \Pi_{.,.}[\zeta_{.}]$ and $\Phi_{.,.}[\mu_.]$,  we are 
now ready to analyse equation \refeq{VlasovWaveEQasFP}.

\subsubsection{Statement of the main fixed point results}
\label{secCONTRACTION}

 So far, \refeq{VlasovWaveEQasFP} has been defined purely formally
as a rewriting of \refeq{VphiEQeps}, \refeq{VvpiEQeps}, \refeq{VfEQeps},
with Cauchy data imposed.
 Our first duty should be to show that \refeq{VlasovWaveEQasFP} 
in fact makes sense, viz. that $F$ maps a relevant, closed subset
of $C^0(\RR,\wGaBpx)$ into itself, indeed.
 We prove this as a byproduct of the auxiliary result that a 
truncated version of $F_{.\,,0}(\,.\,|\Zett_0)$ is a Lipschitz map, 
with Lipschitz constant $<1$, from a closed subset of $C^0(\RR,\wGaBpx)$ 
into itself, where ``closed'' is meant w.r.t. (a suitably weighted) sup-norm. 

 We note that by a density argument for the curves of empirical 
measures $t\mapsto\rho^{\supN}_t$, the a-priori estimates of section 
\ref{secAprioriNboundsINt}  extend to our regularized Vlasov model.
 Hence, any Vlasov solution $t\mapsto \Zett_t$ must be in some subset 
of $C^0(\RR,\wGaBpx)$ satisfying $\norm{\Zett_t} \leq c_0 + c_1 |t| +
c_2t^2$ for some positive constants $c_0$, $c_1$ and $c_2$.
 This suggests to work with the closure of the bounded continuous 
functions from $\RR$ to $\wGaBpx$, denoted $C^0_b(\RR,\wGaBpx)$, 
w.r.t. a \textit{weighted sup-norm} of ${\Zett_.}$ given by 
$\sup_{t \in \RR} \left( [c_0 + c_1 |t|+c_2 t^2]^{-1} \norm{\Zett_t}\right)$; 
however, for technical reasons it is more convenient to close
$C^0_b(\RR,\wGaBpx)$ w.r.t. the weighted sup-norm 
\begin{equation}
  \norm{\Zett_.}_{\mathrm{w}}
  =
  \sup_{t \in \RR} 
 \left( e^{-w|t|} \norm{\Zett_t}\right)
  \label{ea15a}
\end{equation}
for some $w > 0$; eventually we will restrict $w$ to $w >{\wcrit}>0$. 
 The closure of $C^0_b(\RR,\wGaBpx)$ w.r.t. norm \refeq{ea15a}
is a Banach space, denoted  $\Cw^0(\RR,\wGaBpx)$; the subscript 
w can be read as meaning both ``weighted'' and reference to 
the parameter $w$ in the definition \refeq{ea15a}.
 We also introduce the Banach space  $\Cw^0(\RR_+,\wGaBpx|\Zett_0)$ with norm
 $\norm{\Zett_.}_{\mathrm{w}}
   = \sup_{t\geq 0} \left(e^{-wt} \norm{\Zett_t}\right)$.
 By the time reversal symmetry of  \refeq{VlasovWaveEQasFP}
it suffices to limit the discussion to $t\geq 0$.

 Moreover, as regards the argument $\Zett_{.} (=(\mu_{.};\zeta_{.}))$ of 
$ F_{.\,,0}(\,.\,|\Zett_0) (= F_{.\,,0}(\, .\, ;\, .\, |\mu_0;\zeta_0))$
given in \refeq{Fcomponents}, it is not a-priori required that 
$\lim_{t\to 0}\Zett_t$ of $\Zett_{.}$ coincides with the given $\Zett_0$;
in fact, it is not even a-priori necessary that the measure component of 
$\Zett_.$ is in $P_1$ but could as well be in $\wM$.
 However, for the solution of \refeq{VlasovWaveEQasFP} this must be so, 
for $F_{.\,,0}(\, .\, ;\, .\, |\mu_0;\zeta_0)$ has been constructed 
such that 
$F_{0,0}(\mu_{.};\zeta_{.}|\mu_0;\zeta_0) = (\mu_0;\zeta_0)= \Zett_0$,
as is readily verified by inspection of $\Pi_{0,0}$ and $\Phi_{0,0}$.
 Therefore, with the exception of some technical estimates that we will
highlight explicitly, we only need to apply 
$F_{.\,,0}(\,.\,|\Zett_0)$ to those 
$\Zett_{.}\in C^0(\RR,\wGaBpx)$ which satisfy 
$\lim_{t\to 0}\Zett_t = \Zett_0$.
 We denote the corresponding subsets of $\Cw^0(\RR,\wGaBpx)$ and
$\Cw^0(\RR_+,\wGaBpx)$ by 
$\Cw^0(\RR,\wGaBpx|\Zett_0)$ and
$\Cw^0(\RR_+,\wGaBpx|\Zett_0)$, respectively.
 Furthermore we denote the free evolution of the initial data $\Zett_0$ by 
$\Zett_{.}^0:=F_{.,0}(0_{.}|\Zett_0)\in\Cw^0(\RR,\wGaBpx|\Zett_0)$;
here, $0_{.}$ is the trivial constant map 
$(t\mapsto 0)\in
\Cw^0\left(\RR,\wM(\RR^6) \oplus\,\dotH(\RR^3) \oplus L^2 (\RR^3)\right)$.
 We will work with certain closed subsets of 
$\Cw^0(\RR_{(+)},\wGaBpx|\Zett_0)$.
 By $B_{\widetilde{R}}(\Zett_.^0)\subset\Cw^0(\RR_{(+)},\wGaBpx|\Zett_0)$ 
we denote a closed ball of radius $\widetilde{R}$ 
centered at $\Zett_{.}^0$ (with $t\in\RR$ or $\RR_+$).
 Furthermore, we shall need the closed subsets
defined by the condition 
$\sup_{t\geq 0} \norm{\psi(\,.\,,t)}_\dotH \leq C_\psi$
with $C_\psi\geq \norm{\psi_0}_\dotH$.
 In this vein, we also introduce a truncation of 
$F_{.\,,0}(\,.\,|\Zett_0)$, denoted $\overline{F}_{.\,,0}(\,.\,|\Zett_0)$, 
which for each $t>0$ is obtained from $F_{t,0}(\,.\,|\Zett_0)$ by replacing 
$\Phi_{t,0}^\psi[\mu_.](\zeta_0)$ by
\begin{equation}
\overline{\Phi}_{t,0}^\psi[\mu_.](\zeta_0) 
:=
\min\left\{
	1 ,  {C_\psi} \norm{\Phi_{t,0}^\psi[\mu_.](\zeta_0)}_\dotH^{-1}
    \right\}
\Phi_{t,0}^\psi[\mu_.](\zeta_0)
\label{PHItrunc}
\end{equation}

\begin{Prop} 
\label{olFisCONTRACTmap}
{For every $\Zett_0\in \GaBpx$, there exist 
  $\Ccrit_\psi\geq \norm{\psi_0}_\dotH$ and $\wcrit>0$, 
  such that $\olF_{.\,,0}(\,.\,|\Zett_0)$ is a 
  Lipschitz map with Lipschitz constant $<1$ which maps the closed 
  subsets of balls 
  $B_{\widetilde{R}}(\Zett_.^0)\subset\Cw^0(\RR_+,\wGaBpx|\Zett_0)$ 
  for which
  $\sup_{t\geq 0} \norm{\psi(\,.\,,t)}_\dotH \leq C_\psi$
  into themselves whenever 
  $\widetilde{R} \geq\norm{\Zett_{.}^0}_{\mathrm{w}}$, 
  $w>\wcrit$, and $C_\psi\geq\Ccrit_\psi\geq \norm{\psi_0}_\dotH$.}
\end{Prop}

 By the standard contraction mapping theorem, an immediate corollary 
to Proposition \ref{olFisCONTRACTmap} is the existence of a unique fixed 
point $(t\mapsto \Zett_t)\in \Cw^0(\RR_+,\wGaBpx)$, with 
$(t\mapsto\psi(\,.\,,t))\!\in\! C^0_b(\RR_+, \dotH(\RR^3))$,
of the fixed point equation with the truncated $F$,
\beq
\Zett_{.} = \olF_{.\,,0}(\Zett_{.}|\Zett_0).
\label{truncFPeq}
\eeq
 By bootstrapping regularity, fixed points of the untruncated $F$ will 
then be shown to exist and to actually be in $C^1(\RR,\wGaBpx)$,  furnishing 
unique $\wGaBpx$-strong  Vlasov solutions.
 Thus we may state our main existence and uniqueness theorem of this section.
\newpage

\begin{Theo} 
\label{ExistUnique}
 {For every  $\Zett_0 \in \GaBpx$ there exists $\wcrit>0$ such 
	that whenever $w>\wcrit$,
	the Vlasov fixed point equation \refeq{VlasovWaveEQasFP}
	with Cauchy data $\lim_{t\to{0}}\Zett_t = \Zett_0$
        is solved by a unique curve $t\mapsto \Zett_t\in \Cw^0(\RR,\wGaBpx)$;
	since also $\psi_0\in(\dotH\cap\dotHH)(\RR^3)$, the map
	$t\mapsto \Zett_t\in (\Cw^0\cap C^1)(\RR,\wGaBpx)$, 
	and thus it is the unique $\wGaBpx$-strong solution to 
	(\ref{VphiEQeps}), (\ref{VvpiEQeps}), (\ref{VfEQeps}) 
	conserving mass
	\refeq{MfuncVeps}, momentum \refeq{PfuncVeps},
	angular momentum \refeq{JfuncVeps}, and energy \refeq{HfuncVeps}.
}
\end{Theo}

\subsubsection{Proof of Proposition \ref{olFisCONTRACTmap}}
\label{secCONTRACTIONproof}

 We begin with auxiliary results concerning the flow on the particle 
sub-phase space.

\begin{Lemm}
\label{Gestimates}
  Given any curve $\zeta_{.}\in C^k(\RR,(\dotH\oplus L^2)(\RR^3))$,
$k=0,1,...$, we have

\ \ (i) $\sympJ\cdot\pdz\cH(\,.\,,\zeta_{.})\in C^k(\RR \times \RR^6, \RR^6)$,

  \ (ii) $\sympJ\cdot\pdz\cH(\,.\,,\zeta_{t}) \in C^\infty(\RR^6, \RR^6)$;

  (iii) $\pdz\cdot\sympJ\cdot\pdz\cH(\,.\,,\zeta_{t}) \equiv 0$;
  
\  (iv) $\abs{\sympJ\cdot\pdz\cH(\,.\,,\zeta_{t})}    \leq 
	1 + \norm{\fe}_{L^2}\norm{\psi(\,.\,,t)}_{\dotH}$.
\end{Lemm}

\noindent 
\textit{Proof of Lemma \ref{Gestimates}:} 
  Regularity (i), (ii), and incompressibility (iii), are obvious.
  The bound (iv) obtains by using the triangle inequality, 
then $|p|\leq \sqrt{1+\abs{p} {}^2}$ for the momentum part, 
respectively for the space part the Cauchy--Schwarz inequality 
to get 
$\abs{\fe*\nabla\psi}(x) \leq \norm{\fe}_{L^2}\norm{\psi}_{\dotH}$ 
for all $x$; cf. \refeq{forceBOUND}.\QED
\smallskip

 As a straightforward spin-off of Lemma \ref{Gestimates}, we have 

\begin{Coro}
\label{PIsymplecto}
 {If $\zeta_{.}\! \in\! C^k(\RR, (\dotH\oplus L^2)(\RR^3))$,
   then $\Pi_{.,.}[\zeta_\cdot]\!\in\! C^k(\RR\times\RR\times\RR^6,\RR^6)$,
  and $\Pi_{t,t^\prime}[\zeta_\cdot]$ is a symplectomorphism 
  $\forall t,t^\prime\in\RR$;
  in particular, $\det \partial_z \Pi_{t,t^\prime}[\zeta_\cdot](z) = 1$.}
\end{Coro}

\noindent
\textit{Proof of Corollary \ref{PIsymplecto}:} 
 This is a standard corollary. See, e.g. \cite{HirschSmale}.\QED
\bigskip

 Controlling the field space component of $\olF_{.\,,0}$ requires only
the following Lemma:

\begin{Lemm}
\label{LemmPHIdiffable}
 {If $\mu_{.}\in C^0(\RR,\wPpx)$, then
$\Phi_{.,.}[\mu_.] \in 
C^0(\RR\times\RR \times (\dotH \oplus L^2)(\RR^3), (\dotH\oplus L^2)(\RR^3))$.}
\end{Lemm}

\noindent
\textit{Proof of Lemma \ref{LemmPHIdiffable}:}
 For $\zeta_0$ classical: straightforward calculation, 
for \refeq{ea05}, \refeq{ea06} are quite explicit.
 Then apply  the Hahn-Banach theorem.Q.E.D.
\smallskip

\noindent
\textit{Proof of Proposition \ref{olFisCONTRACTmap}:} 
 We first show that, given any $\Zett_0\in\GaBpx$, the map 
$F_{\,.\,,0}(\,.\,|\Zett_0)$ is Lipschitz-continuous from a closed subset of 
$\Cw^0(\RR_+,\wGaBpx|\Zett_0)$, defined by the condition 
$\sup_{t\geq 0} \norm{\psi(\,.\,,t)}_\dotH \leq C_\psi$
with $C_\psi\geq\Ccrit_\psi \geq  \norm{\psi_0}_\dotH$,
to $\Cw^0(\RR_+,\wGaBpx|\Zett_0)$ whenever $w>\wcrit$, with $\wcrit$ depending 
at most on $\fe,C_\psi$, and the Lipschitz constant at most on $\fe,C_\psi,w$. 
 We emphasize that the conditioning $\lim_{t\to 0}\Zett_t =\Zett_0$ and
$\lim_{t\to 0}\widetilde{\Zett}_t =\Zett_0$ implied by the definition
of $\Cw^0(\RR_+,\wGaBpx|\Zett_0)$ do not enter our estimates.

 To break up the proof into two parts, 
we use the triangle inequality in the form
\bea 
\norm{F_{.,0}(\mu_{.};\zeta_{.}|\Zett_0) - 
      F_{.,0}(\tilde\mu_{.};\tilde\zeta_{.}|\Zett_0)}_{\mathrm{w}}
\!\!
 &\leq &
\norm{F_{.,0}(\mu_{.};\zeta_{.}|\Zett_0) - 
      F_{.,0}(\tilde\mu_{.};\zeta_{.}|\Zett_0)}_{\mathrm{w}}
\cr
&& \!\!\!\!
+ 
\norm{F_{.,0}(\tilde\mu_{.};\zeta_{.}|\Zett_0) -
      F_{.,0}(\tilde\mu_{.};\tilde\zeta_{.}|\Zett_0)}_{\mathrm{w}}
\,.
\label{normFtriangleONE}
\eea
 Given $\Zett_0 = (\mu_0,\zeta_0)$ and 
$C_\psi \geq \norm{\psi_0}_\dotH$, we show that, 
(a) given $\zeta_{.} \in C^0(\RR_+,(\dotH\oplus L^2)(\RR^3))$
satisfying $\norm{\psi_t}_\dotH \leq C_\psi$ for all $t>0$, 
for any two $\mu_{.}$ and $\tilde\mu_{.}$ in $C^0(\RR_+,\wPpx)$
and all $w>0$ we have
\beq
\norm{F_{.,0}(\mu_{.};\zeta_{.}|\Zett_0) -
      F_{.,0}(\tilde\mu_{.};\zeta_{.}|\Zett_0)}_{\mathrm{w}}
\leq
L_1[\varrho;w] 
\sup_{t \geq 0}
\left(e^{-wt}\dLIPnorm{\mu_{t} - \tilde\mu_t}\right)
\label{FmuLIP}
,
\eeq
and (b), given $\mu_{.}\in C^0(\RR_+,\wPpx)$,
for any two $\zeta_{.}$ and $\tilde\zeta_{.}$ in
$C^0(\RR_+,(\dotH\oplus L^2)(\RR^3))$, satisfying 
$\max\{\norm{\psi_t}_\dotH,\norm{{\tilde{\psi}}_t}_\dotH\} \leq C_\psi$
for all $t>0$, 
and for all $w>{\wcrit}[\varrho;C_\psi]$ we have
\beq
\norm{F_{.,0}(\mu_{.};\zeta_{.}|\Zett_0) -
      F_{.,0}(\mu_{.};\tilde\zeta_{.}|\Zett_0)}_{\mathrm{w}}
\leq
L_2[\fe;w,{\wcrit}] 
\sup_{t\geq 0}
\left(e^{-wt}\norm{{\zeta_{t} - \tilde\zeta_{t}}}_{HL}\right)
\label{FzetaLIP}
;
\eeq
for then it follows from 
\refeq{normFtriangleONE}, \refeq{FmuLIP}, \refeq{FzetaLIP} that, 
given any $\Zett_0$ and $C_\psi\geq\Ccrit_\psi\geq \norm{\psi_0}_\dotH$,
\beq
\norm{F_{.,0}(\Zett_{.}|\Zett_0) - F_{.,0}(\widetilde{\Zett}_{.}|\Zett_0)}_{\mathrm{w}}
\leq 
L[\varrho;w,\wcrit]
\norm{\Zett_{.}-\widetilde{\Zett}_{.}}_{\mathrm{w}}
\,
\label{FLIPtrajT}
\eeq
whenever $w>{\wcrit}[\varrho,C_\psi]$, with 
$L[\varrho;w,{\wcrit}]:=\max\{L_1[\varrho;w], 
          L_2[\varrho;w,{\wcrit}]\}$.
 
 \emph{Part a)} 
To prove \refeq{FmuLIP}, we fix  $\Zett_0$ and $\zeta_{.}$ 
and note that in this case
\beq
\norm{F_{t,0}(\mu_{.};\zeta_{.}|\Zett_0) - 
      F_{t,0}(\tilde\mu_{.};\zeta_{.}|\Zett_0)}
=
\norm{ \Phi_{t,0}[\mu_{.}] (\zeta_{0}) -  
       \Phi_{t,0}[\tilde\mu_{.}] (\zeta_{0})}_{HL}
\,,
\label{FmuDIFFidA}
\eeq
where, in components,
\bea
\norm{ \Phi_{t,0}[\mu_{.}] (\zeta_{0}) -  
       \Phi_{t,0}[\tilde\mu_{.}] (\zeta_{0})}_{HL}
\!\!\!
&=&
\| {\Phi_{t,0}^\psi[\mu_{.}] (\zeta_{0}) - 
    \Phi_{t,0}^\psi[\tilde\mu_{.}] (\zeta_{0})}\|_{\dotH}
\cr
&&\!\!\!\!
+
\norm{\Phi_{t,0}^\varpi[\mu_{.}] (\zeta_{0}) - 
      \Phi_{t,0}^\varpi[\tilde\mu_{.}] (\zeta_{0})}_{L^2}
\label{HLequalsHandL}
\,.
\eea
 Furthermore, 
using \refeq{ea05} and then the definition of $\norm{\,.\,}_{\dotH}$, we have
\beq
\norm{\Phi_{t,0}^\psi[\mu_{.}] (\zeta_{0}) - 
      \Phi_{t,0}^\psi[\tilde\mu_{.}] (\zeta_{0})}_{\dotH}^2
=
\int\left| \int_{0}^t \SSint
      (t-t^{\prime}) 
\nabla [\fe* (\rho_{t^{\prime}} - \tilde\rho_{t^{\prime}})](x^{\prime}) 
                              \dd \Omega \dd t^{\prime}
\right|^2\!\!\dd{x}
\,,
\label{PHIpsiDIFFnorm}
\eeq
while with \refeq{ea06} and the definition of $\norm{\,.\,}_{L^2}$, we find
\newpage

\bea
\norm{\Phi_{t,0}^\varpi[\mu_{.}] (\zeta_{0}) - 
      \Phi_{t,0}^\varpi[\tilde\mu_{.}] (\zeta_{0})}_{L^2}^2
 &&
\cr
&&\hskip-2.5truecm
=
\int\left| \int_{0}^t \SSint
\left(1 + (t-t^{\prime}) \Omega\cdot\nabla\right)
 [\fe* (\rho_{t^{\prime}} - \tilde\rho_{t^{\prime}})](x^{\prime}) 
                              \dd \Omega \dd t^{\prime}
\right|^2\!\!\dd{x}
\,.\qquad
\label{PHIvpiDIFFnorm}
\eea

 As to \refeq{PHIpsiDIFFnorm}, triangle and Jensen's inequalities, 
and Fubini, yield the estimate
\beq
\norm{\Phi_{t,0}^\psi[\mu_{.}] (\zeta_{0}) - 
      \Phi_{t,0}^\psi[\tilde\mu_{.}] (\zeta_{0})}_{\dotH}^2
\leq
\SSint \int \!\!
\left[\int_{0}^t  (t-t^{\prime})
\left|\nabla \left[\fe\, {\stackrel{\phantom{.}}{*}}\,
		(\rho_{t^{\prime}}^{\phantom{A}} - 
		 \tilde\rho_{t^{\prime}})
	     \right]\!(x^{\prime})
\right|\dd t^{\prime}
\right]^{\!2}\!\!\dd{x}  \dd \Omega 
\,.
\label{PHIpsiDIFFnormESTa}
\eeq
 Now multiply \refeq{PHIpsiDIFFnormESTa} by $e^{-2wt}$,
pull $e^{-2wt}$ under the square in rhs\refeq{PHIpsiDIFFnormESTa} 
and note that  
\bea
&& \hskip-.8truecm
\int \left[ e^{-wt}\!\! \int_{0}^t 
\! (t-t^{\prime})
\left|\nabla \left[\fe\, {\stackrel{\phantom{.}}{*}}\,
		(\rho_{t^{\prime}}^{\phantom{A}} - 
		 \tilde\rho_{t^{\prime}})
	     \right]\!(x^{\prime})
\right|\dd t^{\prime}
\right]^{\!2}\!\!\dd{x}
\cr
\!\!\!&= &\!\!\!
\int\left[\!\int_{0}^t 
\left(e^{-\frac{1}{2}w(t-t^\prime)}  (t-t^{\prime})\right)
\left(e^{-\frac{1}{2}w(t-t^\prime)}  e^{-wt^\prime} 
\left|\nabla \left[\fe\, {\stackrel{\phantom{.}}{*}}\,
		(\rho_{t^{\prime}}^{\phantom{A}} - 
		 \tilde\rho_{t^{\prime}})
	     \right]\!(x^{\prime})
\right|
\right)\dd t^{\prime}
\right]^{\!2}\!\!\dd{x}
\hfill
\cr
\!\!\!&\leq &\!\!\!
\int_{0}^t 
e^{-w(t-t^\prpr)} (t-t^{\prpr})^2 \dd t^{\prpr}
\int_{0}^t e^{- w(t-t^\prime)}  e^{-2wt^\prime}
\int\!
\left|\nabla \left[\fe\, {\stackrel{\phantom{.}}{*}}\,
		(\rho_{t^\prime}^{\phantom{A}} - 
		 \tilde\rho_{t^\prime})
	     \right]\!(x^\prime)
\right|^2\dd{x} \dd t^{\prime}
\cr
\!\!\!&\leq &\!\!\!\!
{\textstyle{\frac{2}{w^4}}}
\sup_{t^\prime\geq 0}
\left(e^{-2wt^\prime}
\int\!
\left|\nabla \left[\fe\, {\stackrel{\phantom{.}}{*}}\,
		(\rho_{t^\prime}^{\phantom{A}} - 
		 \tilde\rho_{t^\prime})
	     \right]\!(x^\prime)
\right|^2\dd{x} 
\right)
\!,
\label{estHoneB}
\eea
the first inequality by Cauchy--Schwarz, followed  by Fubini,
the second inequality by H\"older followed by
$\int_{0}^t e^{-w(t-t^\prpr)} (t-t^{\prpr})^2 \dd t^{\prpr}
 \int_{0}^t e^{-w(t-t^\prime)} \dd t^{\prime}
\leq 
\int_0^\infty  e^{-w\tau} \tau^2 \dd{\tau}
\int_0^\infty e^{-w\tau}\dd{\tau} = 2/w^4$.
 We next estimate the remaining $\dd{x}$ integral by itself.
 For this we first rewrite it with the help of 
one of Green's identities, a change of integration variables 
$x \to x^\prime$, and Fubini's theorem,
exchanging the $\dd{x^\prime}$ integration with one of the convolution 
integrations ($\dd{y}$, say); we then apply  the Kantorovich--Rubinstein 
duality twice to obtain generalized H\"older estimates,
then use the estimate 
$\dLIPnorm{\rho-{\tilde\rho}}\leq \dLIPnorm{\mu-\tilde\mu}$
for $\rho(\dd{x}) = \mu(\dd{x}\times\RR^3)$ (similarly for $\tilde\rho$).
 Thus, independently of $\Omega$, we have
\bea  
\int
\left|
\nabla \left[\fe\, {\stackrel{\phantom{.}}{*}}\,
(\rho_{t^{\prime}}^{\phantom{A}} - 
 \tilde\rho_{t^{\prime}})\right](x^{\prime}
)\right|^2 \!\dd{x}
\!\!\!&= &\!\!\!
- \!\int\!\!
\left( \fe \!\stackrel{\phantom{.}}{*}\! 
\nabla^2 \fe \!\stackrel{\phantom{.}}{*}\!
(\rho_{t^{\prime}} - \tilde\rho_{t^{\prime}})
\right)\!(y)\,
(\rho_{t^{\prime}} - \tilde\rho_{t^{\prime}})
(\dd{y})
\cr
\!\!\!&\leq &\!\!\!
\LipSnorm{ \fe \!\stackrel{\phantom{.}}{*}\! 
\nabla^2 \fe \!\stackrel{\phantom{.}}{*}\!
(\rho_{t^{\prime}} - \tilde\rho_{t^{\prime}})}
\dLIPnorm{\rho_{t^{\prime}}^{\phantom{A}} - \tilde\rho_{t^{\prime}}}
\cr
\!\!\!&\leq &\!\!\!
\LipSnormITERA{ \fe * \nabla^2 \fe} 
\dLIPnorm{\rho_{t^{\prime}}^{\phantom{A}} - 
          \tilde\rho_{t^{\prime}}}^2
\cr
\!\!\!&\leq &\!\!\!
\LipSnormITERA{\fe*\!\nabla^2 \fe} 
\dLIPnorm{\mu_{t^{\prime}}^{\phantom{A}} - 
          \tilde\mu_{t^{\prime}}}^2
\,,
\label{HoneOfNablaVarrhoEST}
\eea
where $\LipSnormITERA{ \fe *\!\nabla^2 \fe} $ is the 
\emph{iterated Lipschitz constant}\footnote{The 
		iterated Lipschitz constant of $f$ is given by
\bea
\LipSnormITERA{f} = 
\sup_{x\neq y}\sup_{\tilde{x}\neq \tilde{y}}
\frac{|f(x-\tilde{x})+f(y-\tilde{y})-f(x-\tilde{y})-f(y-\tilde{x})|}
     {|x-y||\tilde{x}-\tilde{y}|}.
\nonumber
\eea
		 If $f\in C^2(\RR^d)$, then 
$\LipSnormITERA{f} = \sup_{x\in\RR^d} \norm{\nabla^{\otimes 2} f(x)}_\infty$, 
		where $\nabla^{\otimes 2} f(x)$ is the Hessian of $f$ at $x$ 
		and $\norm{M}_\infty$ the sup norm (i.e. spectral radius)
		of a real symmetric matrix $M$.}
of ${\fe *\!\nabla^2 \fe}$.
 We estimate rhs\refeq{estHoneB} with the help of 
\refeq{HoneOfNablaVarrhoEST}, which in turn estimates 
($e^{-2wt}$rhs\refeq{PHIpsiDIFFnormESTa}) in such a way
that the integration over $\dd\Omega$ now factors out,
yielding the factor unity.
 Thus, taking 
$\sup_{t\geq 0}(e^{-2wt}\, \mathrm{l.h.s.}\refeq{PHIpsiDIFFnormESTa})$
and then square roots yields
\beq
\sup_{t\geq 0}
\left(
  e^{-wt} 
\norm{\Phi_{t,0}^\psi[\mu_{.}] (\zeta_{0}) - 
      \Phi_{t,0}^\psi[\tilde\mu_{.}] (\zeta_{0})}_{\dotH}
\right)
\leq
\sqrt{\LipSnormITERA{\fe*\!\nabla^2 \fe}{\textstyle{\frac{2}{w^4}}}}
\sup_{t\geq 0}
\left(
  e^{-wt} \dLIPnorm{\mu_{t}^{\phantom{A}} - 
                    \tilde\mu_{t}}
\right)
\!.\label{PHIpsiDIFFnormSEMIfinalEST}
\eeq

 As for \refeq{PHIvpiDIFFnorm}, we proceed similarly, except that
after the Cauchy-Schwarz and Fubini steps we now use also that
\beq
\int\left|
\left(1 + (t-t^{\prime}) \Omega\cdot\nabla\right)g(x) 
\right|^2\!\dd{x}
=
\int\left(
g(x) \left(1 - (t-t^{\prime})^2(\Omega\cdot\nabla)^2\right)g(x) 
\right)\dd{x}
\label{anAUXid}
\eeq
where 
$g(x) = [\fe\, {\stackrel{\phantom{.}}{*}}\,
		(\rho_{t^{\prime}}^{\phantom{A}} - 
		 \tilde\rho_{t^{\prime}})](x^\prime)$,
and obtain 
\bea
\!\!\!\!\!&
{\displaystyle{\sup_{t\geq 0}}}
&\!\!\!
\left(
  e^{-wt} 
\norm{\Phi_{t,0}^\varpi[\mu_{.}] (\zeta_{0}) - 
      \Phi_{t,0}^\varpi[\tilde\mu_{.}] (\zeta_{0})}_{L^2}
\right)
\leq
\cr
&&
\quad
\sqrt{ \LipSnormITERA{\fe*\fe}\!{\textstyle{\frac{1}{w^2}}}
   +\! \LipSnormSMALLitera{\fe*\left(\Omega_0\cdot\nabla\right)^2\fe}
	               {\textstyle{\frac{2}{w^4}}}
     }\;
\sup_{t\geq 0}
\left(
  e^{-wt}  \dLIPnorm{\mu_{t}^{\phantom{A}} - 
                       \tilde\mu_{t}}
\right),\quad
\label{PHIvpiDIFFnormSEMIfinalEST}
\eea
where $\Omega_0\in\SS^2$ is arbitrary. 
 
 We now recall \refeq{HLequalsHandL}.
 Noting that, by triangle inequality,
$\sup_{t\geq 0} \left(e^{-wt}\,\mathrm{l.h.s}\refeq{HLequalsHandL} \right)$
is not bigger than the sum of \refeq{PHIpsiDIFFnormSEMIfinalEST}
and \refeq{PHIvpiDIFFnormSEMIfinalEST}, we arrive at
\beq
\sup_{t\geq 0}
\left(
  e^{-wt}\norm{ \Phi_{t,0}[\mu_{.}] (\zeta_{0}) -  
                \Phi_{t,0}[\tilde\mu_{.}] (\zeta_{0})}_{HL}
\right)
\leq
L_1[\varrho;w] 
\sup_{t\geq 0}
\left(
  e^{-wt}
  \dLIPnorm{\mu_{t}^{\phantom{A}} - 
            \tilde\mu_{t}}
\right)
,
\eeq
with
\beq
L_1[\varrho;w] 
= 
\sqrt{\LipSnormITERA{\fe*\!\nabla^2 \fe} {\textstyle{\frac{2}{w^4}}}}
\,
+
\sqrt{ \LipSnormITERA{\fe*\fe}\!{\textstyle{\frac{1}{w^2}}}
   +\! \LipSnormSMALLitera{\fe*\left(\Omega_0\cdot\nabla\right)^2\fe}
	               {\textstyle{\frac{2}{w^4}}}
     } \,
\,.
\label{LoneCONST}
\eeq
 Finally recalling \refeq{FmuDIFFidA}, we see that our proof of \refeq{FmuLIP} 
is concluded.

 \emph{Part b)}
To prove \refeq{FzetaLIP}, we fix  $\Zett_0$, 
and $\mu_{.}$ 
 and note that in this case
\beq
\norm{F_{t,0}(\mu_{.};\zeta_{.}|\Zett_0) - 
      F_{t,0}(\mu_{.};\tilde\zeta_{.}|\Zett_0)}
=
\dLIPnorm{ \mu_0\circ\Pi_{0,t}[\zeta_{.}] - 
           \mu_0\circ \Pi_{0,t}^{\phantom{t}}[\tilde\zeta_{.}]}
\label{FzetaDIFFid}
\,.
\eeq
 Recalling \refeq{defLdist} and Corollary \ref{PIsymplecto}, 
we note next that
\bea
\dLIPnorm{ \mu_0\circ\Pi_{0,t}[\zeta_{.}] - 
           \mu_0\circ \Pi_{0,t}^{\phantom{t}}[\tilde\zeta_{.}]}
\!\!\!&= &\!\!\!
    \sup_{{g \in C^{0,1}(\RR^6)}\atop{\LipSnorm{g}\leq 1}}
\Abs{{ \int} g\, 
\dd(\mu_0\circ\Pi_{0,t}[\zeta_{.}] - 
    \mu_0\circ \Pi_{0,t}^{\phantom{t}}[\tilde\zeta_{.}])} 
\cr
&= &\!\!\!
    \sup_{{g \in C^{0,1}(\RR^6)}\atop{\LipSnorm{g}\leq 1}}
	\Abs{{ \int} (g\circ\Pi_{t,0}[\zeta_{.}] 
- g\circ \Pi_{t,0}^{\phantom{t}}[\tilde\zeta_{.}])
\dd\mu_0} 
.
\label{PIdiffESTa}
\eea
 Pulling $\abs{\,.\,}$ under the last integral in \refeq{PIdiffESTa} 
and using $\LipSnorm{g}\leq 1$ gives the estimate
\beq
\dLIPnorm{ \mu_0\circ\Pi_{0,t}[\zeta_{.}] - 
           \mu_0\circ \Pi_{0,t}^{\phantom{t}}[\tilde\zeta_{.}]}
\leq
\int
\abs{
 \Pi_{t,0}[\zeta_{.}](z)
- \Pi_{t,0}^{\phantom{t}}[\tilde\zeta_{.}](z)}
\mu_0(\dd{z})
\,.
\label{PIdiffESTb}
\eeq
 By \refeq{defPARTICLEflow}, for $z_.,{\tilde{z}}_. \in C^1(\RR,{\RR^6})$ 
solving the characteristic equations for given 
$t\mapsto\psi(\,.\,,t)$ and $t\mapsto{\tilde{\psi}}(\,.\,,t)$,
respectively, with\footnote{The initial data condition $z_0={\tilde{z}}_0$
		derives from the $\Zett_0$ in $F_{.,0}(\,.\,;\,.\,|\Zett_0)$.}
$z_0=z={\tilde{z}}_0$, we have
\beq
\Pi_{t,0}[\zeta_{.}](z) - \Pi_{t,0}^{\phantom{t}}[\tilde\zeta_{.}](z)
= 
\int_{0}^t  
\left(
\sympJ\cdot\pdz\cH(z_\tau,\zeta_{\tau}) 
-
\sympJ\cdot\pdz\cH({\tilde{z}}_\tau,{\tilde{\zeta}}_{\tau}) 
\right) 
\dd \tau
  \label{PARTICLEflowDIFF}
\eeq
 We now insert \refeq{PARTICLEflowDIFF} in the \rhs of \refeq{PIdiffESTb}, 
estimate the resulting expression by pulling the absolute bars under the 
$t$-integral and applying the triangle inequality, 
next simplify by noting that $\sympJ$ is an isometry on $\RR^6$, 
and use Fubini's theorem to exchange $\dd{\tau}$ and $\dd\mu_0$ 
integrations.  
 Thus we obtain the estimate
\bea
\hskip-1truecm
\dLIPnorm{\mu_0\circ\Pi_{0,t}[\zeta_{.}] - 
           \mu_0\circ \Pi_{0,t}^{\phantom{t}}[\tilde\zeta_{.}]}
\!\!\! \!\!\!\! && \leq \!
\int_{0}^t\!\! \int\!
\left|
\pdz\cH(z_\tau,\zeta_{\tau}) 
-
\pdz\cH(z_\tau,{\tilde{\zeta}}_{\tau}) 
\right|
\mu_0(\dd{z})
\dd \tau
\cr
+&&\hskip-.7truecm
\int_{0}^t\!\! \int\!
\left|
\pdz\cH(z_\tau,{\tilde{\zeta}}_{\tau}) 
-
\pdz\cH({\tilde{z}}_\tau,{\tilde{\zeta}}_{\tau}) 
\right|
\mu_0(\dd{z})
\dd \tau
  \label{PARTICLEflowDIFFestA}.
\eea
 Since we want an estimate for
$\sup_{t\geq 0}\left(e^{-wt}\,
  \mathrm{l.h.s.}\refeq{PARTICLEflowDIFFestA}\right)$,
we next consider the exponentially weighted suprema of the two integrals 
on r.h.s.\refeq{PARTICLEflowDIFFestA} separately.

 The first integral on r.h.s.\refeq{PARTICLEflowDIFFestA} is
estimated as follows.
 By \refeq{charVecFIELD} with 
$z_\tau = (q_\tau,p_\tau)$ and by
the Cauchy--Schwarz inequality, we have
\bea
\abs{\pdz\cH(z_\tau,\zeta_\tau) - \pdz\cH(z_\tau,{\tilde{\zeta}}_\tau)}
\!\!\!&= &\!\!\!
\left|
\left(
	\fe\!\stackrel{\phantom{.}}{*}\!
\left[\nabla\psi(\,.\,,\tau)-
      \nabla{\tilde{\psi}}(\,.\,,\tau)
\right]
\right)
(q_\tau)\right|
\cr
\!\!\!&\leq &\!\!\!
      \norm{\fe}_{L^2} 
	\norm{\psi(\,.\,,\tau) - 
              {\tilde{\psi}}(\,.\,,\tau)}_{\dotH}
\,.
\label{hamFLOWsquareFIELDSa}
\eea
\newpage

\noindent
 Since r.h.s.\refeq{hamFLOWsquareFIELDSa} is 
independent of $z$, integration w.r.t. $\dd\mu_0$ factors out and yields 1.
 Proceeding now similarly as in estimate \refeq{estHoneB}, we obtain
\bea
\sup_{t\geq 0}\Bigl(
  e^{-wt}{{\int_{0}^t}}
\norm{\psi(\,.\,,\tau)\! -\! 
      {\tilde{\psi}}(\,.\,,\tau)}_{\dotH}
\dd \tau   \Bigr)
\leq
{\textstyle{\frac{1}{w}}}
\sup_{t\geq 0}
\biggl(  
	e^{-w t}
	\norm{\psi(\,.\,,t) - 
	      {\tilde{\psi}}(\,.\,,t)}_{\dotH}
\biggr).
\label{PARTICLEflowDIFFestAa}
\eea
 Multiplying  \refeq{PARTICLEflowDIFFestAa} by $\norm{\fe}_{L^2}$
gives an upper bound for
$\sup_{t\geq 0}\left(\!
e^{-wt}\!\int_0^t\!\int 
\mathrm{r.h.s.}\refeq{hamFLOWsquareFIELDSa} 
	 \dd\mu_0\dd\tau\!
		\right)$, 
which in turn is an upper bound for $\sup_{t\geq 0} (e^{-wt}$ 
first integral on r.h.s.\refeq{PARTICLEflowDIFFestA}).

 As to the second integral on  r.h.s.\refeq{PARTICLEflowDIFFestA},
its integrand is rewritten using \refeq{charVecFIELD}, 
with $z_\tau = (q_\tau,p_\tau)$ and  
${\tilde{z}}_\tau = ({\tilde{q}}_\tau,{\tilde{p}}_\tau)$, then
estimated by the triangle inequality, giving
\bea
\abs{\pdz\cH(z_\tau,{\tilde{\zeta}}_\tau) - 
     \pdz\cH({\tilde{z}}_\tau,{\tilde{\zeta}}_\tau)}
\!\!\!& \leq &\!\!\!
\Bigl|
\left(\fe\!\stackrel{\phantom{.}}{*}\!
      \nabla{\tilde{\psi}}(\,.\,,\tau)\right)(q_\tau) -
\left(\fe\!\stackrel{\phantom{.}}{*}\!
      \nabla{\tilde{\psi}}(\,.\,,\tau)\right)({\tilde{q}}_\tau)
\Bigr|
\cr
&&+ 
\Biggl|\frac{p_\tau}{\sqrt{1 + |p_\tau|{}^2}}
 - \frac{{\tilde{p}}_\tau}{\sqrt{1 + |{\tilde{p}}_\tau|{}^2}}\Biggr|
\,.
\label{hamFLOWsquareFIELDSb}
\eea
 The two expressions on the \rhs of \refeq{hamFLOWsquareFIELDSb} are now
estimated separately.

 The first term on  r.h.s.\refeq{hamFLOWsquareFIELDSb} is 
estimated as follows.
 Spelling out the convolutions and factoring out $\nabla{\tilde{\psi}}$ 
in the integrand, then pulling $\abs{\,.\,}$ into the convolution integral, 
then using the Cauchy--Schwarz inequality, we get
\beq
\big|(\fe*\nabla\tilde\psi(\,.\,,\tau))(q) -
     (\fe*\nabla\tilde\psi(\,.\,,\tau))({\tilde{q}})\big|
\leq
\norm{\fe(\,.\,-q)-\fe(\,.\,-{\tilde{q}})}_{L^2}
\norm{\tilde\psi(\,.\,,\tau)}_{\dotH}
\,.
\label{CAUCHYforceDIFF}
\eeq
 Now recall  that for any two equi-measurable translates 
$\Delta_1$ and $\Delta_2$ of a bounded domain $\Delta$ one has 
$\norm{\chi_{_{\Delta_1}} + \chi_{_{\Delta_2}} - 
       \chi_{_{\Delta_1\cap\Delta_2}}}_{L^2} \leq \sqrt{2|\Delta|}$, 
where
$\chi_{_\Delta}$ is the characteristic function of $\Delta$.
 This, the compact support of $\fe$, and its Lipschitz continuity,
then yield
\beq
\Bigl|
\left(\fe\!\stackrel{\phantom{.}}{*}\!
	\nabla{\tilde{\psi}}(\,.\,,\tau)\right)(q_\tau)
-
\left(\fe\!\stackrel{\phantom{.}}{*}\!
	\nabla{\tilde{\psi}}(\,.\,,\tau)\right)({\tilde{q}}_\tau)
\Bigr|
\leq 
C_\fe C_\psi\, |q_\tau-{\tilde{q}}_\tau| 
\,,
\label{hamFLOWsquareFIELDSbA}
\eeq
where $C_\fe = \sqrt{2|\supp\!(\fe)|}\,\LipSnorm{\fe}$, and where we also
used
$\sup_{\tau \geq 0}\|{\tilde{\psi}}(\,.\,,\tau)\|_{\dotH} \leq C_\psi$.
 We next estimate $|q_\tau-{\tilde{q}}_\tau|$, for which purpose 
we (i) use the integrated characteristic equations for $q_\tau$ and 
${\tilde{q}}_\tau$, with $q(0) = {\tilde{q}}(0)$,  (ii) pull $\abs{\,.\,}$ 
under the time integral, (iii) use that
$\pdp \sqrt{1+|p|^2}\in C^{0,1}_b(\RR^3)$
with $\LipSnorm{\pdp \sqrt{1+|p|^2}}\!=\!1$, and obtain
\bea
|q_\tau-{\tilde{q}}_\tau| 
\!\!\!& \leq &\!\!\! 
\int_0^\tau 
\Biggl|
\frac{p_{\tau^\prime}}{\sqrt{1 + |p_{\tau^\prime}|{}^2}}
 - 
\frac{{\tilde{p}}_{\tau^\prime}}{\sqrt{1 + |{\tilde{p}}_{\tau^\prime}|{}^2}}
\Biggr| 
\dd{\tau^\prime}
\cr
\!\!\!& \leq &\!\!\! 
\int_0^\tau 
\abs{p_{\tau^\prime}  - {\tilde{p}}_{\tau^\prime}}
\dd{\tau^\prime}.
\label{preGRONWALLestAa}
\eea
 Repeating steps (i) and (ii) now for 
$p_{\tau^\prime}$ and ${\tilde{p}}_{\tau^\prime}$ with $p_0 = {\tilde{p}}_0$, 
then applying the triangle inequality to the resulting integrand, followed by 
applications of \refeq{hamFLOWsquareFIELDSbA} and
$\sup_{\tau^\prpr \geq 0} \norm{\psi(\,.\,,\tau^\prpr)}_{\dotH} \leq C_\psi$,
respectively using \refeq{hamFLOWsquareFIELDSa}, 
gives the string of estimates
\bea
\hskip-1truecm
\abs{p_{\tau^\prime}  - {\tilde{p}}_{\tau^\prime}}
\!\!\!& \leq &\!\!\! 
\int_0^{\tau^\prime}\! 
\abs{
\left(\fe\!\stackrel{\phantom{.}}{*}\!
      \nabla\psi(\,.\,,\tau^{\prpr})\right)(q_{\tau^{\prpr}})
- 
\left(\fe\!\stackrel{\phantom{.}}{*}\!
   \nabla{\tilde{\psi}}(\,.\,,\tau^{\prpr})\right)({\tilde{q}}_{\tau^{\prpr}})
}
\dd{\tau^{\prpr}}
\cr
\!\!\!& \leq &\!\!\! 
\int_0^{\tau^\prime}\! 
\abs{
\left(\fe\!\stackrel{\phantom{.}}{*}\!
      \nabla\psi(\,.\,,\tau^{\prpr})\right)(q_{\tau^{\prpr}})
- 
\left(\fe\!\stackrel{\phantom{.}}{*}\!
      \nabla\psi(\,.\,,\tau^{\prpr})\right)({\tilde{q}}_{\tau^{\prpr}})
}
\dd{\tau^{\prpr}}
\cr
&& \qquad \quad + \!
\int_0^{\tau^\prime}\! 
\abs{
\left(\fe\!\stackrel{\phantom{.}}{*}\!
      \nabla\psi(\,.\,,\tau^{\prpr})
- 
\fe\!\stackrel{\phantom{.}}{*}\!
  \nabla{\tilde{\psi}}(\,.\,,\tau^{\prpr})\right)({\tilde{q}}_{\tau^{\prpr}})
     }
\dd{\tau^{\prpr}}
\cr
\!\!\!& \leq &\!\!\! 
C_\fe C_\psi\!\!
\int_0^{\tau^\prime}\!\!\!
|q_{\tau^{\prpr}}\!-\!{\tilde{q}}_{\tau^{\prpr}}|
\dd{\tau^{\prpr}}
\! +\!  
\norm{\fe}_{L^2}\! \! \!
\int_0^{\tau^\prime}\!\!  
 \norm{\psi(\,.\,,\tau^\prpr)\!-\!
       {\tilde{\psi}}(\,.\,,\tau^\prpr)}_{\dotH}
\dd{\tau^{\prpr}}\!,
\label{preGRONWALLestAb}
\eea
with $C_\fe$ given below \refeq{hamFLOWsquareFIELDSbA}, and
$C_\psi \geq \norm{\psi_0}_{\dotH}$ chosen later.
 Inserting \refeq{preGRONWALLestAb} into \refeq{preGRONWALLestAa},
recalling from \refeq{PARTICLEflowDIFFestA} that $\tau \leq t$, 
then employing a second order variant of the Gronwall lemma 
(see Appendix A.2) with $z_0=\tilde{z}_0$,
we find for all $\tau \leq t $ that
\beq
|q_\tau-{\tilde{q}}_\tau| 
\leq 
 \norm{\fe}_{L^2} 
\int_0^\tau
\cosh\left[{\wcrit}(\tau-\tau^\prime)\right]
\int_0^{\tau^\prime}\norm{\psi(\,.\,,\tau^\prpr) - 
      {\tilde{\psi}}(\,.\,,\tau^\prpr)}_{\dotH}
\dd\tau^\prpr
\dd\tau^\prime
\,,
\label{GRONWALLestA}
\eeq
with 
${\wcrit} =  \sqrt{C_\fe C_\psi}$.
 With \refeq{GRONWALLestA} and \refeq{hamFLOWsquareFIELDSbA} 
we have the relevant estimates for the first term on the 
\rhs of \refeq{hamFLOWsquareFIELDSb}. 

 To estimate the second  term on r.h.s.\refeq{hamFLOWsquareFIELDSb}, 
we recall that $\pdp \sqrt{1+|p|^2}\in C^{0,1}_b(\RR^3)$
with $\LipSnorm{\pdp \sqrt{1+|p|^2}}\!=\!1$, then recall 
\refeq{preGRONWALLestAb}.
 Estimating $|q_\tau-{\tilde{q}}_\tau|$ 
by r.h.s.\refeq{GRONWALLestA} and 
inserting this estimate into \refeq{preGRONWALLestAb},
we now find that for all $\tau \leq t$ we have
\bea
\hskip-1truecm
\Biggl|
\frac{p_\tau}{\sqrt{1 + |p_\tau|{}^2}}
 - 
\frac{{\tilde{p}}_\tau}{\sqrt{1 + |{\tilde{p}}_\tau|{}^2}}
\Biggr| 
\!\!\!& \leq &\!\!\! 
\norm{\fe}_{L^2} \int_0^{\tau}\! 
 \norm{\psi(\,.\,,\tau^\prime) - 
       {\tilde{\psi}}(\,.\,,\tau^\prime)}_{\dotH}
\dd{\tau^{\pr}}
\cr
&&\hskip-4.5truecm
+ 
\norm{\fe}_{L^2}\wcrit^2
\int_0^{\tau}\! 
\int_0^{\tau^\prime}\!\!
\cosh\left[{\wcrit}(\tau^\prime-\tau^\prpr)\right]
\int_0^{\tau^\prpr}\!\!
\norm{\psi(\,.\,,\check\tau) - 
      {\tilde{\psi}}(\,.\,,\check\tau)}_{\dotH}
\dd\check\tau
\dd{\tau^\prpr}
\dd{\tau^{\prime}}
,
\label{GRONWALLestB}
\eea
which provides an upper bound to the second  term on 
r.h.s.\refeq{hamFLOWsquareFIELDSb}.

 The bounds on the two terms of r.h.s.\refeq{hamFLOWsquareFIELDSb},
i.e. \refeq{GRONWALLestA} with \refeq{hamFLOWsquareFIELDSbA},  
and \refeq{GRONWALLestB}, 
combine into an estimate of l.h.s.\refeq{hamFLOWsquareFIELDSb}
which is independent of $z_\tau$ and ${\tilde{z}}_\tau$,
the solutions to the characteristic equations for
given fields $\psi$ and ${\tilde{\psi}}$ with initial 
data $z_0 = z = {\tilde{z}}_0$, respectively. 
 We have, independently of $z$, 
\newpage

\bea
\abs{\pdz\cH(z_\tau,{\tilde{\zeta}}_\tau) - 
     \pdz\cH({\tilde{z}}_\tau,{\tilde{\zeta}}_\tau)}
\!\!\!& \leq &\!\!\! 
\norm{\fe}_{L^2} \int_0^{\tau}\! 
 \norm{\psi(\,.\,,\tau^\prime) - 
       {\tilde{\psi}}(\,.\,,\tau^\prime)}_{\dotH}
\dd{\tau^{\pr}}
\qquad\qquad\quad
\cr
&&\hskip-3.5truecm
+
 \norm{\fe}_{L^2}\,\wcrit^2\!
\int_0^\tau
\cosh\left[{\wcrit}(\tau-\tau^\prime)\right]
\int_0^{\tau^\prime}\!\!
\norm{\psi(\,.\,,\tau^\prpr) - 
      {\tilde{\psi}}(\,.\,,\tau^\prpr)}_{\dotH}
\dd\tau^\prpr
\dd{\tau^\prime}
\cr
&&\hskip-5truecm
+ 
\norm{\fe}_{L^2}\,\wcrit^2\!
\int_0^{\tau}\! 
\int_0^{\tau^\prime}\!\!
\cosh\left[{\wcrit}(\tau^\prime-\tau^\prpr)\right]
\int_0^{\tau^\prpr}\!
\norm{\psi(\,.\,,\check\tau) - 
      {\tilde{\psi}}(\,.\,,\check\tau)}_{\dotH}
\dd\check\tau
\dd{\tau^\prpr}
\dd{\tau^{\prime}}
.
\label{hamFLOWsquareFIELDSc}
\eea
 We integrate \refeq{hamFLOWsquareFIELDSc} w.r.t. $\mu_0(\dd{z})$;
due to the $z$-independence of the integrand on the \rhs that integral 
factors out there and equals 1.
 It thus remains to integrate \refeq{hamFLOWsquareFIELDSc}
w.r.t. $\dd\tau$ from $0$ to $t$, to multiply  by $e^{-w t}$ and 
to take the supremum over $t\geq 0$.
 The three terms on r.h.s.\refeq{hamFLOWsquareFIELDSc}
are estimated by repeating the strategy used in \refeq{estHoneB} a 
total of 9 times (however, one of the  estimates is just
\refeq{PARTICLEflowDIFFestAa} again).
 For all $w>{\wcrit}$ we thereby arrive at the desired estimate for 
$\sup_{t\geq 0} (e^{-wt}$ second integral on 
r.h.s.\refeq{PARTICLEflowDIFFestA}),
\bea
&&\hskip-1.5truecm
\sup_{t\geq 0} \left(\!
e^{-wt}\!\!
\int_{0}^t\!\!\int\!
\left|
\pdz\cH(z_\tau,{\tilde{\zeta}}_{\tau}) 
\!-\!
\pdz\cH({\tilde{z}}_\tau,{\tilde{\zeta}}_{\tau}) 
\right| 
\dd\mu_0\dd\tau\!\right)\!
\leq
\cr
&&\hskip-1.5truecm
\norm{\fe}_{L^2}\!
\left(\!
\frac{1}{w^2}  + 
\frac{{\wcrit}^2}{2w^2(w-{\wcrit})}  + 
\frac{{\wcrit}^2}{2w^3(w-{\wcrit})} 
\right)
 \sup_{t\geq 0}\left(e^{-wt}
\norm{\psi(\,.\,,t) - {\tilde{\psi}}(\,.\,,t)}_{\dotH}\right) 
\!,
\label{PARTICLEflowDIFFestAb}
\eea
with ${\wcrit}= \sqrt{C_\varrho C_\psi}$ from \refeq{GRONWALLestA}.
 The estimates given by \refeq{PARTICLEflowDIFFestAb} and
by \refeq{PARTICLEflowDIFFestAa} (and ensuing text),
together with \refeq{PARTICLEflowDIFFestA}, now give the estimate 
\beq
 \sup_{t\geq 0}\left(
e^{-wt}
\dLIPnorm{\mu_0\circ\Pi_{0,t}[\zeta_{.}] - 
          \mu_0\circ \Pi_{0,t}^{\phantom{t}}[\tilde\zeta_{.}]}
\right)
\leq 
L_2
 \sup_{t\geq 0}\left(e^{-wt}
 \norm{\psi(\,.\,,t) - {\tilde{\psi}}(\,.\,,t)}_{\dotH}\right)
\label{preFzetaLIP}
\eeq
whenever $w>{\wcrit}$, with
\beq
L_2[\fe;w,{\wcrit}] 
=
\left(\!
\frac{1}{w}  
 + 
\frac{1}{w^2}  
+
\frac{(1+w){\wcrit}^2}{2w^3(w-{\wcrit})}  
\right)
\norm{\fe}_{L^2}\!
\,.
\label{LtwoCONST}
\eeq
 Since 
$\sup_{t\geq 0}\left(e^{-wt}
 \norm{\psi(\,.\,,\tau) - {\tilde{\psi}}(\,.\,,\tau)}_{\dotH}\right)
\leq 
 \sup_{t\geq 0}\left(e^{-wt}
 \norm{\zeta_{t} - \tilde\zeta_{t}}_{HL}\right)$, 
and because of \refeq{FzetaDIFFid}, we see that \refeq{preFzetaLIP} 
proves \refeq{FzetaLIP}.
 Part b) is completed.

 We have thus proved that $F_{.,0}(\,.\,|\Zett_0)$ is a Lipschitz map 
from a closed subset of $\Cw^0(\RR_+,\wGaBpx|\Zett_0)$, defined by the 
condition $\sup_{t\geq 0} \norm{\psi(\,.\,,t)}_\dotH \leq C_\psi$,
to $\Cw^0(\RR_+,\wGaBpx|\Zett_0)$ whenever 
$C_\psi\geq\Ccrit_\psi \geq  \norm{\psi_0}_\dotH$
and $w>\wcrit=\sqrt{C_\varrho C_\psi}$.
 The Lipschitz constant is 
$L[\varrho;w,{\wcrit}] =\max\{L_1[\varrho;w], L_2[\varrho;w,{\wcrit}]\}$,
with $L_1$ given in \refeq{LoneCONST} and $L_2$ in \refeq{LtwoCONST}.
 Finally, we note that everything proven so far for $F_{.,0}(\,.\,|\Zett_0)$ 
remains valid for its truncation $\olF_{.,0}(\,.\,|\Zett_0)$, obtained 
for all $t>0$ by
replacing $\Phi_{t,0}^\psi[\mu_.]$ by its upper truncation 
$\overline{\Phi}_{t,0}^\psi[\mu_.]$ given in \refeq{PHItrunc}.
 By time reversal symmetry, the same conclusions hold with 
$\Cw^0(\RR_+,\wGaBpx|\Zett_0)$ replaced by $\Cw^0(\RR,\wGaBpx|\Zett_0)$.
 This concludes the Lipschitz continuity part of the proof.

  We show next that for sufficiently large 
$\Ccrit_\psi \geq  \norm{\psi_0}_\dotH$
and $\widetilde{R}\geq \norm{\Zett_{.}^0}_{\mathrm{w}}$, 
the map $F_{.\,,0}(\,.\,|\Zett_0)$ 
sends a closed subset of a closed ball 
$B_{\widetilde{R}}(\Zett_.^0)\subset\Cw^0(\RR_+,\wGaBpx|\Zett_0)$,
satisfying
$\sup_{t\geq 0} \norm{\psi(\,.\,,t)}_\dotH \leq C_\psi$,
into $B_{\widetilde{R}}(\Zett_.^0)$ whenever $C_\psi\geq\Ccrit_\psi$ and 
$w>\wcrit = \sqrt{C_\varrho C_\psi}$; recall that 
$\Zett_{.}^0 = F_{.,0}(0_{.}|\Zett_0)\in\Cw^0(\RR,\wGaBpx|\Zett_0)$
denotes the free evolution of $\Zett_0 (\in\wGaBpx)$, where $0_{.}$ 
is the trivial constant map 
$(t\mapsto 0)\in
\Cw^0\left(\RR,\wM(\RR^6) \oplus\,\dotH(\RR^3) \oplus L^2 (\RR^3)\right)$.
 Note that l.h.s.\refeq{FLIPtrajT} is well-defined if we substitute 
$0_{.}$ for $\widetilde{\Zett}_{.}$;
clearly, $0_{.}\not\in\Cw^0(\RR,\wGaBpx|\Zett_0)$.
 However, since $\mu_.-0_.\neq P_1-P_1$, r.h.s.\refeq{FLIPtrajT} 
cannot be directly used to estimate 
this particular version of l.h.s.\refeq{FLIPtrajT}. 
 One possible way out is to derive an analog of \refeq{FLIPtrajT} 
for more general measures involving our extension $\dLIPnormT{\,.\,}$
of the dual Lipschitz distance, see Appendix A.1.
 A more direct way out is as follows.  
 We invoke the triangle inequality to estimate
$\norm{\Phi_{t,0}[\mu_{.}] (\zeta_{0}) - 
      \Phi_{t,0}[0_{.}] (\zeta_{0})}_{HL}
\leq
 \norm{\Phi_{t,0}[\mu_{.}] (\zeta_{0}) - 
      \Phi_{t,0}[\mu_{.}^0] (\zeta_{0})}_{HL}
+
 \norm{\Phi_{t,0}[\mu_{.}^0] ({0})}_{HL}$,
where we used that $\Phi_{t,0}[\mu_{.}^0] (\zeta_{0})
-\Phi_{t,0}[0_.] (\zeta_{0}) = \Phi_{t,0}[\mu_{.}^0] (0)$,
and also to estimate
\bea
\dLIPnorm{ \mu_0\circ\Pi_{0,t}[\zeta_{.}] - 
           \mu_0\circ \Pi_{0,t}^{\phantom{t}}[0_{.}]}
\!\!&\leq&
\dLIPnorm{ \mu_0\circ\Pi_{0,t}[\zeta_{.}] - 
           \mu_0\circ \Pi_{0,t}^{\phantom{t}}[\zeta_{.}^0]}
\nonumber\\
&&\!\!\!+
\dLIPnorm{ \mu_0\circ\Pi_{0,t}[\zeta_{.}^0] - 
           \mu_0\circ \Pi_{0,t}^{\phantom{t}}[0_{.}]}
\label{triANG}
\eea
 Recalling that $F_{.,0}(0_{.}|\Zett_0)=\Zett_{.}^0$, 
it is straightforward to verify that \refeq{FLIPtrajT} modifies to
\beq
\norm{F_{.,0}(\Zett_{.}|\Zett_0) - \Zett_{.}^0}_{{\mathrm{w}}}
\leq 
L \norm{\Zett_{.}-\Zett_{.}^0}_{\mathrm{w}}  +  K 
\,.
\label{FinBALL}
\eeq
whenever $w\!>\!\wcrit =\!\sqrt{C_\varrho C_\psi}$, with 
$L[\varrho;w,{\wcrit}] \!=\!\max\{L_1[\varrho;w], 
          L_2[\varrho;w,{\wcrit}]\}$ as before,
and 
\beq
K =
\sup_{t\geq 0} \left(  e^{-wt} 
\left( 
\norm{\Phi_{t,0}[\mu_{.}^0] ({0})}_{HL}
+
 \dLIPnorm{ \mu_0\circ\Pi_{0,t}[\zeta_{.}^0] - 
           \mu_0\circ \Pi_{0,t}^{\phantom{t}}[0_{.}]}
\right)\right).
\eeq
 Since
$\norm{\Phi_{t,0}^\psi[\mu_{.}^0] ({0})}_{\dotH}\leq\norm{\fe}_{L^2}t$
and 
$\norm{\Phi_{t,0}^\varpi[\mu_{.}^0] ({0})}_{L^2}\leq\norm{\fe}_{L^2}t$
by \refeq{fieldNORMSboundINt}, and since
\beq
\sup_{t\geq 0} \left(  e^{-wt} 
\dLIPnorm{ \mu_0\circ\Pi_{0,t}[\zeta_{.}^0] - 
           \mu_0\circ \Pi_{0,t}^{\phantom{t}}[0_{.}]}\right)
\leq
L_2
 \sup_{t\geq 0}\left(e^{-wt}
 \norm{\Phi_{t,0}^\psi[\mu_{.}^0] (\zeta_0)}_{\dotH}\right)
\eeq
by \refeq{preFzetaLIP}, with 
$\norm{\Phi_{t,0}^\psi[\mu_{.}^0] (\zeta_0)}_{\dotH}\leq
(2\cE_{\mathrm{W}}(\zeta_0))^{1/2} +  \norm{\varrho}_{L^2} t$
(by \refeq{fieldNORMSboundINt} again), we find that 
\beq
 K \leq {\textstyle{\frac{1}{w}\frac{2}{e}}}\norm{\fe}_{L^2} + L_2 
\max\left\{{\textstyle{\frac{1}{w}}}\norm{\fe}_{L^2} , 
	   (2\cE_{\mathrm{W}}(\zeta_0))^{1/2} \right\}.
\eeq
\newpage

\noindent
 Recalling that $L[\varrho;w,{\wcrit}] =\max\{L_1[\varrho;w], 
          L_2[\varrho;w,{\wcrit}]\}$ 
with $L_1$ given in \refeq{LoneCONST} and $L_2$ in \refeq{LtwoCONST}, 
we see that both $L$ and $K$ are monotonically decreasing functions of 
$w (>\wcrit)$, with asymptotic decay to zero $\propto 1/w$ for large $w$. 
 Now let $w$ be large enough such that $L\leq 1/2$. 
 Pick an $R_*$, independent of $w$, such that $K \leq LR_*$
(clearly such an $R_*$ exists).
 Now, either 
$\norm{\Zett_{.}-\Zett_{.}^0}_{\widetilde{\mathrm{w}}} \leq R_*$ or
$R_* \leq \norm{\Zett_{.}-\Zett_{.}^0}_{\widetilde{\mathrm{w}}}$. 
 In the former case, 
$\norm{F_{.,0}(\Zett_{.}|\Zett_0) - \Zett_{.}^0}_{\mathrm{w}} \leq 2L R_*$, 
i.e.
$F_{.,0}(\,{.}\,|\Zett_0)$ maps any closed subset of the closed ball 
$B_{R_*}(\Zett_.^0)$ satisfying
$\sup_{t\geq 0} \norm{\psi(\,.\,,t)}_\dotH \leq C_\psi$ into 
$B_{2LR_*}(\Zett_.^0)$; clearly, since by assumption $w$ is large
enough so that $L \leq 1/2$, the ball 
$B_{2LR_*}(\Zett_.^0) \subset B_{R_*}(\Zett_.^0)$.
 In the latter case on the other hand we have
$\norm{F_{.,0}(\Zett_{.}|\Zett_0) - \Zett_{.}^0}_{\mathrm{w}}
\leq   2L \norm{\Zett_{.} - \Zett_{.}^0}_{\widetilde{\mathrm{w}}}$,
with $L\leq 1/2$.
 Thus we conclude that, as long as $\widetilde{R} \geq R_*$, 
the fixed point map $F_{.,0}(\,{.}\,|\Zett_0)$ sends 
any closed subset of $B_{\widetilde{R}}(\Zett_.^0)$ which satisfies
$\sup_{t\geq 0} \norm{\psi(\,.\,,t)}_\dotH \leq C_\psi$ into 
$B_{\widetilde{R}}(\Zett_0)$, given that $L \leq 1/2$.
  Again, we note that everything proven in this paragraph 
for $F_{.,0}(\,.\,|\Zett_0)$ remains valid for its truncation 
$\olF_{.,0}(\,.\,|\Zett_0)$.

   It remains to notice that the truncated map 
$\olF_{.\,,0}(\,.\,|\Zett_0)$ sends any closed subset of
$\Cw^0(\RR_+,\wGaBpx|\Zett_0)$ satisfying 
$\sup_{t\geq 0} \norm{\psi(\,.\,,t)}_\dotH \leq C_\psi$,
with $C_\psi \geq \norm{\psi_0}_\dotH$, to itself.
 Hence, for $\widetilde{R} \geq R_*$
and $\Ccrit_\psi \geq  \norm{\psi_0}_\dotH$, the truncated map 
$\olF_{.\,,0}(\,.\,|\Zett_0)$ sends the intersection of any closed ball
$B_{\widetilde{R}}(\Zett_.^0)\subset\Cw^0(\RR_+,\wGaBpx|\Zett_0)$ with any
closed subsets of $\Cw^0(\RR_+,\wGaBpx|\Zett_0)$ satisfying 
$\sup_{t\geq 0} \norm{\psi(\,.\,,t)}_\dotH \leq C_\psi$
to itself whenever $C_\psi\geq\Ccrit_\psi$ and 
$w >\wcrit = \sqrt{C_\varrho C_\psi}$ 
are large enough so that $L\leq 1/2$.
 For instance, this can be achieved as follows.
 Setting $w= 2\wcrit$, the Lipschitz constant becomes
\beq
L 
=\max\Bigl\{
\sqrt{ 
{\textstyle{\frac{\LipSnormSMALLitera{\fe*\nabla^2 \fe}}{16\wcrit^4}}}}
\,+
\sqrt{ {\textstyle{\frac{\LipSnormSMALLitera{\fe*\fe}}{4\wcrit^2}}}
   +\!{\textstyle{\frac{ \LipSnormSMALLitera{\fe*(\Omega_0\cdot\nabla){}^2\fe}
	               }{16\wcrit^4}}}}
\;,\;
 {\textstyle{\left(1 + \frac{1}{2\wcrit}\right)
\frac{5\norm{\fe}_{L^2}}{8\wcrit\ \ }}}
\Bigr\},
\label{Lofwcrit}
\eeq
and also setting now $C_\psi=\Ccrit_\psi$ so that 
$\wcrit = \sqrt{C_\varrho \Ccrit_\psi}$, we see that 
there is a unique $\Ccrit_\psi^*[\fe]$ such that 
r.h.s.\refeq{Lofwcrit}$\leq 1/2$ for 
$\Ccrit_\psi \geq \Ccrit_\psi^*[\fe]$;
hence, choosing 
$\Ccrit_\psi \geq \max\{\Ccrit_\psi^*[\fe]\,,\,\norm{\psi_0}_\dotH\}$ 
will do.
 This completes the proof of  Proposition \ref{olFisCONTRACTmap}.\QED

\subsubsection{Proof of Theorem \ref{ExistUnique}}

 It suffices in the following to continue to work with 
$C_\psi = \Ccrit_\psi$ and $w=2\wcrit$, as done at the end of the 
proof of Proposition \ref{olFisCONTRACTmap}.
 Thus, we need to show that Theorem \ref{ExistUnique} is true 
for sufficiently large 
$\Ccrit_\psi\geq \max\{\Ccrit_\psi^*[\fe]\,,\,\norm{\psi_0}_\dotH\}$.
\smallskip

\noindent
\textit{Proof of Theorem \ref{ExistUnique}:}
 Let 
$C_\psi = \Ccrit_\psi\geq \max\{\Ccrit_\psi^*[\fe]\,,\,\norm{\psi_0}_\dotH\}$,
and $w=2\wcrit$.
 Then, as shown in the proof of Proposition \ref{olFisCONTRACTmap},
$\olF_{.\,,0}(\Zett_{.}|\Zett_0)$ is a contraction map, 
with Lipschitz constant $L \leq 1/2$, from the intersection of 
any closed ball 
$B_{\widetilde{R}}(\Zett_.^0)\subset\Cw^0(\RR,\wGaBpx|\Zett_0)$ of
radius ${\widetilde{R}} \geq R_*$
with any closed subset of $\Cw^0(\RR,\wGaBpx|\Zett_0)$ defined
by the condition
$\sup_{t\geq 0} \norm{\psi(\,.\,,t)}_\dotH \leq C_\psi (=\Ccrit_\psi)$,
to itself.
 The standard contraction mapping theorem now 
guarantees the existence of a unique fixed 
point $t\mapsto \Zett_t\in \Cw^0(\RR,\wGaBpx)$, with 
$t\mapsto\psi(\,.\,,t)\!\in\! C^0_b(\RR, \dotH(\RR^3))$,
of the fixed point equation with the truncated $F$,
\beq
\Zett_{.} = \olF_{.\,,0}(\Zett_{.}|\Zett_0).
\label{truncFPeqAGAIN}
\eeq
 We will now show that for sufficiently large $\Ccrit_\psi$
the solutions of \refeq{truncFPeqAGAIN} are fixed points 
for $F$, moreover of type $C^1(\RR,\wGaBpx)$, 
thus furnishing unique $\wGaBpx$-strong  Vlasov solutions.

 To this effect, choose 
$\Ccrit_\psi\! >\! \max\bigl\{\Ccrit_\psi^*[\fe],\!
	                     \sqrt{2\cE_{\mathrm{W}}(\zeta_0)},
        		     \sqrt{4 + 4E_0  - 8E_\bot}\,\bigr\}$,
where $\cE_{\mathrm{W}}(\zeta_0)$ is the initial field energy,
$E_0 = \cE(\Zett_0)$ is the total energy of the initial state, 
and where $E_\bot$ is the ground state energy of the $N$-body Hamiltonian 
given in \refeq{groundE}; note that the ground state energy is 
$N$-independent and therefore identical to the ground state of the
continuum (Vlasov) limit energy functional \refeq{HfuncVeps}.
 Note that automatically we have
$\Ccrit_\psi > \norm{\psi_0}_\dotH^{\phantom{H}}$, for
it is trivially obvious that
$\norm{\psi_0}_\dotH^{\phantom{H}}\leq \sqrt{2\cE_{\mathrm{W}}(\zeta_0)}$
and easily shown that 
$\norm{\psi_0}_\dotH^{\phantom{H}}\leq \sqrt{4 + 4E_0  - 8E_\bot}$.
 With the so chosen $\Ccrit_\psi$, there then exists at least a small 
neighborhood of $t=0$ such that for all $t$ in this neighborhood,
the fixed points of \refeq{truncFPeqAGAIN}, which are
of type $C^0(\RR,\wGaBpx)$, by continuity satisfy
\beq
\Zett_{t} = F_{t\,,0}(\Zett_{.}|\Zett_0).
\label{FPatTIMEt}
\eeq

 Now recall that $\psi_0\in (\dotH\cap\dotHH)(\RR^3)$, ensuring
a strong solution of the wave equation, by the Hille--Yosida theorem. 
 Next recall the remark after Lemma  \ref{AprioriBOUNDSnormZETA};~viz.
\beq
\norm{\psi(\,.\,,t)}_\dotH \leq 
 (2\cE_{\mathrm{W}}(\zeta_0))^{1/2} + 
 \norm{\varrho}_{L^2}|t|,
\label{psiNORMboundINt}
\eeq
for the strong solution of the wave equation given
\textit{any subluminal} source $\fe*\rho \in C^0(\RR,C^\infty_0(\RR^3))$.
 Clearly, there is a unique $\underline{T}>0$ for which
\beq
\Ccrit_\psi
=
 (2\cE_{\mathrm{W}}(\zeta_0))^{1/2} + 
 \norm{\varrho}_{L^2} \underline{T}
\,,
\label{psiNORMboundEQUALtoCcritpsi}
\eeq
such that $\norm{\psi(\,.\,,t)}_\dotH < \Ccrit_\psi$
strictly for all $|t| < \underline{T}$,  by \refeq{psiNORMboundINt}.
 This now implies that there exist $T\geq \underline{T}$ such that
the fixed point $\Zett_{.}$ of \refeq{truncFPeqAGAIN} satisfies
\refeq{FPatTIMEt} for all $t\in [-T,T]$.
 We now show that 
$\sup \{T : \refeq{FPatTIMEt}$ holds for all $|t|\leq T\} =\infty$.

 Thus, suppose that 
$\sup \{T : \refeq{FPatTIMEt}$ holds for all $|t|\leq T\}
 = T_* <\infty$. 
 Then for either $t = T_*^+$ or $t = - T_*^-$ (or both), 
$\Zett_t$ is given by \refeq{truncFPeqAGAIN} but \textit{not}
by \refeq{FPatTIMEt}.
 For the sake of concreteness, assume that this is so 
for some $t$ in a right neighborhood of $T_*$.
 This then means  that for all $t\in (T_*,T_*+\epsilon)$ 
we have
$\norm{\Phi_{t,0}^\psi}_\dotH \geq \Ccrit_\psi > \sqrt{4 + 4E  - 8E_\bot}$,
which in particular implies that $\lim_{t\downarrow T_*}
\norm{\Phi_{t,0}^\psi}_\dotH \geq \Ccrit_\psi > \sqrt{4 + 4E  - 8E_\bot}$.
  On the other hand, for all $t\in [-T_*,T_*]$, the solution $\Zett_{.}$ of 
\refeq{truncFPeqAGAIN} satisfies \refeq{FPatTIMEt}, and $\zeta_{.}$ is
a strong solution of the wave equation. 
 But then, by Corollary \ref{PIsymplecto}, 
$t\mapsto \Zett_t\in C^1([-T_*,T_*],\wGaBpx)$ is a $\wGaBpx$-strong
solution, which by Theorem \ref{theoKRsolVeps} 
conserves energy. 
 As a consequence of energy conservation, for all $t\in[-T_*,T_*]$,
and in particular for $t=T_*$, we have
\beq
\norm{\Phi_{t,0}^\psi}_\dotH 
\leq 
\sqrt{4 + 4E  - 8E_\bot}
\,.
\label{EPSboundGRADphiZWEInormVLnew}
\eeq
 Since $t\mapsto \Phi_{t,0}^\psi\in C^0(\RR,\dotH(\RR^3))$,
we thus  have a contradiction to the previously concluded
$\norm{\Phi_{t,0}^\psi}_\dotH 
\geq \Ccrit_\psi > \sqrt{4 + 4E  - 8E_\bot}$ for $t>T_*$.
Hence, $T_*=\infty$.\QED

\begin{Rema}
 {By the proof of Theorem \ref{ExistUnique} it 
suffices to work with $w \geq 2 \wcrit^*$, where
\beq
\wcrit^* =
\sqrt{C_\fe \max\biggl\{\Ccrit_\psi^*[\fe]\,, 
	     \sqrt{4 + 4E  - 8E_\bot}\,\biggr\}}.
\label{WORKINGw}
\eeq
}
\end{Rema}

\begin{Rema}
\label{remaBoundsVeps}
  {In the proof of Theorem \ref{ExistUnique} we only made use of 
the a priori bound \refeq{EPSboundGRADphiZWEInormVLnew} following from 
Theorem \ref{theoKRsolVeps} and the analog of the proof of our 
Proposition \ref{propHboundN} for the regularized Vlasov model.
 The other bounds expressed in Proposition \ref{propHboundN} 
are carried over as follows.}

  {Let $t\mapsto \Zett(t)\in C^0(\RR,\wGaBpx)$  be a generalized  solution
  of the wave gravity Vlasov equations which conserves
  energy $E$, momentum $P$, and angular momentum $J$,
  and of course mass $M=1$.
  Then, beside \refeq{EPSboundGRADphiZWEInormVLnew},
uniformly in $t$ we have}
\bea
\norm{\varpi(\,.\,,t)}_{L^2}^2
\!\!\!&\leq&\!\!\!
2E  - 2E_\bot
\,,
\label{EPSboundDOTphiZWEInormVL}
\\
\int \sqrt{1 + \abs{p}{}^2 } \mu_t (\dd z)
\!\!\!&\leq&\!\!\!
1+ E  - E_\bot
\,,
\label{EPSboundPsupVL}
\\
6E_\bot - 3E  - 3
\leq
\int 
 \bigl(\fe*\psi(\,.\,,t)\bigr)(x) \mu_t (\dd z)
\!\!\!
&\leq&
\!\!\!
E  - 1
\label{EPSboundsEpotVL}
\eea
  Moreover, \refeq{PmomBOUND} and \refeq{JposmomBOUND} extend to 
\beq
\Abs{\int p \mu_t (\dd z)}
\leq
\abs{P} 
+ 
\norm{\psi^{\supN}(\,.\,,t)}_{\dotH}
\norm{\varpi^{\supN}(\,.\,,t)}_{L^2}
\,,
\label{PmomBOUNDvlasov}
\eeq
\beq
\Abs{\int p\crprod x \mu_t (\dd z)}
\leq
\abs{J} 
+ 
\left(R+|t|\right)
\norm{\psi^{\supN}(\,.\,,t)}_{\dotH}
\norm{\varpi^{\supN}(\,.\,,t)}_{L^2}
\,.
\label{JposmomBOUNDvlasov}
\eeq
\end{Rema}

\section{The limit $\!N\!\to\!\infty\!$}
\label{secNinfinity}
 
 We prove first that the $\wGaBpx$-strong $N$-body 
generalized solutions of the Vlasov model converge 
$\norm{\,.\,}_{\mathrm{w}}$-strongly to solutions 
when $N\to\infty$.
 We then specify when these limit solutions are continuum solutions.
 Finally we discuss the probabilistic import in terms of a law of
large numbers and a central limit theorem.

\subsection{The $\wGaBpx$-strong limit of the $N$-body generalized solutions}
\label{secVLASOVlim}

  Suppose the family of initial empirical measures converges
$\wGaBpx$-strongly when $N\to\infty$, written
$\veps[{\bzN_0}](\dd{z}) \leadsto \mu_0(\dd{z})$.
    Then the microscopic `density' $\rho^{\supN} (\,.\,, 0)$, 
as given in \refeq{NrhoEQeps} with $t=0$, 
converges strongly (in the marginal measures' subspace) 
to the `density' $\rho (\,.\,, 0)$.
 Finally, assume that
$\zeta[{\bzN_0}] \to \zeta_0\in \dotH(\RR^3) \oplus L^2 (\RR^3)$
satisfying \refeq{phiNULLepsASYMP}, \refeq{varpiNULLepsASYMP} with
$\psi_0\in(\dotH\cap\dotHH)(\RR^3)$.
 Our goal is to  show that, when $N\to\infty$, the generalized solution 
$t\mapsto (\veps[{\bzN_t}]; \zeta^{\supN}_t)\in (\Cw^0 \cap C^1)(\RR,\wGaBpx)$
associated with this converging family of initial data in turn 
converges in $\norm{\ .\ }_{\mathrm{w}}$ norm to a solution 
$t\mapsto  (\mu_t; \zeta_t) \in (\Cw^0 \cap C^1)(\RR,\wGaBpx)$
of the regularized wave gravity Vlasov fixed point equation.
 In the following, $\Zett^{\supN}_t \statelimWIGGLE \Zett_t$ means
$\psi^{\supN}(\,.\,,t)\!\stackrel{\dotH}{\longrightarrow}\!\psi(\,.\,,t)$
satisfying \refeq{phiNULLepsASYMP}, 
$\varpi^{\supN}(\,.\,,t)\!\stackrel{L^2}{\longrightarrow}\!\varpi(\,.\,,t)$
satisfying \refeq{varpiNULLepsASYMP},
and $\veps[{\bzN_t}]\leadsto\mu_t$ in $\wPpx$.

 The main result is an immediate consequence of the following theorem, 
which states that the $\norm{\ .\ }_{\mathrm{w}}$ induced distance between 
any two $\Cw^0(\RR,\wGaBpx)$ solutions of our Vlasov fixed point equation
is controlled by the  $\wGaBpx$ distance of their initial states in $\GaBpx$.

\begin{Prop}
\label{LIPinCAUCHYdata}
 {Let $\II\subset\NN$ or $\II\subset\RR$ be an index set, and let 
	$\{\Zett_{.}^{(\alpha)} \in \Cw^0(\RR, \wGaBpx)\}_{\alpha\in \II}$ 
	be a family of solutions of the Vlasov fixed point equation 
	\refeq{VlasovWaveEQasFP},
	having initial data $\Zett_0^{(\alpha)} \in \GaBpx$ with
	$\psi_0^{(\alpha)}\in(\dotH\cap\dotHH)(\RR^3)$, for which 
	$E^*:=\sup_{\alpha\in\II}\{\cE(\Zett_{0}^{(\alpha)})\}$ exists.
	Define
\beq
\bar{w} =
\sqrt{C_\fe \max\biggl\{\Ccrit_\psi^*[\fe]\,, 
	     \sqrt{4 + 4E^* - 8E_\bot}\,\biggr\}}
\,,
\label{WORKINGww}
\eeq
with $C_\fe = \sqrt{2|\supp\!(\fe)|}\,\LipSnorm{\fe}$ 
(cf. text below \refeq{hamFLOWsquareFIELDSbA}).
 Then for all $w \geq 2\bar{w}$ there exists a constant $L_0[\bar{w}]$ 
such that for any $(\alpha,\tilde\alpha)\in\II^2$,}
\beq
     \norm{\Zett_{.}^{(\alpha)} - \Zett_{.}^{(\tilde\alpha)}}_{\mathrm{w}}
\leq
     L_0 \norm{\Zett_{0}^{(\alpha)} - \Zett_{0}^{(\tilde\alpha)}}
\,.
\eeq
\end{Prop}

 Before we prove this proposition, we state and prove its main corollary.

\begin{Theo} 
\label{LargeN}
        {Let $t\mapsto \Zett_t^{\supN} \in (\Cw^0\cap C^1)(\RR,\wGaBpx)$
  be the $\wGaBpx$-strong $N$-body solution 
  of the regularized wave gravity Vlasov equations 
  \refeq{VphiEQeps}, \refeq{VvpiEQeps}, \refeq{VfEQeps}
  with Cauchy data  $\Zett^{\supN}_0 =\lim_{t\to  0} \Zett^{\supN}_t$
  described in Theorem  \ref{theoKRsolVeps}.
  Suppose $\Zett_0^{\supN} \statelimWIGGLE \Zett_0$, with $\Zett_0$
  having mass $M(=1)$, energy $E$, momentum $P$, and angular momentum $J$,
  and with $\psi_0\in(\dotH\cap\dotHH)(\RR^3)$.
   Then $\norm{\Zett_{\,.}^{\supN} - \Zett_{\,.}}_{\mathrm{w}}\to 0$, 
   where $t\mapsto \Zett_t\in \Cw^0(\RR,\wGaBpx)$ is the unique
  solution of \refeq{VlasovWaveEQasFP} described 
  in Theorem \ref{ExistUnique}. 
  Beside mass, $t\mapsto \Zett_t$ also conserves
  energy, momentum, and angular momentum.
  Furthermore, since $\psi_0\in(\dotH\cap\dotHH)(\RR^3)$, we also have
	$\Zett_{.}\in C^1(\RR,\wGaBpx)$.
}
\end{Theo}

\noindent \textit{Proof of Theorem \ref{LargeN}:} 
  By Theorem \ref{ExistUnique}, there exist unique 
type $\Cw^0(\RR,\wGaBpx)$ solutions $\Zett_{.}^{\supN}\,,\,\Zett_{.}$
 of the fixed point equation \refeq{VlasovWaveEQasFP} for each 
Cauchy data $\Zett_0^{\supN}\,,\,\Zett_0\in \GaBpx$, with
$\psi_0^{\supN}\,,\, \psi_0$ in $(\dotH\cap\dotHH)(\RR^3)$, respectively.  
 The latter restriction upgrades the solutions to the
wave equation to be strong, which by Lemma \ref{Gestimates}
implies  solutions of type $(\Cw^0\cap C^1)(\RR,\wGaBpx)$ which
conserve  mass, momentum, angular momentum, and energy.

 Let $\Zett^{(\infty)}_{.}\equiv \Zett_{.}$, and set $\II=\NN\cup\{\infty\}$.
 Since	energy is conserved by each solution, 
$E^*=\sup_{\alpha\in\II}\{\cE(\Zett_{0}^{(\alpha)})\}$ exists.
 Thus, $\bar{w}$ exists. 
 Pick any $w>2\bar{w}$.
 Now Proposition \ref{LIPinCAUCHYdata} applies to our family
$\{\Zett_{.}^{(\alpha)}\}_{\alpha\in \II}$, and 
since  $\norm{\Zett^{\supN}_0 - \Zett_0}\to{0}$ by hypothesis,
Proposition \ref{LIPinCAUCHYdata} now implies that
$\norm{\Zett_{\,.}^{\supN}-\Zett_{\,.}^{(\infty)}}_{\mathrm{w}}\to{0}$.\QED

 To prepare the proof of Proposition \ref{LIPinCAUCHYdata},
we will need the following lemmata.

\begin{Lemm}\label{Liapunov}
{Let $\zeta_{.}\!\in\!C^0_b(\RR, (\dotH\oplus L^2)(\RR^3))$,
with $\sup_{t\geq 0} \norm{\psi(\,.\,,t)}_\dotH \leq C_\psi$,
and let ${\wcrit} =  \sqrt{C_\fe C_\psi}$.
 Then $\Pi_{t,t^\prime}[\zeta_\cdot]\in C^{0,1}(\RR^6,\RR^6)$, with
Lipschitz constant}\footnote{Incidentally, by Lemma \ref{Liapunov}, 
		the largest Liapunov exponent for 
		$\Pi_{t,t^\prime}[\zeta_\cdot]$ is bounded above by $\wcrit$.}
\beq
\LipSnorm{\Pi_{t,t^\prime}[\zeta_\cdot]}
=
{\textstyle{\frac{1}{\sqrt{2}}}}
(2 + \max\{\wcrit \,,\, 1/\wcrit\}) e^{\wcrit\, |t-t^\prime|}.
\label{LipPI}
\eeq
\end{Lemm}

\noindent
\textit{Proof of Lemma \ref{Liapunov}:}
 Let $\psi(\,.\,,t)\!\in\!C^0_b(\RR, \dotH(\RR^3))$ be given,
with $\sup_{t\geq 0} \norm{\psi(\,.\,,t)}_\dotH \leq C_\psi$.
 To unburden notation, let $\psi_t(\,.\,)$ stand for $\psi(\,.\,,t)$.
 Let $t\mapsto z_t\in\RR^6$ and $t\mapsto \tilde{z}_t\in\RR^6$
 be two distinct solutions of \refeq{qDOT}, \refeq{pDOT} for this 
$\psi_{.}$.
 Proceeding analogously to the steps taken in
\refeq{preGRONWALLestAa} and \refeq{preGRONWALLestAb},
this time for $\psi =\tilde\psi$, but now allowing 
$z_0\neq \tilde{z}_0$, we find
\bea
|\tilde{q}_t-q_t| 
\!\!\!& \leq &\!\!\! 
|\tilde{q}_{t^\prime}-q_{t^\prime}| 
+
\int_{t^\prime}^t
\Biggl|
\frac{\tilde{p}_\tau}{\sqrt{1 + |\tilde{p}_\tau|{}^2}}
 - 
\frac{p_\tau}{\sqrt{1 + |p_\tau|{}^2}}
\Biggr| 
\dd{\tau}
\cr
\!\!\!& \leq &\!\!\! 
|\tilde{q}_{t^\prime}-q_{t^\prime}| 
+
\int_{t^\prime}^t 
\abs{\tilde{p}_\tau  - p_\tau}
\dd{\tau},
\label{preLIAPUNOVestAa}
\eea
respectively
\bea
\hskip-.5truecm
\abs{\tilde{p}_t  - p_t}
\!\!\!& \leq &\!\!\! 
|\tilde{p}_{t^\prime}-p_{t^\prime}| 
+
\int_{t^\prime}^t\! 
\abs{
\left(\fe\!\stackrel{\phantom{.}}{*}\!
      \nabla\psi_{\tau}\right)(\tilde{q}_\tau)
- 
\left(\fe\!\stackrel{\phantom{.}}{*}\!
      \nabla\psi_{\tau}\right)(q_\tau)
}
\dd{\tau}
\cr
\!\!\!& \leq &\!\!\! 
|\tilde{p}_{t^\prime}-p_{t^\prime}| 
+
C_\fe C_\psi\!\!
\int_{t^\prime}^t\!\!\!
|\tilde{q}_\tau\!-\!q_\tau|
\dd{\tau}
\,,
\label{preLIAPUNOVestAb}
\eea
with $C_\fe$ and $C_\psi$ as stated in the lemma.
 Inserting \refeq{preLIAPUNOVestAb} into \refeq{preLIAPUNOVestAa}
and using the second order variant of Gronwall's lemma gives
\beq
|\tilde{q}_t-q_t| 
\leq
|\tilde{q}_{t^\prime}-q_{t^\prime}| 
\cosh\left[{\wcrit}(t-t^\prime)\right]
+ 
\abs{\tilde{p}_{t^\prime}  - p_{t^\prime}}
{\textstyle{\frac{1}{\wcrit}}}
\sinh\left[{\wcrit}|t-t^\prime|\right],
\label{LIAPUNOVestA}
\eeq
with ${\wcrit} =  \sqrt{C_\fe C_\psi}$.
 Back-substituting \refeq{LIAPUNOVestA} into \refeq{preLIAPUNOVestAb}
and integrating then gives
\beq
|\tilde{p}_t-p_t| 
\leq 
|\tilde{p}_{t^\prime}-p_{t^\prime}| 
\cosh\left[{\wcrit}(t-t^\prime)\right] 
+ 
\abs{\tilde{q}_{t^\prime}  - q_{t^\prime}}
{\textstyle{\wcrit}}
\sinh\left[{\wcrit}|t-t^\prime|\right]
.
\label{LIAPUNOVestB}
\eeq
 To get from \refeq{LIAPUNOVestA} and \refeq{LIAPUNOVestB} to 
the conclusion of Lemma \ref{Liapunov}, use $\cosh(x)\leq e^{|x|}$
and $\sinh(|x|)\leq e^{|x|}/2$, as well as the familiar
$\norm{\vec{v}}_2\leq\norm{\vec{v}}_1\leq\sqrt{2}\norm{\vec{v}}_2$
for $\vec{v}\in\RR^n$.\QED

 The next lemma transfers control about the flow $\Pi_{t,t^\prime}$
on $\RR^6$ to the flow $\Pi^\dag_{t,t^\prime}$ on $\wPpx$.

\begin{Lemm}
\label{LemAdLip}
  For any symplectomorphism $\Pi$ on $\RR^6$ which in addition 
  is a Lipschitz map with Lipschitz constant $\Lambda$, the adjoint map 
  $\Pi^\dagger\! : M(\RR^6)\mapsto M(\RR^6)$,
  defined by
  $\Pi^\dagger(\sigma):=\sigma \circ \Pi^{-1}$,
  is a positivity- and $\norm{\,.\,}_{\mathrm{TV}}$-preserving smooth 
  automorphism of $M(\RR^6)$, and it is a Lipschitz homeomorphism 
  on $\wM(\RR^6)$ for $\dLIPnormT{\,.\,}$ with Lipschitz constant $\Lambda$.
\end{Lemm}

\noindent 
\textit{Proof of Lemma \ref{LemAdLip}:}
 First, since $\Pi$ is a symplectomorphism, by way of the 
definition of its adjoint, $\Pi^\dagger$ maps $M(\RR^6)$ smoothly
onto $M(\RR^6)$, and it preserves 
  (a) $\norm{\sigma}_{\mathrm{TV}}$ for $\sigma\in M$ and 
  (b) the positivity of $\mu\in M_+$. 
 Furthermore, since $\Pi$ is invertible, so is $\Pi^\dagger$. 

 To see that $\Pi^\dagger$ is a homeomorphism of $\wM(\RR^6)$, 
we only need to show that $\Pi^\dagger$
maps $\wM$ into $\wM$.
 Thus, let $z_*\in\RR^6$ be the unique element of $\ker\Pi$. 
 Then note that by the definition of $\Pi^\dagger$ and the Lipschitz property 
of $\Pi$ we have
\beq
  \Abs{\int \abs{z} \sigma \circ \Pi^{-1} (\dd{z})}
   =
  \Abs{\int \abs{\Pi(z)-\Pi(z_*)} \sigma (\dd{z})}
\leq 
  \Lambda \int \abs{z-z_*}  \abs{\sigma}(\dd{z})
\,,
\eeq
where $\abs{\sigma}$ is the total variation of $\sigma$; 
the last integral exists for $\sigma \in \wM$. 

 As for the Lipschitz continuity of the adjoint flow, let 
$\hat\sigma, \check\sigma \in \wM(\RR^6)$.  
 We have
\newpage

\bea
\dLIPnorm{\Pi^\dagger(\hat\sigma) - \Pi^\dagger(\check\sigma)}
\!\!\!&=&\!\!\!
\sup_{g \in C^{0,1}(\RR^6)}\left\{
     \Abs{\int\! g\, \dd\!\left( \hat\sigma \circ \Pi^{-1} -  
                                 \check\sigma \circ \Pi^{-1}\right)}\! :
     \LipSnorm{g} \leq 1\right\}
\cr
\!\!\!& =&\!\!\! 
\sup_{g \in C^{0,1}(\RR^6)}\left\{ 
   \Abs{\int\! g \circ \Pi\, \dd(\hat\sigma - \check\sigma) }:
     \LipSnorm{g} \leq 1\right\}
\cr
\!\!\!& =&\!\!\! 
 \Lambda \sup_{g \in C^{0,1}(\RR^6)}\left\{ 
\Abs{\int\! {\small\frac{1}{\Lambda}}\, g \circ \Pi\, 
     \dd(\hat\sigma - \check\sigma) }:
     \LipSnorm{g} \leq 1\right\}
\cr
\!\!\!&\leq&\!\!\!
  \Lambda \dLIPnorm{\hat\sigma - \check\sigma}
\,.
\eea
 In the last step, we used that $\Lambda^{-1}g \circ \Pi\in C^{0,1}(\RR^6)$ 
with $\LipSnorm{\Lambda^{-1}g \circ \Pi}\leq 1$.\QED
\medskip

\noindent \textit{Proof of Proposition \ref{LIPinCAUCHYdata}:} 
 Pick $w>2\bar{w}$, with $\bar{w}$ defined in 
\refeq{WORKINGww}, and pick
$\Zett_{.}, \widetilde{\Zett}_{.} \in \Cw^0(\RR, \wGaBpx)$
from the family of solutions $\Zett_{.}^{(\alpha)}$ 
of the Vlasov fixed point equation \refeq{VlasovWaveEQasFP} specified
in  Proposition \ref{LIPinCAUCHYdata}.
 Then 
\beq
\norm{\Zett_{.} - \widetilde{\Zett}_{.}}_{\mathrm{w}}
=
\norm{F_{.,0}(\Zett_{.}|\Zett_0) - 
      F_{.,0}(\widetilde{\Zett}_{.}|\widetilde{\Zett}_0)}_{\mathrm{w}}
\,.
\label{wDISTofYYisDISTofFF}
\eeq
 By the triangle inequality,
\bea
\norm{F_{.,0}(\Zett_{.}|\Zett_0) - 
      F_{.,0}(\widetilde{\Zett}_{.}|\widetilde{\Zett}_0)}_{\mathrm{w}}
\!\!\!&\leq&\!\!\!
\norm{F_{.,0}(\Zett_{.}|\Zett_0) - F_{.,0}({\Zett}_{.}|\widetilde{\Zett}_0)}_{\mathrm{w}}
\cr
&&\hskip-.3truecm
+\,
\norm{F_{.,0}({\Zett}_{.}|\widetilde{\Zett}_0) - 
      F_{.,0}(\widetilde{\Zett}_{.}|\widetilde{\Zett}_0)}_{\mathrm{w}}
\,.
\label{normFtriangleYY}
\eea
  Now,
$\norm{F_{.,0}(\Zett_{.}|\widetilde{\Zett}_0) - 
       F_{.,0}(\widetilde{\Zett}_{.}|\widetilde{\Zett}_0)}_{\mathrm{w}}$ 
was estimated already in the proof of Proposition \ref{olFisCONTRACTmap},
see \refeq{FLIPtrajT} (recall that the conditioning
$\lim_{t\to 0}\Zett_t =\Zett_0 = \lim_{t\to 0}\widetilde{\Zett}_t$ 
that entered the statement of  Proposition \ref{olFisCONTRACTmap} did not 
enter the estimates for \refeq{FLIPtrajT} themselves).
 Furthermore, with $w >2\bar{w}$ it follows that the 
parameter conditions in the proof of Theorem \ref{ExistUnique} 
are fulfilled for each $\widetilde{\Zett}_0$; hence, in
\refeq{FLIPtrajT} we have  $L[\fe;w,\wcrit]\leq 1/2$ 
for each $\widetilde{\Zett}_0$. 
 Thus, by \refeq{wDISTofYYisDISTofFF}, \refeq{normFtriangleYY},
and \refeq{FLIPtrajT}, and with $(1-L[\fe;w,\wcrit])^{-1} \leq 2$, 
we arrive at the estimate
\beq
\norm{\Zett_{.} - \widetilde{\Zett}_{.}}_{\mathrm{w}}
\leq
2\norm{F_{.,0}(\Zett_{.}|\Zett_0) - F_{.,0}({\Zett}_{.}|\widetilde{\Zett}_0)}_{\mathrm{w}}
\,,
\label{wDISTofYYfirstESTIM}
\eeq
uniformly for all 
$\widetilde{\Zett}_0\in \{\Zett_{0}^{(\alpha)}\}_{\alpha\in \II}$.

  The proof of Proposition \ref{LIPinCAUCHYdata} has thus been reduced 
to proving Lipschitz continuity of $F_{.,0}$ in its second argument, 
given the first.
 Since, by the triangle inequality,
\bea 
\norm{F_{.,0}(\Zett_{.}|\mu_0;\zeta_0) 
- 
F_{.,0}(\Zett_{.}|\tilde{\mu}_0;\tilde{\zeta}_0)}_{\mathrm{w}}
\!\!\! &\leq &\!\!\!
\norm{F_{.,0}(\Zett_{.}|\mu_0;\zeta_0) 
- 
F_{.,0}(\Zett_{.}|\tilde{\mu}_0;\zeta_0)}_{\mathrm{w}}
\cr
&& \!\!\!\! + 
\norm{F_{.,0}(\Zett_{.}|\tilde{\mu}_0;\zeta_0) 
- 
F_{.,0}(\Zett_{.}|\tilde{\mu}_0;\tilde{\zeta}_0)}_{\mathrm{w}}
\,,
\label{normFtriangleTWO}
\eea
it suffices to show that for given $\Zett_{.}$ and $\zeta_0$, we have
\beq
\norm{F_{.,0}(\Zett_{.}|\mu_0;\zeta_0) - 
      F_{.,0}(\Zett_{.}|\tilde{\mu}_0;\zeta_0)}_{\mathrm{w}}
\leq
L_1^*
\dLIPnorm{\mu_0 - \tilde{\mu}_0}
\label{FmuLIPnull}
\,,
\eeq
and for given $\Zett_{.}$ and $\tilde{\mu}_0$, 
\beq
\norm{F_{.,0}(\Zett_{.}|\tilde{\mu}_0;\zeta_0) - 
      F_{.,0}(\Zett_{.}|\tilde{\mu}_0;\tilde{\zeta}_0)}_{\mathrm{w}}
\leq
L_2^*
\norm{\zeta_0 - \tilde{\zeta}_0}_{HL}
\,,
\label{FzetaLIPnull}
\eeq
with $L_1^*,L_2^*$ depending at most on $\bar{w}$.
 For then it follows from 
\refeq{normFtriangleTWO}, \refeq{FmuLIPnull}, \refeq{FzetaLIPnull}
that 
\beq
\norm{F_{.,0}(\Zett_{.}|\Zett_0) - F_{.,0}(\Zett_{.}|\widetilde{\Zett}_0)}_{\mathrm{w}}
\leq L^*
\norm{\Zett_0-\widetilde{\Zett}_0}
\,,
\label{FLIPnull}
\eeq
with $L^*[\bar{w}] := \max\{L_1^*, L_2^*\}$, completing the proof of
Proposition \ref{LIPinCAUCHYdata}, with $L_0 = 2L^*$.

 As to \refeq{FmuLIPnull},  for all $t\in\RR$ we have
\bea
\norm{F_{t,0}(\Zett_{.}|\mu_0;\zeta_0) - 
      F_{t,0}(\Zett_{.}|\tilde{\mu}_0;\zeta_0)}
\!\!\!& = &\!\!\!
\norm{\Pi_{t,0}^\dagger[\zeta_.](\mu_0)
     - 
      \Pi_{t,0}^\dagger[\zeta_.](\tilde{\mu}_0 )}_\Ld
\cr &\leq&\!\!\!
{\phantom{\biggl(\!\!} }\!\!
{\textstyle{\frac{2 + \max\{\bar{w} , 1/\bar{w} \}}
	         {\sqrt{2}}
}} \,e^{\bar{w} |t|}\, \norm{\mu_0 - \tilde{\mu}_0}_\Ld
\,,
\label{preLoneSTAR}
\eea
the inequality by Lemma \ref{Liapunov} and Lemma \ref{LemAdLip}.
 Since $w\geq 2\bar{w}$, the
$\sup_{t\in \RR}\Bigl( e^{-w |t|} \refeq{preLoneSTAR}\Bigr)$ 
exists.
 Estimating it further with $w -\bar{w}\geq \bar{w}$ for 
$w\geq 2\bar{w}$ now gives \refeq{FmuLIPnull}, with 
\beq
L_1^*[\bar{w}] 
= 
\frac{2 + \max\{ {\textstyle{\bar{w}}}, {\textstyle{1/\bar{w}}} \}}
     {\sqrt{2}}
.
\label{LoneSTAR}
\eeq

 As to  \refeq{FzetaLIPnull},  for all $t\in\RR$ we have
\beq
\norm{F_{t,0}(\Zett_{.}|\tilde{\mu}_{0};\zeta_{0}) - 
      F_{t,0}(\Zett_{.}|\tilde\mu_{0};\tilde{\zeta}_{0})}
=
\| {\Phi_{t,0}^\psi[0_{.}] (\zeta_{0} - \tilde\zeta_{0})}\|_{\dotH}
+
\norm{\Phi_{t,0}^\varpi[0_{.}] (\zeta_{0} - \tilde\zeta_{0})}_{L^2}
\label{HLequalsHandLzak}
\,,
\eeq
where $\Phi_{t,0}^\psi[0_{.}] (\,.\,)$ 
and $\Phi_{t,0}^\varpi[0_{.}] (\,.\,)$
are the free propagators obtained from Kirchhoff's formulas 
\refeq{ea05} and \refeq{ea06} by replacing $\mu_{.}\to 0_{.}$;
note that $\Phi_{t,0}^\psi[0_{.}] (\,.\,)$ 
and $\Phi_{t,0}^\varpi[0_{.}] (\,.\,)$
are linear operators. 
 For initial data $\psi_0\in (\dotH\cap\dotHH)(\RR^3)$, 
the freely propagating wave
is a $HL$-strong solution of the homogeneous wave equation and its
field energy $\cE_{\mathrm{W}}(\zetaFREE_{.})$ is conserved. 
 This implies the bounds
\beq
\| {\Phi_{t,0}^\psi[0_{.}] (\zeta_{0} - \tilde\zeta_{0})}\|_{\dotH}
+
\norm{\Phi_{t,0}^\varpi[0_{.}] (\zeta_{0} - \tilde\zeta_{0})}_{L^2}
\leq
2\sqrt{\cE_{\mathrm{W}}(\zeta_0-\tilde\zeta_0)}
\leq 
\sqrt{2}
 \norm{\zeta_0 - \tilde\zeta_0}_{HL}
\,.
\label{FREEfieldNORMSbounded}
\eeq
 Hence, 
\beq
	 L_2^* = \sqrt{2}.
\label{LtwoSTAR} 
\eeq

  Estimates \refeq{FmuLIPnull} with 
\refeq{preLoneSTAR},
 and \refeq{FzetaLIPnull}
with \refeq{LtwoSTAR} now combine to
\beq
\norm{F_{.,0}(\Zett_{.}|\mu_0;\zeta_0) 
- 
F_{.,0}(\Zett_{.}|\tilde{\mu}_0;\tilde{\zeta}_0)}_{\mathrm{w}}
\leq
L^*
\norm{\Zett_0-\widetilde{\Zett}_0 }
\eeq
with 
\beq
L^*[\bar{w}]  
= 
\max
\left\{
{\textstyle{
 \sqrt{2}
\,,\,
\frac{2 + \max\{ \bar{w}, 1/\bar{w} \}}
     { \sqrt{2}}}}
\right\}
\label{Lstar}
\eeq
for all $w>2\bar{w}$.

 The proof of
Proposition \ref{LIPinCAUCHYdata} is complete, with 
$L_0[\bar{w}]   = 2L^*[\bar{w}]$.\QED

\subsection{The continuum limit}
\label{secCONTINUUMlim}

 Note that so far nothing prevents the measure $\mu_0(\dd{z})$, 
which obtains as limit of the $\veps[{\bzN_0}](\dd{z})$ when $N\to\infty$, 
from being as singular as the $\veps[{\bzN_0}](\dd{z})$ are.
 In particular, we may even allow 
$\veps[{\bzN_0}](\dd{z}) \leadsto \delta_{z_0}(\dd{z})$.
 Since in physical applications of Vlasov theory one is
typically interested in continuum  solutions, we
now suppose that when $N\to\infty$, the familiy of initial empirical 
measures $\veps[{\bzN_0}](\dd{z})$ converges $\wGaBpx$-strongly
to a measure $\mu_0(\dd{z})$ 
which is absolutely continuous \wrt Lebesgue measure.
 We write  $\mu(\dd{z}) = \mu^{f}(\dd{z}) = f(z)\dd{z}$ for
the absolutely continuous measures in ${P_1}(\RR^6)$.
 The set of their Radon--Nikodym derivatives is denoted $\probLpx(\RR^6)$;
thus $f\in\probLpx(\RR^6)$.
  We now show that when $\mu_0(\dd{z}) = \mu^{f_0}(\dd{z})$, then
$\mu_t = \mu_t^{f}$, with $f(\,.\,,\,.\,,t)\in \probLpx(\RR^6)$ 
for all $t\in\RR$.

\begin{Prop}
\label{LPsolutions}
	If $(\mu_{.},\zeta_{.})\in (\Cw^0\cap C^1)(\RR,\wGaBpx)$ solves 
	the Vlasov fixed point equation \refeq{VlasovWaveEQasFP} 
	with $\mu_0 = \mu^{f_0}$, $f_0\in (\probLpx\cap L^p)(\RR^6)$ 
	for some $p\geq 1$, then $\mu_{.}=\mu^{f(.,.,t)}$ with 
	$f(\,.\,,\,.\,,t)\in (\probLpx\cap L^p)(\RR^6)$ for all $t\in\RR$;
	note that  $p\geq 1$ includes the case that
	$f_0\in \probLpx(\RR^6)$ while $f_0\not\in L^p(\RR^6)$ for any $p> 1$.
\end{Prop}

\noindent
\textit{Proof of Proposition \ref{LPsolutions}:}
  Suppose $\mu_{.} \in C^1(\RR,\wPpx)$ is a strong generalized solution 
of the Vlasov continuity equation \refeq{VfEQeps} for given
$\zeta_{.}\in (C_b^0\cap C^1)(\RR,(\dotH\oplus L^2)(\RR^3))$,
with Cauchy data $\mu_0 = \mu^{f_0}$, $f_0\in (\probLpx\cap L^p)(\RR^6)$ 
for some $p\geq 1$. 
  Then $\mu_{.}=\mu^{f(.,.,t)}$ with 
$f(\,.\,,\,.\,,t)\in (\probLpx\cap L^p)(\RR^6)$ for all $t\in\RR$.
 But this follows from the definition of a generalized solution,
a straightforward change of variables from $z$ to 
$\Pi_{t,0}[\zeta_.](z)$ under the integral, noting
the properties of the flow $\Pi_{.,.}$ summarized in 
Corollary \ref{PIsymplecto}.\QED
\newpage

\subsubsection{Additional conservation laws for continuum solutions}

 The argument used to prove Proposition \ref{LPsolutions} has the
useful corollary that continuum solutions $\Zett^f_{.}$ with 
$f_0\in (\probLpx\cap L^p)(\RR^6)$ for some $p> 1$ enjoy 
additional conservation laws.
 Here we wrote $\Zett^f_{.}$ for $\Zett_{.}=(\mu_{.},\zeta_{.})$ 
with $\mu_t = \mu^{f(.,.,t)}$.

 For any $g:\RR_+\to \RR$, we define the $g$-Casimir functional
of $\Zett^f$ by
\beq
\cC^{(g)}\left(\Zett^f\right)
=
{\small\int} g\circ f \, \dd{z}\, ,\qquad
{\mathrm{whenever}}\quad g\circ f \in L^1(\RR^6) \,.
\label{CASIMIRfuncVeps} 
\eeq 
 For 
$g =\, \mathrm{id}$, we obtain the mass functional \refeq{MfuncVeps}
for absolutely continuous $\mu_t=\mu_t^f$;
for 
$g(\,.\,) =\, ({\mathrm{id}}(\,.\,))^p$, $p>1$, we get the $p$-th power
of the $L^p$ norm of $f$; 
the case
$g(\,.\,) = -{\mathrm{id}}(\,.\,)  \log({\mathrm{id}}(\,.\,) /f_*)$, 
gives the entropy of $f$ relative to $f_*$,
\beq
 \cC^{(-{\mathrm{id}} \log({\mathrm{id}}/f_*))}\left(\Zett^f\right)
=
- \int f\ln ({f}/{f_*}) \dd{z}
\equiv 
 \cS(f|f_*); 
\label{relENTROPY}
\eeq
here, $f_*\in \probLpx(\RR^6)$
is an otherwise arbitrary probability density function.
 In particular, $\cS(f|f_{*})$  is well-defined if 
$f\in(\probLpx\cap L^{1+\epsilon})(\RR^6)$ for some $\epsilon>0$.

\begin{Prop}\label{propClawsVeps} 
Let $t\mapsto \Zett_t\in (\Cw^0\cap C^1)(\RR,\wGaBpx)$ be a 
  generalized solution of the regularized wave gravity Vlasov model 
  for which $\mu_t=\mu_t^f$ is absolutely continuous.
Then, beside the conservation laws
	\refeq{PconstVeps},
	\refeq{JconstVeps},
	\refeq{EconstVeps},
whenever $\cC^{(g)}\bigl(\Zett_0^f\bigr)$ exists, also
\beq
\cC^{(g)}\left(\Zett_{.}^f\right) = \cC^{(g)}\bigl(\Zett_0^f\bigr)
\,.
\label{CASIMIRconstVeps}
\eeq
 In particular, if $f_0\in(\probLpx\cap L^{1+\epsilon})(\RR^6)$ 
for some $\epsilon>0$, then the relative entropy of 
$f(\,.\,,\,.\,,t)\in \probLpx(\RR^6)$ is conserved, i.e.
\beq
 \cS(f|f_*) =  \cS(f_0|f_*) 
\,.
\eeq
\end{Prop}

\subsection{Law of large numbers and central limit theorem}
\label{subsecLLNCLT}

\subsubsection{The law of large numbers}
\label{subsubsecLLN}

 Convergence in KR topology of probability measures $\mu^{\supN}$
as $N\to\infty$ implies the convergence in probability of a family 
of random variables with laws $\mu^{\supN}$; see \cite{Dudley}.
 Since at time $t$ the {\textit{empirical}} 
measures $\veps[{\bzN_t}]$ do converge in
KR topology to $\mu_t$ if they do so at $t=0$, 
our theorem about the $N\to\infty$ 
limit of the $N$-body generalized solutions to the regularized 
wave gravity Vlasov model is equivalent to the following law of 
large numbers. 

\begin{Theo}
\label{FirstLLN}
 For $N\in\NN$,  let $\mu^{\supN}_0\in {P_1}(\RR^6)$ 
be given, with $\supp(\mu_0(\dd{x}\times\RR^3))\subset B_{R}$,
and suppose $\mu^{\supN}_0\leadsto \mu_0\in {P_1}(\RR^6)$.
 Moreover, let 
${\mathbf z}_0^{\supN}\in \RR^{6N}$ 
be a random vector whose components in $\RR^6$ are (not necessarily 
independent)  random variables $z^{\supN}_{1}(0), ..., z^{\supN}_{N}(0)$ 
with common law $\mu^{\supN}_0$.
 To each  $\bz_0^{\supN}$ assign a unique
$\zeta[{\bz_0^{\supN}}] \in ((\dotH\cap\dotHH)\oplus L^2)(\RR^3)$
satisfying \refeq{phiNULLepsASYMP},\refeq{varpiNULLepsASYMP}, such that 
$\zeta[{\bz_0^{\supN}}] \to \zeta_0=(\psi(\,.\,,0),\varpi(\,.\,,0))
\in ((\dotH\cap\dotHH)\oplus{L}^2)(\RR^3)$ $HL$-strongly  when 
$N\to\infty$, with $\psi(\,.\,,0)$, $\varpi(\,.\,,0)$
satisfying   \refeq{phiNULLepsASYMP}, \refeq{varpiNULLepsASYMP}.

 Let $(\mu_{.}, \zeta_{.}) \in (\Cw^0\cap C^1)(\RR,\wGaBpx)$ 
be the unique generalized strong solution of the regularized wave 
gravity Vlasov equations for initial data 
$(\mu_{0}, \zeta_0)$.
 Let $t\mapsto \bzN_t\in \RR^{6N}$ be the 
path in particle phase space whose $N$ components in $\RR^6$
jointly solve the Einstein--Newton equations of motion 
(\ref{NqDOTeps}), (\ref{NpDOTeps}) with the initial data 
$\bz_0^{\supN}$, and with $t\mapsto\psi^{\supN} (\,.\,,t)$ and 
$t\mapsto\varpi^{\supN} (\,.\,,t)$ solving the regularized 
wave gravity equations
\refeq{NphiEQeps} and \refeq{NvpiEQeps} with initial data 
$\zeta[{\bz_0^{\supN}}]$.
 Then, for any $g\in C^{0,1}(\RR^{6})$ and for all $t$, 
in probability we have 
\beq
 \frac{1}{N}{\textstyle\sum\limits_{i=1}^N}
g\!\left(z_i^{\supN}(t)\right) 
\stackrel{N\to\infty}{\longrightarrow}
 \int g(z)\mu_t(\dd{z}).
\label{firstLLN}
\eeq
\end{Theo}

\begin{Rema}
 By invoking the extremal decomposition of permutation invariant
probability measures on $({\RR^6})^\NN$ \cite{HewittSavage}, 
our LLN \refeq{firstLLN} can be generalized to 
\beq
 {\textstyle{\left({N\atop n} \right)^{-1}}}
{\textstyle\sum\limits_{1\leq i_1 <...<i_n\leq N}^{}}
g\!\left(z_{i_1}^{\supN}(t),...,z_{i_n}^{\supN}(t)\right) 
\stackrel{N\to\infty}{\longrightarrow}
 \int g(z_1,...,z_n)\mu_t^{\times n}(\dd{z_1}...\dd{z_n})
\label{secondLLN}
\eeq
for any permutation invariant $g\in C^{0,1}(\RR^{6n})$ and for all $t$.
\end{Rema}

\begin{Rema}
 One actually should also allow the field initial data $\zeta^{\supN}_0$
to be random variables independently of the particle
random variables for each $N$, but this would require a whole new setup 
involving probability measures on field space,
the choice of an adequate topology on that space, beyond what has been
developed in this paper.
\end{Rema}

 Even though our LLN does not demand that $\mu_.$ be a continuum 
solution, in applications this is typically so. 
 While our LLN implies that the Vlasov continuum approximation to 
the sampling of $N$ body systems becomes exact for all $t$ in the 
limit of infinite $N$, for any particular physical system $N$ is 
fixed and may vary only from system to system.
 Thus, take $\mu_0^{\supN} = \mu_0$ for all $N$, with $\mu_0=\mu^{f_0}$.
 By assumption $\veps[{\bzN_0}]\leadsto\mu_0$ when $N\to\infty$.
 Yet for any finite $N$ we have $\dLIPnorm{\veps[\bzN_0] - {\mu}_0}>0$,
and then our estimates of section 4.1 show that at any other time $t$ 
we only have 
$\dLIPnorm{\veps[\bzN_t] - {\mu}_t} 
	\leq e^{C|t|}\dLIPnorm{\veps[\bzN_0] - {\mu}_0}$. 
 Hence, we can only conclude that the physical mean values 
l.h.s.\refeq{firstLLN} 
at time $t$ can be computed in acceptable approximation by their Vlasov 
continuum analog, i.e. r.h.s.\refeq{firstLLN} with $\mu_t=\mu^{f(.,.,t)}$, 
if $|t|$ is ``not too large,'' a notion which depends on $N$ and on 
how good the approximation is initially.
\newpage

\subsubsection{The central limit theorem}
\label{subsubsecCLT}

 Having obtained the law of large numbers, we next inquire into the
fluctuations around the deterministic mean.
 Our goal is to derive a central limit theorem for the dynamical variables 
$(\bzN_t, \zeta^{\supN}_t)$, which are random variables through their 
dependence on the random initial data for the $N$ particles, viz. $\bzN_0$.

 We adapt the technique of \cite{BraunHepp}, who studied
fluctuations of particle motions for  non-relativistic Vlasov equations. 
 This is done in two steps.
 First we study the differences of (primarily) test particle motions generated 
by the finite $N$ versus the continuum flows, and of similar type differences 
of field evolutions.
 The attribute ``primarily'' in parentheses refers to the fact that almost 
all (w.r.t. Lebesgue measure) initial conditions launch a test particle
evolution, with finitely many exceptions which are upgraded and follow the 
proper evolution.
 In our case all these particle evolutions are generated by the adjoint
flows on particle phase space of the Vlasov flows on $\wPpx$ whose fixed 
points are the proper Vlasov evolutions (both of course coupled to the 
same wave gravity equations).
  We prove the convergence of the characteristic function of the fluctuation 
process to a Gaussian characteristic function in a suitable norm.
 In the second step we extract from this analysis the fluctuations for
the proper evolutions.

 We recall that initial data 
$\bzN_0 = (z_1^{\supN}(0),...,z_N^{\supN}(0))\in\RR^{6N}$ 
with $q_k^{\supN}(0)\in B_R$
for $k=1,2,...,N$ uniquely define initial data 
$(\veps[{\bzN_0}],\zeta[{\bzN_0}])\in \GaBpx$ which launch 
the unique strong solution 
$(\veps[{\bzN_.}],\zeta^{\supN}_.)\in (\Cw^0\cap C^1)(\RR,\wGaBpx)$ 
of our regularized wave gravity Vlasov equations. 
 By \refeq{defPARTICLEflow}, \refeq{waveEqFlow}, the
solution $(\veps[{\bzN_.}],\zeta^{\supN}_.)$ generates a single 
particle flow $\Pi_{.,.}[\zeta^{\supN}_.] (.)$ on ${\RR^6}$ giving
single particle evolutions
\begin{equation}
  z_t(z_0;\bzN_0)
=
  \Pi_{t,0}[\zeta^{\supN}_.]( {z}_0)\,,
  \label{flow1p}
\end{equation}
and a flow $\Phi_{.,.}[\veps[{\bzN_.}]] (.)$ on field space giving 
field evolutions
\begin{equation}
  \zeta_t(\zeta_0;\bzN_0)
=
  \Phi_{t,0}[\veps[{\bzN_.}]]({\zeta}_0)\,,
  \label{field1p}
\end{equation}
with data $z_0\in \RR^6$ and $\zeta_0\in (\dotH\oplus  L^2)(\RR^3)$;
note, however, that we are \emph{exclusively} considering data
$(z_0,\zeta_0)\in \GaB^{(1)}$.
 As for the notation, by the r.h.s. of \refeq{flow1p} the dynamics 
$z_.$ depends on $z_0$ and on $\zeta^{\supN}_.$, but the latter in turn 
is implicitly fixed by $\bzN_0$ (and our Vlasov equations); similarly the 
notation in \refeq{field1p} is explained.
 As for their dynamical significance, the dynamics $z_.$
solves the \emph{characteristic equations} of the Vlasov continuity 
equation given the fields $\zeta^{\supN}_.$.
 For almost all data $z_0\in\RR^6$ this is a
\emph{test particle} dynamics, the exception being when 
$z_0\in \{z_1^{\supN}(0),...,z_N^{\supN}(0)\}$, in which case
 $z_.$ coincides with one of the components $z_k^{\supN}(\,.\,)$
of the unique solution $(\bzN_., \zeta^{\supN}_.)$
of equations \refeq{NphiEQeps}, \refeq{NvpiEQeps},
\refeq{NrhoEQeps}, \refeq{NqDOTeps}, \refeq{NpDOTeps}
with Cauchy data
 $(\bz_0^{\supN},\zeta[{\bz_0^{\supN}}])$.
 The wave dynamics ${\zeta}_.$ in turn solves the linear inhomogeneous
wave equation given the source term $\fe*\rho^{\supN}_.$ obtained from 
$\veps[{\bzN_.}]$
 Note that the dynamics of ${z}_.$ and ${\zeta}_.$ are in general
independent of each other; the exception occurs when $\zeta_0=\zeta[{\bzN_0}]$,
in which case $\zeta_. = \zeta^{\supN}_.$.

 When $N\to\infty$ such that 
$\veps[{\bzN_t}] \leadsto \veps[{\bz_t^{(\infty)}}]$ 
and $\zeta^{\supN}_t \to \zeta^{(\infty)}_t$ ($HL$-strongly) for all $t$,
the flows for \refeq{flow1p}, \refeq{field1p} converge to the corresponding
flows generated by the Vlasov solution 
$t\mapsto(\veps[{\bz_t^{(\infty)}}],\zeta^{(\infty)}_t)$; 
note that flows analogous to those for
\refeq{flow1p}, \refeq{field1p} are defined
for any admissible solution $\in (\Cw^0\cap C^1)(\RR,\wGaBpx)$ 
of our Vlasov equations.
  The discussion of the finite-$N$ fluctuations amounts to analyzing 
the difference of the evolutions  \refeq{flow1p}, \refeq{field1p} 
for finite $N$ versus $N=\infty$.
 Note that in either case the Vlasov dynamics that generates
\refeq{flow1p} and \refeq{field1p} is determined by $\bzN_0$,
respectively $\bz^{(\infty)}_0$.

 As to our notation, 
above we use the symbol $\zeta^{(\infty)}_.$ to distinguish the limit
fields of $\zeta^{\supN}_.$ from the fields $\zeta_.$ solving \refeq{field1p}
in general.
 However, eventually we will choose $\zeta_0 \equiv \zeta^{(\infty)}_0$ 
for the sake of concreteness in the CLT.
 Note furthermore that $\bz^{(\infty)}_0$ could be anything from a single
point $z_*$ to a continuous function, i.e. we may have
$\bz^{(\infty)}_0 = f_0(z)$ (in distribution).
 Therefore, $\veps[{\bz_0^{(\infty)}}]$ could be any probability measure
$\mu_0\in P_1$ with  $\supp(\mu_0(\dd{x}\times\RR^3))\subset B_{R}$;
in particular, if $\bz^{(\infty)}_0 = f_0(z)$ is
an empirical continuum density,  the empirical continuum measure 
$\mu^{f_0}(\dd{z})=\veps[\bz^{(\infty)}_0] (\dd{z})$.

  We stipulate further notation.
  Recall that $(z,\zeta) = \zett\in \GaB^{(1)}$ denotes generic points
in $\GaB^{(1)}$. 
 In this vein, for solutions of \refeq{flow1p}, \refeq{field1p} 
we use whenever possible the shorthand
\beq
 (z_t,\zeta_t) = \zett_t 
\eeq
but when necessary to discuss the components in more detail, we write
\beq
 (z_t,\zeta_t) = 
(q_t,p_t,\psi_t,\varpi_t)
\,.
\eeq
 Note that $\zett_t$ is a function of $\zett_0$ and $\bzN_0$; in components, 
$q_t$ and $p_t$ are points in $\RR^3$ which are functions of $z_0$ and
$\bzN_0$ while $\psi_t$ and $\varpi_t$ are points in $\dotH\cap\dotHH$, 
respectively $L^2$, which are functions of $\zeta_0$ and $\bzN_0$.
 We also recycle some of our previously stipulated abbreviations. 
 Thus,
$\int .. \mu(\dd{z})$ stands for $\iint_{\RR^6}.. \mu(\dd{x}\dd{p})$
and 
$\iint .. \mu(\dd{z})\nu(\dd{z'})$ for 
$\iint_{\RR^6\times\RR^6} .. \mu(\dd{x}\dd{p})\nu(\dd{x}'\dd{p}')$.
 So much for notation.

 As a technical primer which will allow us to perform estimates needed
for the main theorem, we will first show that $\zett_t(\zett_0,\bzN_0)$
is regular as a function of $\bzN_0$, with bounded derivatives.
 Since $\bzN_0$ is uniquely identified with the empirical measure
$\veps[{\bzN_0}]$ and similarly $\bz^{(\infty)}_0$ is uniquely
identified with the empirical measure $\veps[\bz^{(\infty)}_0]$
(with the possibility $\veps[\bz^{(\infty)}_0](\dd{z})=\mu^{f_0}(\dd{z})$,
the continuum case), we really mean regularity of $\zett_t$ as a function 
of $\veps[{\bzN_0}]$, respectively $\veps[{\bz_0^{(\infty)}}]$, 
with bounded derivatives w.r.t. $\veps[{\bzN_0}]$
(respectively $\veps[{\bz_0^{(\infty)}}]$).
 Actually, it will be necessary to take derivatives not just
restricted to the subset $\wPpx$ of $\wM$. 
 For this purpose, we note that by a simple scaling of our 
wave gravity Vlasov equations 
\refeq{VphiEQeps}, \refeq{VvpiEQeps}, \refeq{VfEQeps},
we can first generalize  the initial data $\in \wPpx(\RR^6)$ to
those $\in \wMplus(\RR^6)$, and by the linearity  of \refeq{VfEQeps} 
in $f$ given $\psi$ together with the linearity of the map $f\mapsto\rho$,
we can even allow the initial data for \refeq{VfEQeps} to be a signed measure
$\sigma_0\in \wM(\RR^6)$; however, we always demand that 
$\supp(\sigma_0(\dd{x}\times\RR^3))\subset B_{R}$.
 As to the initial data for \refeq{VvpiEQeps}, \refeq{VfEQeps},
we suitably extend the unique map $\bzN_0\mapsto \zeta[{\bzN_0}]$ to 
$\sigma_0\mapsto\zeta[{\sigma_0}]
   \in (\dotH\cap\dotHH)(\RR^3) \oplus L^{2}(\RR^3)$
(slightly abusing notation), obeying 
 \refeq{phiNULLepsASYMP}, \refeq{varpiNULLepsASYMP}, and we write
$\zett_t(\zett_0,\bzN_0)$ by $\zett_t(\zett_0,\sigma_0)$.
 Note that $\zett_0$ itself does \emph{not} depend on $\sigma_0$.

 The (Gateaux) derivative $D^1 g(\sigma,.):\RR^6\to \mathfrak{B}$ 
with respect to a finite measure $\sigma\in\wM(\RR^6)$
of a function $g(\sigma)\in\mathfrak{B}$ (any Banach space) 
is defined by the identity
\begin{equation}
  \int D^1 g(\sigma, z) \nu(\dd z)
  =
  \lim_{s \to 0}
    \frac {g(\sigma +s\nu) - g(\sigma)} {s}
\,,
\label{drm}
\end{equation}
valid for all $\nu\in \wM(\RR^6)$; 
we here will restrict $\nu$ to satisfy
$\supp(\nu_0(\dd{x}\times\RR^3))\subset B_{R}$.
 Analogously one defines the higher derivatives $D^jg(\sigma,.)$ on
$\RR^{6j}$. 
  To have a shorthand we write $\bzj$ for generic points 
in $\RR^{6j}$; to achieve a more uniform notation in the presentation 
we will also write $\bz^{(1)}$ for $z$ in \refeq{drm} and other first 
derivatives.

 Next we define several auxiliary norms. 
 Below, whenever we take the sup over $z_0,\zeta_0,\sigma_0,\bzj$, 
it is understood that $z_0 \in  B_R\times\RR^3$, that
$\zeta_0 \in (\dotH\cap\dotHH)(\RR^3) \oplus L^{2}(\RR^3)$
obeying \refeq{phiNULLepsASYMP}, \refeq{varpiNULLepsASYMP},
that $\sigma_0\in \wM$ has 
$\supp(\sigma_0(\dd{x}\times\RR^3))\subset B_{R}$ and $\abs{\sigma_0}\leq 1$,
and that $\bzj \in (B_R\times\RR^3)^j$.
 Thus, we define (noticing that $D^1 \nabla \psi_. = \nabla D^1 \psi_.$)
\begin{eqnarray}
  \norm{D^j z_t}_{(\mathrm{u})}
  \!\!&:=&\!\! 
  \sup_{z_0,\sigma_0,\bzj}
  |D^j z_t(z_0,\sigma_0,\bzj)|
\label{defDjz}
\\
  \norm{D^j \psi_t }_{(L)} 
  \!\!&:=&\!\!
  \sup_{\zeta_0,\sigma_0,\bzj}
  \norm{ D^j \psi_t(\zeta_0,\sigma_0,\bzj) }_{L^2}
\label{defDjphi}
\\
  \norm{D^j \varpi_t }_{(L)}  
  \!\!&:=&\!\!
  \sup_{\zeta_0,\sigma_0,\bzj}
  \norm{ D^j \varpi_t(\zeta_0,\sigma_0,\bzj) }_{L^2}
\label{defDjvarpi}
\\
  \norm{D^j \psi_t }_{(H)} 
  \!\!&:=&\!\!
  \sup_{\zeta_0,\sigma_0,\bzj}
  \norm{ D^j \psi_t(\zeta_0,\sigma_0,\bzj) }_\dotH
\,,
\label{defDjDphi}
\end{eqnarray}
and we define $\norm{D^j \zeta_t }_{(HL)}$ by setting
\beq
  \norm{D^j \zeta_t }_{(HL)}^2 
:=
  \norm{ D^j \psi_t }_{(H)}^2 
  + \norm{ D^j \varpi_t }_{(L)}^2 
\label{defDjzeta}
\eeq
\begin{Prop}
\label{bd}
  \textit{Given Cauchy data $\sigma_0 \in \wM ({\RR^6})$ with
    $\supp(\sigma_0(\dd{x}\times\RR^3))\subset B_{R}$ for \refeq{VfEQeps},
    and associated with it Cauchy data 
    $\zeta[{\sigma_0}]\in ((\dotH\cap\dotHH)\oplus L^2)(\RR^3)$
    for \refeq{VphiEQeps}, \refeq{VvpiEQeps},
    satisfying \refeq{phiNULLepsASYMP}, \refeq{varpiNULLepsASYMP},
    this Vlasovian Cauchy problem has a unique strong solution 
    $(\sigma_.,\zeta_.^{(\sigma)})
      \in (\Cw^0\cap C^1)(\RR,\wM\oplus((\dotH\cap\dotHH)\oplus L^2)(\RR^3))$.
    This solution generates evolutions $(z_.,\zeta_.)=\zett_.$
    defined by \refeq{flow1p}-\refeq{field1p} which $\forall k \in \NN$
    are $k$ times continuously differentiable
    with respect to $\sigma_0$.
    Moreover, for all $\sigma_0\mapsto\zeta[\sigma_0]$ such that
    there exist functions $B^j(.)\in C^0(\RR^+)$, $j=1,...,k$,
    depending on $\sigma_0$ only through $\cE(\sigma_0,\zeta[\sigma_0])$ 
    and $|\sigma_0|$ such that
    $\norm{D^j \psi^{(\sigma)}(.,t^\prime,.) }_{(H)}\leq B^j(t)$,
    there exist functions $B^j_\ell(.)\in C^0(\RR^+)$, $j=1,\ldots,k$, 
    and $\ell=1,2,3$, depending 
    on $\sigma_0$ only through $\cE(\sigma_0,\zeta[\sigma_0])$ 
    and $|\sigma_0|$, such that for all $t \in \RR$,}
    \bea
      \norm{D^j z_t }_{(\mathrm{u})}
      \!\!\!&\leq&\!\!\!
      B^j_1 (t)\,,
    \label{boco}\\
      \norm{D^j \zeta_t }_{(HL)} 
      \!\!\!&\leq&\!\!\!
      B^j_2 (t)\,,
    \label{bopo}\\
      \norm{D^j \psi_t }_{(L)} 
      \!\!\!&\leq&\!\!\!
      B^j_3 (t)\,.
    \label{bode}
    \eea
\end{Prop}

\noindent 
\textit{Proof}. 
 By inspection one verifies that our existence and uniqueness proof for the 
Vlasov model carries over to these more general data with only miniscule
changes.

 As to the existence and uniqueness of $D^j \zett_.$, this
follows in the standard way \cite{Dieudonne}. 
 The first variation of the 
evolution equations yields $D^1 \zett_.$ thus
\begin{eqnarray}
  D^1 q_t(z_0,\bzu )
  \!\!\!&=&\!\!\!\!
  \int_{0}^{t} \Bigl[
      \frac {D^1 p_{t^\prime}(z_0,\bzu )}
            {(1 + \abs{p_{t^\prime}(z_0)}^2)^{1/2}}
      -
      \frac {p_{t^\prime}(z_0) \otimes p_{t^\prime}(z_0) 
	     \cdot D^1 p_{t^\prime}(z_0,\bzu )}
            {(1 + \abs{p_{t^\prime}(z_0)}^2)^{3/2}}
      \Bigr] \dd t^\prime
  \label{deriv1}
    \\
  D^1 p_t(z_0,\bzu)
  \!\!\!&=&\!\!\!\!
  \int_{0}^{t} \int
    \Bigl( [ \nabla \fe (y - q_{t^\prime}(z_0))
             \cdot D^1 q_{t^\prime}(z_0,\bzu)]
           \nabla \psi^{(\sigma)}(y,t^\prime)
  \cr
  && \hskip3truecm
 - \fe (y -q_{t^\prime}(z_0)) D^1 \nabla\psi^{(\sigma)}(y,t^\prime,\bzu)\Bigr)
    \dd y \dd t^\prime
  \label{deriv2}
    \\ 
  D^1 \psi_t(\zeta_0,\bzu)(x)
  \!\!\!&=&\!\!\!\!
\int_{0}^{t} (t-t^\prime)  
  \SSint
\Bigl[
      \int [ \nabla \fe (x^\prime - q_{t^\prime}(z_0))
 \cdot D^1 q_{t^\prime}(z_0,\bzu) ]
    \sigma_0 (\dd z_0 )
\Bigr.
\cr 
&&\hskip2.3truecm
\Bigl.\Bigl.
- \fe \left(x^\prime - q_{t^\prime}(\bzu)\right)
\Bigr] \dd\Omega \dd t^\prime
  \label{deriv4}
    \\
  D^1 \varpi_t (\zeta_0,\bzu)(x)
  \!\!\!&=&\!\!\!\!
 \int_{0}^{t} \!\SSint\!\left[\int
     ( D^1 q_{t^\prime}(z_0,\bzu)\! \cdot\! \nabla)
 (1 \!+\! (t-t^\prime) \Omega\! \cdot\! \nabla)
       \fe (x^\prime\! -\! q_{t^\prime}(z_0))
     \sigma_0 (\dd z_0)\right.
   \cr
  && \hskip1truecm
\Bigl.  - [1 + (t-t^\prime) \Omega \cdot \nabla]
      \fe (x^\prime \!-\! q_{t^\prime}(\bzu))
\Bigr]      \dd\Omega \dd t^\prime
  \label{deriv5}
    \\
  D^1 \nabla \psi_t(\zeta_0,\bzu)(x)
  \!\!\!&=&\!\!\!\!
 \int_{0}^{t} (t-t^\prime) \SSint \!\left[\int
   ( D^1 q_{t^\prime}(z_0,\bzu)\! \cdot\! \nabla )
 \nabla
        \fe (x^\prime \!-\! q_{t^\prime}(z_0))
     \sigma_0 (\dd z_0) \right.
     \cr
  &&  \hskip3truecm
\Bigl.  - \nabla \fe (x^\prime \!-\! q_{t^\prime}(\bzu))\Bigr]
    \dd\Omega \dd t^\prime\, ,
  \label{deriv6}
\end{eqnarray}
\newpage

\noindent
where once again we used the notation
$x^{\prime} =x + (t-t^{\prime})\Omega$, with $\Omega\in\SS^2$,
and $\SSint$ to denote $\frac{1}{4\pi}\int_{\SS^2}$;
moreover, to shorten the presentation, we have not displayed
the dependence on $\sigma_0$ of $(z_t,\zeta_t)$.
 Note also that in \refeq{deriv4}, \refeq{deriv5}, \refeq{deriv6} we have
transferred the time-dependence from $\sigma_{t^\prime}$ to the adjoint 
time-dependence of $q_{t^\prime}$ by change of variables.

 Next, as to \refeq{deriv4}, one easily obtains the bound
\begin{eqnarray}
  \Norm{ \int_0^t (t-t^\prime) \SSint
        \fe ( .^\prime - q_{t^\prime}(\bzu))
     \dd\Omega \dd t^\prime
  }_{(L)}
 \!\!\!&\leq&\!\!\!\!
 \int_0^t (t-t^\prime)\!\!\!\!\!\!
 \sup_{|\sigma_0| \leq 1\atop \bzu\in B_R\times\RR^3}
    \Norm{ \SSint
         \fe ( .^\prime - q_{t^\prime}(\bzu))
         \dd\Omega
    }_{L^2}\!
 \dd t^\prime
  \cr
  \!\!\!&\leq&\!\!\!
  \int_0^t (t-t^\prime) \norm{ \fe }_{L^2} \, \dd t^\prime
  =
  t^2{\textstyle{\frac{1}{2}}} 
\, \norm{ \fe }_{L^2},
\end{eqnarray}
where $.^{\prime} = . + (t-t^{\prime})\Omega$, and similarly
easily, using \refeq{defDjz}, one obtains
\begin{eqnarray}
  &&
  \Norm{ \int_0^t (t-t^\prime) \SSint \int
        \nabla \fe ( .^\prime - q_{t^\prime}(\bzu))
        \cdot D^1 q_{t^\prime}(z_0,\bzu) \sigma_0(\dd z_0)
     \dd\Omega \dd t^\prime
  }_{(L)}
  \cr
  \!\!\!&\leq&\!\!\! |\sigma_0|
  \int_0^t (t-t^\prime)
\!\!\! \sup_{|\sigma_0| \leq 1\atop\bzu\in B_R\times\RR^3}
    \Norm{ \SSint \int
         \nabla \fe ( .^\prime - q_{t^\prime}(\bzu))
         \cdot D^1 q_{t^\prime}(z_0,\bzu) \sigma_0(\dd z_0 )
         \dd\Omega
    }_{L^2}
   \dd t^\prime
  \cr
  \!\!\!&\leq&\!\!\!
  |\sigma_0| \int_0^t (t-t^\prime)
 \norm{ \nabla \fe }_{L^2} \norm{ D^1 q_{t^\prime}}_{(\mathrm{u})}\dd t^\prime
\,.
\end{eqnarray}
 These bounds and similar ones for
\refeq{deriv5} and \refeq{deriv6}, and some obvious estimates for
\refeq{deriv1} and \refeq{deriv2}, yield
\begin{eqnarray}
  \norm{D^1 q_t }_{(\mathrm{u})}
  \!\!\!&\leq&\!\!\!
  \int_{0}^{t} \norm{D^1 p_{t^\prime} }_{(\mathrm{u})} \dd t^\prime
  \label{dbound1}
  \\ \norm{D^1 p_t }_{(\mathrm{u})}
  \!\!\!&\leq&\!\!\!
  \int_{0}^{t}
  \Bigl[ \norm{\nabla\fe}_{L^2} \norm{ \psi^{(\sigma)}(\,.\,,t^\prime) }_\dotH 
         \norm{D^1 q_{t^\prime} }_{(\mathrm{u})}\cr
&&\hskip1truecm
    + \norm{\fe}_{L^2} \norm{D^1 \psi^{(\sigma)}(.,t^\prime,.) }_{(H)}
  \Bigr]
    \dd t^\prime
  \label{dbound2}
  \\
  \norm{D^1 \psi_t }_{(H)}
  \!\!\!&\leq&\!\!\!
t^2{\textstyle{\frac{1}{2}}} 
\norm{ \nabla \fe}_{L^2}
  + \norm{\nabla^{\otimes 2} \fe}_{L^2} \, |\sigma_0|
\int_{0}^{t} (t-t^\prime) \norm{D^1 q_{t^\prime} }_{(\mathrm{u})} \dd t^\prime
  \label{dbound4}
  \\
  \norm{D^1 \varpi_t }_{(L)}
  \!\!\!&\leq&\!\!\!
  t \norm{\fe}_{L^2}
  + t^2{\textstyle{\frac{1}{2}}} \norm{\nabla \fe}_{L^2}
  \cr
  &&
  + \, |\sigma_0|
    \int_{0}^{t}
     \Bigl[ \norm{\nabla \fe}_{L^2}
       + (t-t^\prime) \norm{\nabla^{\otimes 2} \fe}_{L^2} \Bigr] \
      \norm{ D^1 q_{t^\prime} }_{(\mathrm{u})} \dd t^\prime
  \label{dbound5}
  \\
  \norm{D^1 \psi_t }_{(L)}
  \!\!\!&\leq&\!\!\!
t^2{\textstyle{\frac{1}{2}}} 
\norm{\fe}_{L^2}
  + \norm{\nabla \fe}_{L^2} \, |\sigma_0|
\int_{0}^{t} (t-t^\prime) \norm{D^1 q_{t^\prime} }_{(\mathrm{u})} \dd t^\prime
  \, ,
  \label{dbound6}
\end{eqnarray}
where $ \norm{\nabla^{\otimes 2} \fe}_{L^2}^2 = 
\int \abs{{\nabla^{\otimes 2} \fe}}^2(x) \dd{x}$, with $\abs{M}$ the 
familiar Euclidean norm of a real symmetric $3\times 3$ matrix as an
element of $\RR^9$.

 We now recall that
$\norm{\psi^{(\sigma)}(\,.\,,t^\prime)}_\dotH$ is bounded uniformly 
for $t \in \RR$ by a constant depending on $\sigma_0$ only through
$\cE(\sigma_0,\zeta[\sigma_0])$; cf. Remark \ref{remaBoundsVeps}.  

 Next, as a special case, assume first that $\zeta[\sigma_0] = \zeta_0$ 
independent of $\sigma_0$.
 In this case $\psi^{(\sigma)}(.,t)=\psi_t$, and 
the bounds \refeq{boco}, \refeq{bopo} for $j=1$ follow from 
\refeq{deriv1}--\refeq{deriv5} by variants of the Gronwall lemma.
 The bound \refeq{bode} for $j=1$ then follows immediately. 
  The bounds \refeq{boco}, \refeq{bopo} 
for general values of $j$ follow by applying $D^{j-1}$
to \refeq{deriv1}, \refeq{deriv2}, \refeq{deriv4},
\refeq{deriv5}, \refeq{deriv6}.
 In particular, the bound
$\norm{D^j \psi^{(\sigma)}(.,t^\prime,.) }_{(H)}\leq B^j(t)$, $j=1,...,k$,
holds, with $B^j(.)\in C^0(\RR^+)$ depending on 
$\sigma_0$ only through $\cE(\sigma_0,\zeta[\sigma_0])$ and $|\sigma_0|$, 
for $\psi^{(\sigma)}(.,t) = \psi_t$ in this case. 

 Finally, the bounds clearly generalize to field data 
$\sigma_0\mapsto\zeta[\sigma_0]$
for which there exist functions $B^j(.)\in C^0(\RR^+)$, $j=1,...,k$,
depending on $\sigma_0$ only through $\cE(\sigma_0,\zeta[\sigma_0])$ 
and $|\sigma_0|$ such that
    $\norm{D^j \psi^{(\sigma)}(.,t^\prime,.) }_{(H)}\leq B^j(t)$.
 That this hypothesis is legitimate we just showed, for its supposition 
is valid in particular when $\zeta[\sigma_0] = \zeta_0$.\QED

\begin{Rema}\label{ignoramis}
	Unfortunately we do not yet know how general the class
of initial data $\sigma_0\mapsto\zeta[\sigma_0]$ is which validate
our hypothesis  that
$\norm{D^j \psi^{(\sigma)}(.,t^\prime,.) }_{(H)}\leq B^j(t)$, $j=1,...,k$,
with $B^j(.)\in C^0(\RR^+)$ depending on $\sigma_0$ only through 
$\cE(\sigma_0,\zeta[\sigma_0])$ and $|\sigma_0|$.
\end{Rema}

 Because of Remark \ref{ignoramis}, in the following we will restrict
the initial conditions for the fields to the special case 
$\zeta[\bzN_0]=\zeta_0$ independent of $\bzN_0$; the case $N=\infty$
is included.
 In this vein, we may now analyze the limit $N\to\infty$ of the
finite-$N$ fluctuations
\begin{eqnarray}
  \Delta_z (t,z_0;{\bzN_0},\mu_0)
  \!\!\!&:=&\!\!\!
  \sqrt{N} [z_t(z_0;{\bzN_0}) - z_t( z_0 ,\mu_0)]
\,,
\label{defzN}
\\
  \Delta_\zeta(t,\zeta_0;{\bzN_0},\mu_0)
  \!\!\!&:=&\!\!\!
  \sqrt{N}[\zeta_t(\zeta_0;{\bzN_0}) - \zeta_t(\zeta_0,\mu_0)]
\,.
\label{defzetaN}
\end{eqnarray}
  We write $\Delta_\zett =(\Delta_z, \Delta_\zeta)$.
  Recall that by Theorem \ref{LargeN}, if $\veps[{\bzN_0}]
\leadsto \mu_{0}$ as $N\to\infty$
and $\zeta[\bzN_0] = \zeta^{(\infty)}_0\equiv \zeta_0$ for all $N$, then
\begin{eqnarray}
  \veps[{\bzN_t}]
  & \leadsto &
  \mu_{t}
  \\
  \psi_t(\zeta_0;{\bzN_0})
  & \stackrel{\dotH}{\longrightarrow} &
  \psi_t(\zeta_0;\mu_0)
  \\
  \varpi_t(\zeta_0;{\bzN_0})
  & \stackrel{L^2}{\longrightarrow} &
  \varpi_t(\zeta_0;\mu_0)
\end{eqnarray}
and, due to Corollary \ref{PIsymplecto},
\begin{equation}
  | z_t( z_0,{\bzN_0}) - z_t(z_0 ,\mu_0) |
  \longrightarrow
  0 \, .
\end{equation}

  We now are ready to state the central limit theorem.

\begin{Theo}
\label{teorema}
  \textit{For $\mu_0 \in {P_1}({\RR^6})$,
define the sequence of particle product measures
$\mu_0^{\times N}(\dd \bzN_0) = \prod_{i=1}^{N} \mu_0 (\dd z^{(N)}_i (0))$,
and consider field initial data
$\zeta_0 \in (\dotH\cap\dotHH)(\RR^3) \oplus L^{2}(\RR^3)$.
  For any $K \geq 1$ and  $1 \leq k \leq K$, let
$\Zet_k\in \Gamma^{(1)}$, $\zet_k \in {\RR^6}$,  $t_{k} \in \RR$
and define 
${\VZet}=(\Zet_1,\ldots, \Zet_K)\in (\Gamma^{(1)})^K$, 
${\Vzet}=(\zet_1,\ldots, \zet_k)\in \RR^{6K}$, and
${\vec{t}}=(t_1,\ldots, t_K)\in\RR^K$. 
 Moreover, let
$\vec{\Delta_\zett}(\vec{t},\Vzet, \zeta_0; {\bzN_0})\in (\Gamma^{(1)})^K$ 
have $k$-th component $\Delta_\zett(t_k,\zet_k, \zeta_0; {\bzN_0})$.
 Then}
  \begin{equation}
    \lim_{N\to\infty} \int
      e^{ {\mathrm i}  
\langle\VZet |\vec{\Delta_\zett}(\vec{t},\Vzet,\zeta_0;{\bzN_0})\rangle}
      \mu_0^{\times N} (\dd \bzN_0)
    =
 e^{- \frac{1}{2} 
\langle\VZet|\opQ|\VZet\rangle}
\qquad
\label{clt}
\end{equation}
  \textit{where $\langle\,.\,|\,.\,\rangle$ 
is the scalar product in the Hilbert space 
   $(\Gamma^{(1)})^K$ (i.e. the sum over $K$ scalar products 
   $\langle\,.\,,\,.\,\rangle$ in $\Gamma^{(1)}$ indexed by $_k$),
and the operator $\opQ$ has $k,k^\prime$ component}
\begin{eqnarray}
  Q_{k,k^{'}} 
  \!\!&=&\!\!
  \int \bigotimes_{\kappa \in \{k,k'\}}
    D^1 \zett_{t_\kappa} (\zet_\kappa, \zeta_0, \mu_0, \bzu )
    \mu_0 (\dd \bzu )
%
  \cr
  &&
 - \bigotimes_{\kappa \in \{k,k'\}}
       \int D^1 \zett_{t_\kappa}( \zet_\kappa, \zeta_0, \mu_0, \bzu_\kappa)
         \mu_0 (\dd \bzu_\kappa)
  \, .
  \label{defQkk}
\end{eqnarray}
  \textit{For $\Vt$ and $\VZet$ in bounded sets, 
	the convergence is uniform in $\Vzet$.}
  \\
  \textit{The stochastic process
    $\eta = \lim_{N \to \infty} \Delta_\zett$ on $\Gamma^{(1)}$,
    with vanishing expectation and covariance \refeq{defQkk},
    can be represented as
\begin{equation}
  \eta_t(z_0 ,\mu_0)
  =
  \int D^1 \zett_t(z_0 ,\mu_0, \bzu ) \varphi(\mu_0,\dd \bzu )
\end{equation}
    and satisfies the equations obtained integrating
    \refeq{deriv1}, \refeq{deriv2}, \refeq{deriv4}, \refeq{deriv5}
    with respect to $\varphi(\mu_0,\dd \bzu )$
    with initial conditions $\eta_0 (z_0 ,\mu_0)=0$.
 Here, the random measure
    $\varphi(\mu,\dd \bzu ) \in M({\RR^6})$
    with Gaussian law is defined by
\begin{eqnarray}
  {\EE}_{\mu_0^{\times N}} \varphi(\mu_0,\triangle_1)
  \!\!\!&=&\!\!\! 0  \, ,
  \\
  {\EE}_{\mu_0^{\times N}}
    [ \varphi(\mu_0,\triangle_1) \varphi(\mu_0,\triangle_2) ]
  \!\!\!&=&\!\!\!
  \mu_0 (\triangle_1 \cap \triangle_2) - \mu_0 (\triangle_1)\mu_0 (\triangle_2)
  \, ,
\end{eqnarray}
    for measurable $\triangle_1, \triangle_2 \subset {\RR^6}$.}
\end{Theo}

\noindent \textit{Proof}. Apart from the different kind of
convergence for the fluctuations of the potential, the proof is
analogous to the one in~\cite{BraunHepp}. We carry out some
calculations to clarify the procedure.
  Let $\nu[{\bzN_0}] := \veps[{\bzN_0}] - \mu_0$. The
following calculations are valid whenever $D^1 \zett_.$ exists.

Obviously,
\begin{equation}
  \zett_t(z_0 ,\veps_{\bzN_0 }) - \zett_t(z_0 ,\mu_0)
  =
  \int_0^1 \int
    D^1 \zett_t( z_0 , \zeta_0, \mu_0 + r \nu[{\bzN_0}], \bzu  )
    \nu[{\bzN_0}] (\dd \bzu )
    \dd r
  \, .
\end{equation}
 We now define
\begin{eqnarray}
  \Xi_k^{(1)}(\sigma , \bzu ) 
  \!\!&=&\!\! \sqrt{N}  D^1 \zett_{t_k} (\zet_k, \zeta_0, \sigma, \bzu ) 
  \, , 
  \\
   \Xi_k^{(2)}(\sigma, \bzd)  
  \!\!&=&\!\! \sqrt{N} D^2 \zett_{t_k} (\zet_k, \zeta_0, \sigma , \bzd )
  \, ,
  \\
  \xi^{(1)}_k (\sigma)
  \!\!&=&\!\!
  \int
   \Xi_k^{(1)}(\sigma, \bzu)
    \nu[{\bzN_0}](\dd \bzu )
  \, ,
  \\
\xi^{(2)}_k (\sigma)
  \!\!&=&\!\!
  \iint
   \Xi_k^{(2)}(\sigma, \bzd)
    \nu[{\bzN_0}]^{\times 2} (\dd \bzd ) 
  \, ,
\end{eqnarray}
and group the corresponding $K$ components into the vectors
$\vec{\Xi}^{(1)}$, $\vec{\Xi}^{(2)}$, $\vec{\xi}^{(1)}$
and $\vec{\xi}^{(2)}$. 
 We also define
\beq
  \tilde \xi_{kj} (\sigma)
=
\frac{1}{\sqrt{N}}\left( \Xi_k^{(1)}(\sigma ,  z^{\supN}_j(0)) 
-   \int \Xi_k^{(1)}(\sigma, \bzu)\sigma  (\dd \bzu )\right)
\eeq
where $z^{\supN}_j(0)$ is the $j$-th component of $\bzN_0$, and note that
\begin{equation}
  \xi^{(1)}_k
  =
   {\textstyle{\frac{1}{\sqrt{N}} }}\sum_{j=1}^{N} \tilde \xi_{kj}
\end{equation}
and that
\begin{eqnarray}
  {\EE}_{\mu_0^{\times N}} (\tilde \xi_{kj})
  \!\!\!&=&\!\!\!
  0  \, ,
  \\
  {\EE}_{\mu_0^{\times N}} ( \tilde \xi_{k j} \tilde \xi_{k' j'} )
  \!\!\!&=&\!\!\!
  Q_{k,k'} \delta_{j,j'}  \, ,
\end{eqnarray}
where $\delta_{j,j'}$ is the Kronecker symbol. Hence the central
limit theorem applies to $\xi^{(1)}_k$.

 We may now write 
\begin{equation}
  e^{{\rm i} 
     \langle\VZet | \vec{\Delta_\zett}\rangle}
  =
  e^{{\rm i} 
 \langle \VZet | \vec{\xi}^{(1)}\rangle} 
+
  \int_{0}^{1}  {\textstyle{\frac{\dd}{\dd s}}}
    \exp\left({\rm i}   
  \int_{0}^{1} \langle \VZet |
		\vec{\xi}^{(1)} (\mu_0 + r s \nu[{\bzN_0}])\rangle
%
%
       \dd{r}\right) \dd s
  \, .
\label{conto}
\end{equation}
  When $N \to \infty$, the expectation of the first term in the
right-hand side converges to the right-hand side of \refeq{clt}.

  The integrand in the second term becomes, after differentiation \wrt $s$,
\begin{eqnarray}
 &&{\rm i} 
 \int_{0}^{1} \langle\VZet | \vec{\xi}^{(2)}(\mu_0 + rs \nu[{\bzN_0}]) \rangle
       r \dd{r}
\exp\left({\rm i} 
 \int_{0}^{1} \langle\VZet |\vec{\xi}^{(1)}(\mu_0 + r' s \nu[{\bzN_0}])\rangle
       \dd r'\right)
  \cr
  \!\!&=&\!\!
   {\textstyle{\frac{\rm i}{N \sqrt{N}}}} \sum_{i,j=1}^{N}
  \Bigl[ \cG(z^{(N)}_i (0) , z^{(N)}_j (0),\veps[{\bzN_0}] )
         - \int \cG(z^{(N)}_i (0), \bzu ,\veps[{\bzN_0}] ) \mu_0(\dd \bzu)
  \cr
  &&     - \int \cG(\bzu ,z^{(N)}_j (0),\veps[{\bzN_0}] ) \mu_0(\dd \bzu)
         + \iint \cG(\bzd,\veps[{\bzN_0}] )
           \mu_0^{\times 2}(\dd \bzd )
  \Bigr]
  \cr
  &&  \exp\left(\frac{\rm i}{\sqrt{N}} \sum_{i'=1}^{N}
    \left[\cF (z^{\supN}_{i'}(0), \veps[{\bzN_0}] )
     - \int \cF (\bzuu , \veps[{\bzN_0}] ) \mu_0(\dd \bzuu )\right]\right)
 \label{puffo}
\end{eqnarray}
where (for given ${\VZet}$, ${\Vzet}$, ${\Vt}$, $\zeta_0, \mu_0, s$)
\begin{eqnarray}
  \sqrt{N} \cF (\bzu , \sigma_0)
  \!\!\!&=&\!\!\!
 \int_0^1 
\langle \VZet | \VXi^{(1)}(\mu_0 + r s(\sigma_0 - \mu_0), \bzu) \rangle
       \dd{r}
     \, ,
  \label{defcF}
  \\
  \sqrt{N} \cG (\bzd, \sigma_0 )
  \!\!\!&=&\!\!\!
     \int_0^1\langle \VZet
        \VXi^{(2)}(\mu_0 + r s (\sigma_0 - \mu_0), \bzd )\rangle
       r \dd{r}
  \, .
  \label{defcG}
\end{eqnarray}
 The regularization ensures that $\cG$ and $\cF$ are differentiable
to any order with respect to $\sigma_0$, with bounded
derivatives in the sense of Proposition~\ref{bd}. 
  We may then evaluate the size of the expectation of \refeq{puffo} in the
following way.

Expression \refeq{puffo} can be split in two components~: the
``diagonal'' part, which is obtained from the terms in the sum
such that $i=j$, gives with trivial estimates a contribution of
order $N^{-1/2}$ to the expectation of \refeq{puffo}~; an estimate
of the size of the ``non-diagonal'' component needs some more
manipulation.

Consider the measure (positive with mass $1-2/N$)
\begin{equation}
  \mu^{ij}[{\bzN_0}]
  =
   {\textstyle{\frac{1}{N}}}
 \sum_{\stackrel{k=1}{k \neq i,j}}^{N} \delta_{z^{(N)}_k (0)}
  =
  \veps[{\bzN_0}]
  -
   {\textstyle{\frac{1}{N}}}(\delta_{z^{(N)}_i (0)} + \delta_{z^{(N)}_j (0)})
  \, .
\end{equation}
Given the result in Proposition~\ref{bd}, we may write
\begin{eqnarray*}
  \cF(\bzu ,\veps[{\bzN_0}])
  \!\!\!&=&\!\!\!
  \cF (\bzu,\mu^{ij}[{\bzN_0}])
    +  {\textstyle{\frac{1}{N}}}
 {\textstyle{\sum\limits_{k \in \{i,j\} }}}
 D^1 \cF (\bzu ,\mu^{ij}[{\bzN_0}], z^{(N)}_k (0))
    + {\mathrm{O}} ({\textstyle{\frac{1}{N^2}}})
  \, ,
  \\
  \cG (\bzd,\veps[{\bzN_0}])
  \!\!\!&=&\!\!\!
  \cG (\bzd,\mu^{ij}[{\bzN_0}])
    +  {\textstyle{\frac{1}{N}}}
{\textstyle{\sum\limits_{k \in \{i,j\} }}}
 D^1 \cG (\bzud,\mu^{ij}[{\bzN_0}],z^{(N)}_k(0))
    + {\mathrm{O}}({\textstyle{\frac{1}{N^2}}})
  \, .
\end{eqnarray*}

  For a given pair $\{i,j\}$, we then obtain 
\begin{eqnarray}
  &&\hskip-1truecm
 \exp\left(\frac{\rm i}{\sqrt{N}} \sum_{n=1}^{N}
    \left[\cF(z^{(N)}_n (0) ,\veps[{\bzN_0}])
     -\int \cF(\bzu ,\veps[{\bzN_0}] ) \mu_0(\dd \bzu)\right]\right)
  \cr
  &&=
  \left[1 +  {\textstyle{\frac{\rm i}{\sqrt{N}}}}
 \left[\cF_1(\hat{z}^{(N)}_i (0) ) + \cF_1(\hat{z}^{(N)}_j (0))\right]
     + {\mathrm{O}}\left({\textstyle{\frac{1}{N}}}\right)\right]
  \cF_2(\hat{z}^{(N)}_i (0),\hat{z}^{(N)}_j (0))
\end{eqnarray}
where we denote by $g(\hat{x})$ a function such that $\partial_x g = 0$, 
which is bounded according to Proposition~\ref{bd} and need
not be further specified. 
  Moreover,
\begin{eqnarray}
\hskip-1truecm
  \cG (z^{(N)}_i (0), z^{(N)}_j (0),\veps[{\bzN_0}])
  \!\!\!& - &\!\!\!
 \int \cG (z^{(N)}_i (0),\bzuu,\veps[{\bzN_0}]) \mu_0(\dd \bzuu)
   \cr
  \!\!\!& - &\!\!\!
 \int \cG (\bzu , z^{(N)}_j (0),\veps[{\bzN_0}]) \mu_0 (\dd \bzu)
\cr
  \!\!\!& + &\!\!\!
 \iint \cG (\bzd,\veps[{\bzN_0}])
     \mu_0^{\times 2} (\dd \bzd)
=
  \cD_{ij} + {\mathrm{O}}\left({\textstyle{\frac{1}{N}}}\right)
\end{eqnarray}
where
\begin{eqnarray*}
  \cD_{ij}
 =
  \cG (z^{(N)}_i (0) ,z^{(N)}_j (0) ,\mu^{ij}[{\bzN_0}])
  \!\!\!&-&\!\!\!
 \int \cG (z^{(N)}_i (0), \bzuu ,\mu^{ij}[{\bzN_0}]) \mu_0(\dd \bzuu)
   \cr
  \!\!\!&-&\!\!\!
 \int \cG (\bzu ,z^{(N)}_j (0),\mu^{ij}[{\bzN_0}]) \mu_0 (\dd \bzu)
\cr
  \!\!\!&+&\!\!\! 
\iint \cG (\bzd,\mu^{ij}[{\bzN_0}])] \mu_0^{\times 2} (\dd \bzd)
\end{eqnarray*}
so that
\begin{eqnarray}
  {\EE}_{\mu_0^{\times N}}[\cD_{ij}]
  \!\!\!&=&\!\!\!
  0  \, ,
  \\
  {\EE}_{\mu_0^{\times N}}[\cD_{ij} g(\hat{z}^{(N)}_k (0))]
  \!\!\!&=&\!\!\!
  0  \, ,
\end{eqnarray}
for any bounded function $g(\hat{z}^{(N)}_k (0) )$ with $k \in \{i,j\}$. The
expectation of the non-diagonal component of \refeq{puffo} is then
given by
\begin{eqnarray}
 {\textstyle{ \frac{\rm i}{N\sqrt{N}}}}
 \sum_{\stackrel{i,j=1}{i \neq j}}^{N} \int
    \left[\cD_{ij} + {\mathrm{O}}\left({\textstyle{\frac{1}{N}}}\right)\right]
    \left[1
    +  {\textstyle{\frac{\rm i}{\sqrt{N}}}}
\left(\cF_1(\hat{z}^{(N)}_i (0)) + \cF_1(\hat{z}^{(N)}_j (0))\right)
    + {\mathrm{O}}\left({\textstyle{\frac{1}{N}}}\right)\right]
\times&&\cr  
\times \cF_2\left(\hat{z}^{(N)}_i (0),\hat{z}^{(N)}_j (0)\right)
\mu_0^{\times N}(\dd \bzN_0 )
=
{\textstyle{\frac{N(N-1)}{N\sqrt{N}}}} 
{\mathrm{O}}\left({\textstyle{\frac{1}{N}}}\right)&&
\end{eqnarray}
so that the second term in \refeq{conto} is
${\mathrm{O}}(N^{-1/2})$.

The identification of the limit stochastic process is obtained
from
\refeq{deriv1}-\refeq{deriv2}-\refeq{deriv4}-\refeq{deriv5}.
\QED

 \newpage
\noindent \textbf{Acknowledgement}
The work of M.K. was supported in parts by the National Science
Foundation under Grant No. DMS-0103808, and
in parts by CNRS through a poste rouge to M.K. while visiting
CNRS-Universit\'e de Provence;
the work of Y.E. by Universit\'e de
Provence through a cong\'e pour recherche, 
the work of V.R., when the collaboration started,
by the Foundation BLANCEFLOR Boncompagni Ludovisi n\'ee Bildt. 
The participation by A. Nouri in the early stages of 
this work is gratefully acknowledged.
M.K. thanks  C. Lancellotti, M. Kunze, 
and H. Spohn for valuable discussions. 

\newpage


\appendix 
\centerline{\textbf{Appendix}}
\resetseq

\subsection*{A.1\ \  Nested modes of convergence of probability measures}
\label{secBLmetric} 

 A certain frustration about the absence of an authoritative 
survey of the relationships of various important notions of 
convergence that are
used in the probability literature has already been expressed 
   \cite{GibbsSu},
where that gap has been filled to some extent.
  Unfortunately,
   \cite{GibbsSu}
does not cover all our needs.
 Furthermore, when addressing a mixed readership of mathematical physicists, 
analysts and probabilists, the frustration can get compounded by the various 
`competing' terminologies and notations that are in use in these areas of 
activity.
 In view of this, it seems advisable to be more explicit about 
how the notions of convergence that we use.
 The following general notions hold (and are formulated) for any dimension 
$d\geq 1$.

 We recall that, if $\{\mu_n\}_{n\in\NN}$ is a sequence of Borel probability 
measures on $\RR^d$ and $\mu \in P(\RR^d)$, too, and if 
$\int f \dd\mu_n \to\int f \dd\mu$ for every bounded continuous function 
$f\in C^0_b(\RR^d)$, then one says that $\mu_n$ converges to $\mu$ 
\textit{in law},\footnote{In the probability literature, 
                          convergence in law is usually called 
		   ``weak convergence'' of probability measures; 
		     however, this notion generally differs from the 
		     analysts' notion of weak convergence on $M$.}
written $\mu_n\inlawto\mu$; see  p. 292 of
               \cite{Dudley}.
 Clearly, since $C^0_0(\RR^d)\subset C^0_b(\RR^d)$, 
convergence in law $\mu_n\inlawto\mu$
implies vague convergence $\mu_n \rightharpoonup\mu$.  
 Moreover, convergence in law $\mu_n\inlawto\mu$ is equivalent to 
\textit{convergence in probability} of the underlying family of 
random variables having laws $\mu_n$ to a random variable with law $\mu$, 
a notion we need for our law of large numbers.

 Convergence in law can be metrized as follows.
 Let $C_{b}^{0,\alpha}(\RR^d)$ denote the subset of the bounded continuous 
functions on $\RR^d$ which are also H\"older continuous with exponent 
$\alpha \in (0,1]$.
 Now $C_{b}^{0,\alpha}(\RR^d)$ is not a closed subspace of $C_{b}^0(\RR^d)$ 
w.r.t. $\norm{\,.\,}_{\mathrm{u}}$, but 
\begin{equation}
  \norm{g}_{{\mathrm{u}},\alpha}
  \equiv
  \max\left\{\norm{g}_{\mathrm{u}}\,,\, \HalphaSnorm{g} \right\} \,,
\end{equation}
where
\beq
\HalphaSnorm{g} \equiv
    \sup_{\xi \neq \xi' \in \RR^d}
           \frac{\abs{g(\xi)-g(\xi')}}{\abs{\xi-\xi'}^\alpha}
\eeq
is the $\alpha$-H\"older seminorm of $g$, 
turns $C_{b}^{0,\alpha}(\RR^d)$ into a (non-separable) Banach space.
  The positive cone in $C_{b}^{0,\alpha}(\RR^d)$ is denoted by
$C_{{b},+}^{0,\alpha}(\RR^d)$.
 If the suffix $_{b}$ is replaced by the suffix $_0$, 
we mean the corresponding subsets of these functions that 
vanish at infinity.
 In much of what follows, we will need $C_{b}^{0,1}(\RR^d)$,
the space of bounded Lipschitz functions on $\RR^d$, and we 
write\footnote{Since
              $\LipSnorm{\,.\,}$ is a seminorm, we 
              prefer this notation over $\norm{\,.\,}_{\mathrm{L}}$, 
	     which is also in use in the literature.}
$\LipSnorm{g}$ for $\HalphaSnorm{g}$ when 
$\alpha=1$.\footnote{We recall that if $g\in C^1(\RR^d)$, then 
	$\LipSnorm{g} = \sup_{x\in\RR^d} |\nabla g(x)|$.}

 Now let $\mu_1\in P(\RR^d)$ and  $\mu_2\in P(\RR^d)$ be two Borel 
probability measures on $\RR^d$. 
 We define the dual bounded-Lipschitz distance between $\mu_1$ and $\mu_2$ 
as\footnote{The * at $\dbL{\,,\,}$ refers to the Kantorovich--Rubinstein 
duality theorems; see below.}
  \begin{equation}
    \dbL{\mu_1,\mu_2}
    :=
    \sup_{g \in C_{{b}}^{0,1}(\RR^d)}
   \left\{\Abs{ \int g\, \dd(\mu_1-\mu_2)} : \norm{g}_{{\mathrm{u}},1}\leq 1
      \right\}\,.
  \label{defDISTbL}
  \end{equation}
 Our dual bounded-Lipschitz distance, though not identical to, is equivalent 
to the Fortet--Mourier $\beta$-distance (p.395 of 
               \cite{Dudley}), 
which instead of $\norm{g}_{{\mathrm{u}},1}\leq 1$ works with the equivalent
condition $\norm{g}_{{\mathrm{u}}} + \LipSnorm{g}\leq 1$. 
 Therefore, by Proposition 11.3.2 of 
               \cite{Dudley}, 
$\dbL{\,,\,}$ is a metric on the convex set $P(\RR^d)$, 
and by Corollary 11.5.5 of 
               \cite{Dudley}, 
$P(\RR^d)$ is complete for $\dbL{\,,\,}$. 
 Furthermore, by Theorem 11.3.3 of 
               \cite{Dudley}, 
if $\{\mu_n\}_{n\in\NN}$ is a sequence of Borel probability measures on 
$\RR^d$, and $\mu \in P(\RR^d)$, too, then
$\dbL{\mu_n,\mu}\to{0}$ as $n\to\infty$ is equivalent to 
$\mu_n\inlawto\mu$ as $n\to\infty$.
 Hence, $\dbL{\,,\,}$ metrizes convergence in law of the Borel probability 
measure on $\RR^d$.

 Our dual bounded-Lipschitz distance $\dbL{\,,\,}$ 
 is equivalent, but not identical, to the distance obtained
 by restricting $g$ to $C_{b,+}^{0,\alpha}(\RR^d)$, here 
 denoted $\dd_\bLd(\,,\,)$
 (following [Spo91], Def. 2.2; actually, Spohn writes $\dd_\bL(\,,\,)$,
 but we here better keep the *).
 Clearly, $\dbL{\mu_n,\mu}\to 0$ implies $\dd_\bLd(\mu_n,\mu)\to 0$. 
 The converse of this follows from three simple observations: first, 
 the integral on the r.h.s. of \refeq{defDISTbL} 
 is invariant under $g\to g + \norm{g}_{{\mathrm{u}}}$, so that in our 
 definition of 
 $\dbL{\,,\,}$ we can  replace $C_{b}^{0,1}(\RR^d)$ by $C_{b,+}^{0,1}(\RR^d)$ 
 and simultaneously replace the condition $\norm{g}_{{\mathrm{u}},1}\leq 1$ 
 with the condition 
 $\max\{\frac{1}{2}\norm{g}_{{\mathrm{u}}}, \LipSnorm{g}\}\leq 1$;  second, 
 $\{g\in C_{b,+}^{0,1}(\RR^d): 
  \norm{g}_{{\mathrm{u}}}\leq 2, \LipSnorm{g}\leq 1\}$
 is a strict subset of 
$\{g\in C_{b,+}^{0,1}(\RR^d):
  \norm{g}_{{\mathrm{u}}}\leq 2,\LipSnorm{g}\leq 2\}$;
 third, the simple scaling $g\to 2g$ reveals that the sup of 
 $\Abs{ \int g\,\dd(\mu_1-\mu_2)}$ over 
 $\{g\in C_{b,+}^{0,1}(\RR^d): 
   \norm{g}_{{\mathrm{u}}}\leq 2, \LipSnorm{g}\leq 2\}$
 is twice the sup of 
 $\Abs{ \int g\, \dd(\mu_1-\mu_2)}$ over 
 $\{g\in C_{b,+}^{0,1}(\RR^d): 
  \norm{g}_{{\mathrm{u}}}\leq 1, \LipSnorm{g}\leq 1\}$.
 These three facts together imply that 
 $\dbL{\mu_1,\mu_2}\leq 2\dd_\bLd(\mu_1,\mu_2)$, and this means that
 $\dbL{\mu_n,\mu}\to 0$ whenever $\dd_\bLd(\mu_n,\mu)\to 0$.

 Recall that the general Kantorovich--Rubinstein distance\footnote{Also 
                    associated with the names of Monge and Wasserstein.}
is defined as 
\begin{equation}
 \dKRc{\mu_1,\mu_2}
 :=
\inf_{\mu\in P_c(\RR^{2d}|\mu_1,\mu_2)}
    \left\{
	\int {\mathrm{cost}}( \xi_1,\xi_2) \mu(\dd \xi_1\dd\xi_2) 
    \right\}\,,
\label{defKRdistGENERAL}
\end{equation}
 where 
${\mathrm{cost}}( \xi, \xi^\prime) = \dKRc{\delta_{\xi}, \delta_{\xi^\prime}}$
 for $\xi, \xi^\prime \in {\RR^d}$
 is the ``cost (per transport unit) function,'' and where
 $P_c(\RR^{2d}|\mu_1,\mu_2)$ is the set of 
 Borel probability measures $\mu$ on $\RR^d\times\RR^d$ satisfying 
 $\mu(\dd\xi_1\times\RR^d)=\mu_1(\dd\xi_1)$ and 
$\mu(\RR^d\times\dd\xi_2)=\mu_2 (\dd\xi_2)$, 
 with $\mu_1$ and $\mu_2$ satisfying 
 $\int\!{\mathrm{cost}}( \xi_1,\xi) \mu_1(\dd\xi_1)<\infty$ and
 $\int\!{\mathrm{cost}}( \xi,\xi_2) \mu_2(\dd\xi_2)<\infty$ for 
some $\xi\in\RR^d$.

\hskip-.7pt
   By the Kantorovich--Rubinstein theorem 
          (\cite{Dudley}, Theorem 11.8.2),
 $\dbL{\mu_1,\mu_2}$ is identical to the Kantorovich--Rubinstein distance for
 ${\mathrm{cost}}( \xi_1,\xi_2) = \min \{ 2, |\xi_1-\xi_2| \}$.
   Incidentally, ${\mathrm{cost}}( \xi_1,\xi_2) = \min \{1, |\xi_1-\xi_2| \}$
   is the 
cost function for the particular Kantorovich--Rubinstein distance identical 
to $\dd_\bLd(\,,\,)$.
 The dual bounded-Lipschitz distance ($\dd_\bLd$) is used in 
 \cite{NeunzertA, BraunHepp, NeunzertB, SpohnBOOKa}
 and 
     \cite{FirpoElskens}.

 However, if one is only interested, as we are, in the subset 
${P_1}(\RR^d)\subset P(\RR^d)$,
it is rather prudent to work with the dual Lipschitz distance in 
${P_1}(\RR^d)$, given by
  \begin{equation}
    \dLip{\mu_1,\mu_2}
    :=
    \sup_{g \in C^{0,1}(\RR^d)}
      \left\{
\Abs{ \int g\, \dd(\mu_1-\mu_2)} : \LipSnorm{g}\leq 1
      \right\}\,,
  \label{defLdistAPP}
  \end{equation}
which is identical with the standard\footnote{The word
	 ``standard'' refers to the custom in the probability community that,
          by default, the cost function is identified with the metric of the 
   underlying complete metric space on which the Borel probability measures 
	  are defined; in standard Euclidean $\RR^d$ this gives 
	  ${\mathrm{cost}}( \xi_1,\xi_2)=\abs{\xi_1-\xi_2}$.}
Kantorovich--Rubinstein distance, given by
  \begin{equation}
    \dKR{\mu_1,\mu_2}
    :=
    \inf_{\mu\in {P_1}(\RR^{2d}|\mu_1,\mu_2)}
      \left\{
   \int \abs{ \xi_1 -\xi_2} \mu(\dd \xi_1\dd\xi_2) 
      \right\}\,,
  \label{defKRdistAPP}
  \end{equation}
where ${P_1}(\RR^{2d}|\mu_1,\mu_2)$ is the set of Borel probability 
measures $\mu$ on $\RR^d\times\RR^d$ satisfying 
$\mu(\dd\xi_1\times\RR^d)=\mu_1(\dd\xi_1)\in{P_1}(\RR^d)$
and $\mu(\RR^d\times\dd\xi_2)=\mu_2 (\dd\xi_2)\in{P_1}(\RR^d)$.
 We write $\mu_n\leadsto\mu$ if $\dLip{\mu_n,\mu}\to 0$.
 Clearly, $\dLip{\mu_n,\mu}\to 0$ implies\footnote{The converse is not true.
          In particular, Dudley gives the following counterexample for $d=1$:
	 $\mu_n = (1- n^{-1})\delta_0 +n^{-1}\delta_n$ and $\mu=\delta_0$, 
	for which 
 $\dLip{\mu_n, \mu} = 1$ while $\dbL{\mu_n, \mu} \leq 2n^{-1}\downarrow{0}$.}
$\dbL{\mu_n,\mu}\to 0$.

 We note that the metric $\dLip{\,.\,,\,.\,}$ 
defines a norm $\dLIPnorm{\,.\,}$ on $(P_1 - P_1)\subset M$ 
by\footnote{In particular, if $\sigma = \mu_1 - \mu_2$ with 
		$\mu_1,\mu_2\in  P_1$, then 
	$\dLip{\sigma_+,\sigma_-} = \dLip{\mu_1,\mu_2}$;
 	note, however, that generally $\mu_1\neq (\mu_1-\mu_2)_+$ and
	$\mu_2\neq (\mu_1-\mu_2)_-$.}
$\dLIPnorm{\sigma}:=\dLip{\sigma_+,\sigma_-}$.
 This definition extends identically to $\lambda (P_1 - P_1)$ for any 
$\lambda\in \RR$. 
 To extend $\dLIPnorm{\,.\,}$ to the linear span of $P_1$
for $\sigma\in$ lsp $P_1$ we define 
\beq
 \dLIPnormT{\sigma} := 
\dLip{(\sigma - \sigma(\RR^d)\tilde\mu)_+,(\sigma - \sigma(\RR^d)\tilde\mu)_-}
 + \abs{\sigma(\RR^d)}
\label{extLIPnorm}
\eeq
where $\tilde\mu\in P_1(\RR^d)$ is arbitrary but fixed; 
e.g. $\tilde\mu = \delta_0$.
 Clearly, for $\sigma\in P_1-P_1$, such that $\sigma(\RR^d)=0$, 
\refeq{extLIPnorm} reduces to $\dLIPnormT{\sigma}=\dLip{\sigma_+,\sigma_-}$,
i.e. $\dLIPnormT{\sigma}= \dLIPnorm{\sigma}$ whenever $\sigma(\RR^d)=0$.
 It is straightforward to verify that $\dLIPnormT{\,.\,}$ 
is a norm on lsp $P_1$. 
  The completion of the linear span of 
${P_1}(\RR^d)$ w.r.t. \refeq{extLIPnorm}, denoted $\wM(\RR^d)$, 
is a Banach space with norm  $\dLIPnormT{\,.\,}$ given in 
\refeq{extLIPnorm}.
  We write $\wPpx(\RR^d)$ for ${P_1}(\RR^d)\hookrightarrow\wM(\RR^d)$.

\newpage
\subsection*{A.2\ \ The second order variant of the Gronwall lemma}
\label{secGRONWALL} 

 The standard Gronwall lemma provides a simple upper bound on a function 
$t\mapsto u(t)$ satisfying the first order differential inequality 
\beq
\ddt{u} \leq f(t) u + g(t)
\label{firstOineq}
\eeq
for all $t\in \RR_+$, with $u(0)=u_0>0$, and with $f(t)$ and $g(t)$ given
positive continuous functions; namely, with the help of an integrating 
factor one finds right away that $u$ is bounded by
\beq
 u(t) \leq 
u_0\exp\left(\int_0^t f(\tau)\dd\tau\right)
+
\int_0^t \exp\left(\int_{\tau}^t f(\tilde\tau)\dd\tilde\tau\right)
g(\tau)\dd\tau
\,.
\label{GWLfirst}
\eeq
 In particular, if $f(t)\equiv \gamma >0$ is a constant, then
\beq
 u(t) \leq 
u_0\exp(\gamma t)
+
\int_0^t \exp[\gamma(t-\tau)]g(\tau)\dd\tau
\,.
\label{GWLfirstSPECIAL}
\eeq

 However,  \refeq{GWLfirst} does not suit our purposes; instead, we
need the following second order variant of \refeq{GWLfirst}:

\smallskip
\noindent
{\sc Lemma} A1: 
{\textit{Let $\gamma >0$ be a given constant and $g(t)$ a given 
positive continuous function.
 Suppose $t\mapsto u(t)$ satisfies the second order differential inequality 
\beq
\ddtsq{u} \leq \gamma^2 u + g(t)
\label{secondOineq}
\eeq
for all $t\in \RR_+$, with $u(0)=u_0\geq 0$ and $u^\prime(0)=v_0\geq 0$.
 Then  $u$ is bounded by
\beq
 u(t) \leq 
u_0 \cosh({\gamma}t) + 
v_0{\textstyle{\frac{1}{\gamma}}} \sinh({\gamma}t) + 
\int_0^t \cosh[\gamma(t-\tau)]\int_0^\tau\! g(\tilde\tau)\dd\tilde\tau\,
\dd\tau
\,
\label{GWLsecond}
\eeq
for all $t\in \RR_+$.}}

\noindent
\textit{Proof of Lemma A.1:}
 Denote r.h.s.\refeq{GWLsecond} $=U(t)=U_{hom}(t)+U_{inh}(t)$, 
where $U_{inh}(t)$ is the term linear in $g$.
  By direct computation one verifies that the function 
$t\mapsto U(t)$ satisfies \refeq{secondOineq}
with ``$=$'' instead of $\leq$, and $U(0)=u_0$ and $U^\prime(0)=v_0$. 
 Since the Cauchy problem for \refeq{secondOineq} with positive data
has a unique positive solution, it follows that $u(t)\leq U(t)$
by the usual subsolution argument.\QED

\bigskip 
\bigskip 
\bigskip 
\bigskip 

\newpage

\subsection*{A.3\ \  Proof of the conservation laws}
\label{AppConsLaws}

  We prove the  conservation laws for the regularized wave gravity 
Vlasov equations.
        The laws for the microscopic regularized field \&\ $N$-body
systems are included as a special case.
        For general background material on conservation laws,  see
\cite{SudarshanBOOK}.
\smallskip

\noindent 
\textit{Proof of Proposition \ref{propClawsVeps}.}
  The conservation of $\cC^{(\alpha)}(\Zett)$ holds because
\refeq{VfEQeps} is a continuity equation in $\RR^6$, and 
because a Hamiltonian vector field is divergence-free.\QED
\smallskip

\noindent 
\textit{Proof of the conservation laws of Theorems
\ref{theoVAGUEsolVeps} and \ref{theoKRsolVeps}.} 
  As to the conservation of $\cE(\Zett)$, for the time derivative of the 
matter energy (\ie kinetic plus rest) we have
\bea
\ddt \iint\!
 \sqrt{1+ |p|^2} f(x,p,t)\, \dd{x}\,\dd{p}\,
\!\!\!&=&\!\!\!
\iint\!  \sqrt{1+ |p|^2} (\fe *\nabla\psi) (x,t)\cdot \pdp
f(x,p,t) \,\,\dd{p}\,\dd{x} \nonumber
\\
\!\!\!&=&\!\!\!
-
\iint\!  (\fe *\nabla\psi) (x,t)\cdot v  f(x,p,t)
\,\dd{p}\,\dd{x} 
\,, 
\label{matterEdot} 
\eea 
where we first pulled the time derivative into the integral,
then used \refeq{VfEQeps} to rewrite the integrand, noted that $x$ 
divergences integrate to zero, then integrated by parts \wrt $p$, 
and finally used that $\pdp \sqrt{1+|p|^2} = v$ is the velocity of 
a particle with unit mass, having momentum $p$.
        On the other hand, for the wave field energy, we have
\bea
\ddt
\frac{1}{2}
\int\! \left(\abs{\pdx\psi}^2 + |\varpi|^2\right)\!(x,t)
\dd{x}
\!\!\!&=&\!\!\!
\int\! \left(-\pdxsq\psi + \pdt\varpi\right)\!(x,t)
\varpi\!(x,t)
\dd{x}
\nonumber
\\
\!\!\!&=&\!\!\!
- \int\!
        \left(\fe *\int{f}(\,.\,,p,t) \,\dd p\right)\!(x)
\,\varpi(x,t) \,\dd{x}
\nonumber
\\
\!\!\!&=&\!\!\!
- \iint\! f(x,p,t)(\fe *\varpi) (x,t)
\,\dd{x}\,\dd{p}\, \,, \label{fieldEdot}
\eea
where we pulled the time derivative into the integral, 
used \refeq{VvpiEQeps} to rewrite the integrand, and invoked Fubini.
        Finally, for the regularized coupling energy, we have
\bea
\ddt \iint\!
 (\fe * \psi) (x,t)  f(x,p,t)
\dd{x}\,\dd{p}\,
\!\!\!&=&\!\!\!
\iint\!  (\fe *\varpi) (x,t) f(x,p,t)
\dd{x}\,\dd{p}\,
\nonumber
\\
&&\!\!+\!
\iint\!
 (\fe * \psi) (x,t)\, \pdt f(x,p,t)
\dd{x}\,\dd{p}\,
\,.
\label{coupleEdot}
\eea
   The last expression in  \refeq{fieldEdot}
cancels against the first term on r.h.s.\refeq{coupleEdot}.
        It remains to show that the second term on 
r.h.s.\refeq{coupleEdot} cancels against the final expression in
\refeq{matterEdot}.
        We use \refeq{VfEQeps}
to rewrite the integrand of the second term on r.h.s.\refeq{coupleEdot}, 
note that $p$ divergences integrate to zero, 
then invoke Fubini and integrations by parts.
 Thus
\bea
\iint\!
 (\fe * \psi) (x,t) \pdt f(x,p,t)
\dd{x}\,\dd{p}\,
\!\!\!&=&\!\!\!
- \int\!
\left(
\fe * \int\! v\cdot\nabla f(\,.\,,p,t)\,\dd{p}\right)\!(x)
\,\psi (x,t)
\,\dd{x}
\nonumber
\\
\!\!\!&=&\!\!\!
-
\iint\! v\cdot\pdx f(x,p,t)\,\dd{p}\,
 \left(\fe * \psi\right) (x,t) \,\dd{x}
\nonumber
\\
\!\!\!&=&\!\!\!
- \int\!
\left(\pdx\!\cdot\!\int\! v f(x,p,t)\,\dd{p} \right)\,
 (\fe * \psi) (x,t) \,\dd{x}
\nonumber
\\
\!\!\!&=&\!\!\!
\iint\! v f(x,p,t)\,\dd{p}
\cdot\pdx (\fe * \psi) (x,t) \,\dd{x}
\nonumber
\\
\!\!\!&=&\!\!\!
\iint\! v f(x,p,t)\,\dd{p}
\cdot (\fe * \nabla\psi) (x,t) \,\dd{x}
\,.
\eea
  Thus conservation of the energy $\cE$ is proved.

  The conservation of $\cP(\Zett)$ is shown similarly.
  For the matter momentum, we have
\bea
\ddt \iint\!
  p f(x,p,t) \,\dd{p}\,\dd{x}
\!\!\!&=&\!\!\!
\iint\! p (\fe *\nabla\psi) (x,t)\cdot \pdp  f(x,p,t)
\,\dd{p}\,\dd{x}
\nonumber
\\
\!\!\!&=&\!\!\!
-
\iint\!  (\fe *\nabla\psi) (x,t) f(x,p,t)
\,\dd{p}\,\dd{x}
\,,
\label{matterPdot}
\eea
the last step through integration by parts, using the identity
$(\pdx{u}(x)\cdot\pdp)p =\pdx{u}(x)$.
        On the other hand, for the field momentum we have
\bea
\ddt\left(
 -\!\!\int\!\pdx\psi(x,t)\varpi (x,t) \dd{x}\right)
\!\!\!&=&\!\!\!
- \int\!  \pdt\varpi(x,t) \pdx\psi\!(x,t) \dd{x}
\nonumber
\\
\!\!\!&=&\!\!\!
- \int\! \left( \pdt\varpi
- \pdxsq\psi \right)(x,t) \pdx\psi\!(x,t) \dd{x}
\nonumber
\\
\!\!\!&=&\!\!\!
 \int\!
\left(\fe *\int{f}(\,.\,,p,t) \dd p\right)\!(x)
\,\pdx\psi(x,t) \,\dd{x}
\nonumber
\\
\!\!\!&=&\!\!\!
 \iint\! f(x,p,t)(\fe *\nabla\psi) (x,t) \,\dd{p}\,\dd{x}
\,,
\label{fieldPdot}
\eea
where we used the identity
$ 2 \varpi \pdx{\varpi} = \pdx \left(|\varpi|^2\right)$
in the first step, and the identity
\beq
 \pdx{\psi}\, \pdxsq{\psi}
=
\pdx\cdot\bigl( \pdx{\psi}\otimes \pdx{\psi}\bigr) -
 {\textstyle\frac{1}{2}} \pdx |\pdx{\psi}|^2
\eeq
in the second step, and 
noting the vanishing of ``surface integrals at infinity.'' 
    Adding \refeq{matterPdot} and \refeq{fieldPdot} we see that
total momentum $\cP$ is conserved.
\newpage

    As to the conservation of $\cJ(\Zett)$, 
for the orbital matter angular momentum we have
\begin{eqnarray}
  \ddt \iint  x \crprod p \, f(x,p,t)
    \,\dd{p}\,\dd{x}
\!\!\!&=&\!\!\!
  - \iint (x\crprod p)\,
      v \cdot \pdx  f(x,p,t) \,\dd{p}\,\dd{x}
  \cr
&&
\hskip-.2truecm
+\!
 \iint (x\crprod p)\,
      (\fe *\nabla\psi) (x,t)\cdot \pdp  f(x,p,t)
    \,\dd{p}\,\dd{x}
  \cr
\!\!\!&=&\!\!\!
  \iint (\fe *\nabla\psi) (x,t)\crprod x f(x,p,t)
    \,\dd{p}\,\dd{x}
  \cr
\!\!\!&=&\!\!\!
  \int\! \pdx\psi(x,t)\crprod{x}\,
    \left(\fe *\int{f}(\,.\,,p,t) \dd p\right)\!(x)  \,\dd{x}
  \cr
&&
\hskip-.2truecm
-\!
 \int\! \pdx \psi (x,t)\crprod\!
     \left(\fe \Id * \int\! f(\,.\,,p,t)\,\dd{p}\right)\!\!(x)
     \,\dd{x} \,.
\label{matterLdot}
\end{eqnarray}
        In the first step we used the continuity equation \refeq{VfEQeps}, 
in the second step integrations by parts and the identities
$(v\cdot\pdx) (x\crprod p) = v\crprod p =0$ and
$(\pdx{u}(x)\cdot\pdp)(x\crprod p) = x\crprod \pdx{u}(x)$;
the last step is Fubini and a trivial rewriting.
  The last integral in \refeq{matterLdot} gives
\begin{eqnarray} 
\hskip-.2truecm
  -\!\!\int\! \pdx \psi (x,t)\crprod\!\!
    \left(\!\fe \Id\! *\!\! \int\!\! f(\,.\,,p,t)\dd{p}\!\right)\!\!(x)
    \dd{x} 
=
  \! \iint\! (\fe \Id\! *\!\crprod\nabla\psi(\,.\,,t)) (x) f(x,p,t)
    \dd{p}\dd{x}
\label{matterSdot}
\end{eqnarray}
where again we used Fubini.
 Finally, by some standard identities of vector analysis and the radial
symmetry of $\fe$, the (negative of the) field torque evaluates to
\beq
\left(\fe\Id *\crprod\nabla\psi (\,.\,,t)\right)(x)
= 
\int\pdy\crprod \Bigl((y-x)\fe(y-x) \psi(y,t)\Bigr)\dd{y}
= 0
\,.
\eeq
 The last integral vanishes by one of Green's theorems and the 
compact support of $\fe$.

    Lastly, for the field angular momentum we have
\begin{eqnarray}
  \ddt \int (- x\crprod \pdx\psi(x,t) ) \varpi (x,t)
    \dd{x}
\!\!\!&=&\!\!\!
  - \int  (x\crprod \pdx\psi(x,t)) \pdt\varpi(x,t) \dd{x}
  \cr
\!\!\!&=&\!\!\!
  \int ( x\crprod\pdx\psi(x,t) )
    \left(\pdxsq\psi - \pdt\varpi\right)\!(x,t) \dd{x}
  \cr
\!\!\!&=&\!\!\!
  -\! \int\! \pdx\psi(x,t)\crprod{x}\!
  \left(\fe *\!\!\int\!\!{f}(\,.\,,p,t) \dd p\right)\!\!(x)\dd{x}
\label{fieldJdot}
\end{eqnarray}
where we used the identity
$2\varpi x\crprod  \pdx{\varpi}
 = \pdx \crprod \left(x \abs{\varpi}^2\right)$
in the first step, and the identity
\begin{equation}
  ( x \crprod \pdx{\psi} )\, \pdxsq{\psi}
  =
  \pdx \cdot \bigl( ( x\crprod \pdx{\psi})
    \otimes \pdx{\psi}\bigr)
  - \pdx \crprod \left( {\textstyle\frac{1}{2}}
      \, x \,\abs{\pdx{\psi}}^2\right) 
\label{fieldTORQUEnull}
\end{equation}
in the second, and  \refeq{VvpiEQeps} in the third,
noting the vanishing of ``surface integrals.''

        Adding \refeq{matterLdot} and \refeq{fieldJdot}, noting
\refeq{matterSdot}, \refeq{fieldTORQUEnull}, proves conservation
of $\cJ$.\QED


\newpage


%
\end{document}